\documentclass[utf8]{frontiersSCNS} 

\usepackage{url,hyperref,lineno,microtype,subcaption}

\usepackage[onehalfspacing]{setspace}

\def\keyFont{\fontsize{8}{11}\helveticabold }
\def\firstAuthorLast{Rivilla {et~al.}} 
\def\Authors{V\'ictor M. Rivilla\,$^{1,*}$, 
Juan Garc\'ia de la Concepci\'on\,$^{1}$,
Izaskun Jim\'enez-Serra\,$^{1}$,
Jes\'us Mart\'in-Pintado\,$^{1}$,
Laura Colzi\,$^{1}$,
Bel\'en Tercero\,$^{2}$,
Andr\'es Meg\'ias\,$^{1}$,
\'Alvaro L\'opez-Gallifa\,$^{1}$,
Antonio Mart\'inez-Henares\,$^{1}$,
Sara Massalkhi\,$^{1}$,
Sergio Mart\'in\,$^{4,5}$,
Shaoshan Zeng\,$^{3}$,
Pablo de Vicente\,$^{2}$,
Fernando Rico-Villas\,$^{1}$,
Miguel A. Requena-Torres\,$^{6,7}$,
Giuliana Cosentino\,$^{8}$
}
\def\Address{$^{1}$Centro de Astrobiolog\'{\i}a (CSIC-INTA), Ctra Ajalvir km 4, 28850, Torrej\'{o}n de Ardoz, Madrid, Spain
$^{2}$Observatorio Astronómico Nacional (OAN-IGN), Calle Alfonso XII, 3, 28014 Madrid, Spain
$^{3}$Star and Planet Formation Laboratory, Cluster for Pioneering Research, RIKEN, 2-1 Hirosawa, Wako, Saitama, 351-0198, Japa
$^{4}$European Southern Observatory, ALMA Department of Science, Alonso de Córdova 3107, Vitacura 763 0355, Santiago, Chile
$^{5}$Joint ALMA Observatory, Department of Science Operations, Alonso de Córdova 3107, Vitacura 763 0355, Santiago, Chile
$^{6}$University of Maryland, Department of Astronomy, College Park, ND 20742-2421, USA
$^{7}$Department of Physics, Astronomy and Geosciences, Towson University, Towson, MD 21252, USA
$^{8}$Space, Earth and Environment Department, Chalmers University of Technology, SE-412 96 Gothenburg, Sweden
}
\def\corrAuthor{Corresponding Author}

\def\corrEmail{vrivilla@cab.inta-csic.es}

\begin{document}
\onecolumn
\firstpage{1}

\title[Ionise hard: interstellar PO$^{+}$ detection]{Ionise hard: interstellar PO$^{+}$ detection} 


\author[\firstAuthorLast ]{\Authors} 
\address{} 
\correspondance{} 

\extraAuth{}

\maketitle

\begin{abstract}


We report the first detection of the phosphorus monoxide ion (PO$^{+}$) in the interstellar medium. Our unbiased and very sensitive spectral survey towards the G+0.693$-$0.027 molecular cloud covers four different rotational transitions of this molecule, two of which ($J$=1$-$0 and $J$=2$-$1) appear free of contamination from other species. The fit performed, assuming Local Thermodynamic Equilibrium conditions, yields a column density of $N$=(6.0$\pm$0.7)$\times$10$^{11}$ cm$^{-2}$. 
The resulting molecular abundance with respect to molecular hydrogen is 4.5$\times$10$^{-12}$. 
The column density of PO$^{+}$ normalised by the cosmic abundance of P is larger than those of NO$^{+}$ and SO$^{+}$, normalised by N and S, by factors of 3.6 and 2.3, respectively.
The $N$(PO$^{+}$)/$N$(PO) ratio is 0.12$\pm$0.03, more than one order of magnitude higher than those of $N$(SO$^{+}$)/$N$(SO)
and $N$(NO$^{+}$)/$N$(NO).
These results indicate that P is more efficiently ionised in the ISM than N and S. We have performed new chemical models that confirm that the PO$^+$ abundance is strongly enhanced in shocked regions with high values of cosmic-ray ionisation rates (10$^{-15}-$10$^{-14}$ s$^{-1}$), as occurs in the G+0.693$-$0.027 molecular cloud. The shocks sputter the interstellar icy grain mantles, releasing into the gas phase most of their P content, mainly in the form of PH$_3$, which is converted into atomic P, and then ionised efficiently by cosmic rays, forming P$^+$. Further reactions with O$_2$ and OH produce PO$^{+}$. 
The cosmic-ray ionisation of PO might also contribute significantly, which would explain the high $N$(PO$^{+}$)/$N$(PO) observed. The relatively high gas-phase abundance of PO$^{+}$ with respect to other P-bearing species stresses the relevance of this species in the interstellar chemistry of P.

\tiny
 \keyFont{ \section{Keywords:} Phosphorus, Interstellar: abundances, Interstellar: clouds, astrochemistry, interstellar: ions} 
\end{abstract}

\section{Introduction}

Phosphorus (P) is, along with carbon (C), hydrogen (H), oxygen (O), nitrogen (N),  and sulfur (S), one of the key elements for the development of life (CHONPS). This is because P-bearing compounds, and in particular phosphates (PO$_4^{3-}$), are unique in forming large biomolecules, such as deoxyribonucleic acid (DNA) and ribonucleic acid (RNA), phospholipids (the structural components of cellular membranes), and the adenosine triphosphate (ATP) molecule, which stores the chemical energy within cells. 
However, while the P abundance in living organisms is relatively high, e.g. P/H$\sim$10$^{-3}$ in bacteria (\citealt{fagerbakke1996}), the abundance of P in the Universe is several orders of magnitude lower, e.g. P/H$\sim$3$\times$10$^{-7}$ in the solar photosphere (\citealt{asplund2009}), much lower than those of other biogenic elements.
Due to this low cosmic availability of P, the detection of P-bearing molecules beyond our planet is challenging.

In our Solar System, phosphine (PH$_3$) has been observed in the atmospheres of Jupiter and Saturn (\citealt{Bregman1975,ridgway1976}). P has been identified in meteorites in the form of the mineral schreibersite (\citealt{pasek2005}) and phosphoric acids (\citealt{schwartz2006}). Recently, the measurements of the Rosetta spacecraft detected P in the comet 67P/Churyumov–Gerasimenko (\citealt{altwegg2016}), which is predominantly in phosphorus monoxide, PO (\citealt{rivilla2020a}). %
In the circumstellar envelopes of evolved stars, six different simple P-bearing molecules, up to 4 atoms, have been detected: PN, PO, CP, HCP, C$_2$P, and PH$_3$ (e.g. \citealt{Guelin1990,Agundez2007,Halfen2008,tenenbaum_identification_2007,Agundez2014}). However, in diffuse and molecular clouds and star-forming regions only P$^+$, PN, and PO have been detected. The ion P$^+$ was discovered in several diffuse clouds (\citealt{jura_observations_1978}), although no P-bearing molecules have been identified towards these regions despite deep observational searches (\citealt{chantzos2020}).
In molecular clouds and star-forming regions, only PN and PO have been identified in different environments
(\citealt{turner_detection_1987,Ziurys1987,Fontani2016,Fontani2019,rivilla_first_2016,rivilla2018,rivilla2020a,lefloch2016,Bergner2019,Bernal2021}). Recently PN has been detected towards molecular cloud complexes within the central region of the starburst galaxy NCG 253 (\citealt{haasler2021}).

In recent years, several efforts have been made to better understand  the chemistry of P in the ISM, based on the comparison of the molecular abundances derived from observations and the predictions from dedicated chemical models (e.g. \citealt{Thorne1987,Fontani2016,Rivilla2016,Jimenez-Serra2018,chantzos2020}), and quantum chemical calculations (\citealt{mancini2020,garcia_de_la_concepcion2021}). However, the small number of P-bearing species so far detected prevents us from establishing strong constraints on the chemical networks, in which many possible chemical pathways are still unexplored and poorly characterized. For this reason, it is needed to further expand the census of P-bearing species in the ISM, and their molecular abundances. 

One of the most promising candidates is the phosphorus monoxide ion, PO$^+$. This molecule presents a large dipole moment of $\sim$3.44 Debye (from calculations of \citealt{peterson1990}), which might allow its detection through rotational spectroscopy even with relatively low abundance.
Moreover, other monoxide ions of biogenic elements such as SO$^+$, CO$^+$, and NO$^+$ have already been detected previously in the ISM (\citealt{turner1992,latter1993,cernicharo2014}, respectively). 
Previous detections of the monoxide ion SO$^+$ have shown that this species is strongly enhanced in shocked gas, following the release of S and S-bearing molecules from the grain mantles, and by subsequent ionisation, being the main formation route the reaction S$^+$ + OH $\rightarrow$  SO$^+$ + H (\citealt{neufeld1989,herbst_leung1989,turner1992,turner1996,podio2014}).

Similarly to S, P is considered to be (even more) depleted in interstellar dust grains (\citealt{Rivilla2016,lefloch2016}), and shocks have been invoked as a key agent for the formation of P-bearing species by gas-phase chemistry (\citealt{lefloch2016,mininni2018,Jimenez-Serra2018,rivilla2020a}). Therefore, the best targets to search for PO$^+$ are regions with the presence of shocks and sources of ionisation such as high cosmic-ray ionisation rates, and with previous detections of P-bearing species.

The molecular cloud G+0.693$-$0.027 (hereafter G+0.693), located in the Sgr B2 complex in the Central Molecular Zone (CMZ) of our Galaxy, fullfills all of these conditions. Its chemistry is strongly affected by large-scale shocks (e.g. \citealt{requena-torres_largest_2008,martin_tracing_2008,zeng2020}), likely produced by a cloud-cloud collision (\citealt{zeng2020}). Moreover, it is well-known that the cosmic-ray ionisation rate in the CMZ can be above $\zeta\sim$10$^{-15}$ s$^{-1}$ (\citealt{goto2014}), more than two orders of magnitude higher than the standard local Galactic value (\citealt{Padovani2009}).
All this and the detection of PO and PN (\citealt{rivilla2018}) towards G+0.693, makes this cloud one of the best targets to search for new P-bearing species, and in particular PO$^+$.


In this work we report the first detection of PO$^+$ in the ISM towards G+0.693. The molecular ions SO$^+$ and NO$^+$ have also been detected, the latter for the second time in the ISM.  We have also analysed their neutral counterparts PO, SO, and NO, and computed the relative ratio among neutrals and ions. Finally, we have discussed about their possible chemistry in the ISM, including new chemical modelling of PO$^{+}$.

%
%
%




\section{Observations}

We have used an unbiased and sensitive spectral line survey  towards G+0.693 carried out with three telescopes: Yebes 40m telescope (Guadalajara, Spain), IRAM 30m telescope (Pico Veleta, Spain) and APEX telescope (Atacama, Chile).
The observations were centred at $\alpha$(J2000.0)$\,$=$\,$17$^h$47$^m$22$^s$, and $\delta$(J2000.0)$\,$=$\,-\,$28$^{\circ}$21$'$27$''$. The position switching mode was used in all the observations with the off position located $\Delta\alpha$~=~$-885$'', $\Delta\delta$~=~$290$'' from the source position. The line intensity of the spectra was measured in units of $T_{\mathrm{A}}^{\ast}$ as the molecular emission toward G+0.693 is extended over the beam (\citealt{requena-torres_organic_2006,requena-torres_largest_2008,zeng2018}). 

\vspace{2mm}

\subsection{IRAM 30m observations}

We used the dual polarisation receiver EMIR connected to the fast Fourier transform spectrometers (FFTS), which provided a channel width of 200 kHz in the following
spectral windows: 71.76 to 116.72 GHz, 124.77 to 175.5 GHz, and 223.307$-$238.29 GHz. 
Detailed information of the observational survey is presented in \citet{rivilla2021a} and \citet{rivilla2021b}.
The half power beam width (HPBW) of the telescope ranges from $\sim$10.3$^{\prime\prime}$ to  $\sim$34.3$^{\prime\prime}$ in the spectral range covered.

\vspace{2mm}

\subsection{Yebes 40m observations}

We have performed new observations with the Yebes 40m telescope. A total of 29 observing sessions between March and June 2021 (total observing time $\sim$100 hr) were performed, as part of the project 21A014 (PI: Rivilla).
The Nanocosmos Q-band (7$\,$mm) HEMT receiver was used, which enables ultra broad-band observations ($\sim$18 GHz) in two linear polarizations
(\citealt{tercero2021}). The receiver was connected to 16 FFTS providing a channel width of 38 kHz and a bandwidth of 18.5 GHz per polarisation.
We observed two different spectral setups centered at 41.400 and 42.300 GHz. The total frequency range is 31.07$-$50.42 GHz. 
We perform an initial data inspection and reduction using the CLASS module of the GILDAS package\footnote{https://www.iram.fr/IRAMFR/GILDAS}, and our own Python-based scripts\footnote{https://www.python.org}.
The comparison of the spectra of the two spectral setups was used to identify possible spurious lines in the IF. 
Telescope pointing and focus were checked every one or two hours through pseudo-continuum observations towards the red hypergiant star VX Sgr. 
The final spectra were smoothed to 2 km s$^{-1}$.
The half power beam width (HPBW) of the telescope was $\sim$39$^{\prime\prime}$ at 44 GHz.

\vspace{2mm}

\subsection{APEX observations}

We have also used data from two projects observed in service mode with the APEX telescope: O-0108.F-9308A-2021 (PI: Rivilla) and E-0108.C-0306A-2021 (PI: Rivilla). We used the NFLASH230 receiver connected to two FFTS backends, which provide a simultaneous coverage of two sidebands of 7.9 GHz each separated by 8 GHz. The spectral resolution was 250 kHz.  
The project O-0108.F-9308A-2021 was observed during ten different observing runs from July 11 to September 26 2021, with a total observing time of 23.4 hrs. We used two different frequency setups slightly shifted in frequency (262.0 and 262.3 GHz) to check for possible spurious lines from the image band. The total spectral range covered was 243.94$-$252.13 GHz and 260.17$-$268.38 GHz.
The precipitable water vapour (pwv) during the observations was in the range of 0.7 to 3.6 mm.
The project E-0108.C-0306A-2021 was observed during two different observing runs: October 31 and November 1 2021. The total observing time was 5.8 hrs. We used two different frequency setups slightly shifted in frequency (224.0 and 224.1 GHz) to check for possible spurious lines from the image band. The total spectral range covered was 217.93$-$225.93 GHz and 234.18$-$242.18 GHz.
The precipitable water vapour (pwv) during the observations was in the range of 0.9 to 2.2 mm.
The final spectra from both projects were smoothed to 1 MHz, which translates into velocity resolution of 1.1-1.4 km s$^{-1}$ in the frequency ranges observed. The half power beam width (HPBW) of these APEX observations is $\sim$23$^{\prime\prime}-$28$^{\prime\prime}$ in the frequency range observed.


\section{Analysis and results}

The identification and fitting of the molecular lines were performed using the SLIM (Spectral Line Identification and Modeling) tool within the MADCUBA package{\footnote{Madrid Data Cube Analysis on ImageJ is a software developed at the Center of Astrobiology (CAB) in Madrid; http://cab.inta-csic.es/madcuba/}} (version 09/11/2021; \citealt{martin2019}).
SLIM uses the molecular databases of the Cologne Database for Molecular Spectroscopy (CDMS, \citealt{endres2016}) and the Jet Propulsion Laboratory (JPL; \citealt{pickett1998}) to generate synthetic spectra under the assumption of Local Thermodynamic Equilibrium (LTE). 
In the Table \ref{tab:spectroscopy} of the Appendix A we list the details of the molecular spectroscopy of all the molecules analysed in this work.
To evaluate if the transitions are blended with emission from other species, we have also considered the LTE model that predicts the total contribution of more than 120 species than have been identified so far towards G+0.693 (e.g., \citealt{requena-torres_largest_2008,zeng2018,rivilla2019a,rivilla2020b,jimenez-serra2020,rivilla2021a,rivilla2021b,zeng2021,rodriguez-almeida2021,rodriguez-Almeida2021b}). To derive the physical parameters of the molecular emission, we used the AUTOFIT tool of SLIM, which finds the best fit between the observed spectra and the predicted LTE model (see details in \citealt{martin2019}). The free parameters of the LTE model were: molecular column density ($N$), excitation temperature ($T_{\rm ex}$), linewidth (or full width at half maximum, $FWHM$), and velocity ($v_{\rm LSR}$).



\begin{figure}[]
\begin{center}
\includegraphics[width=15cm]{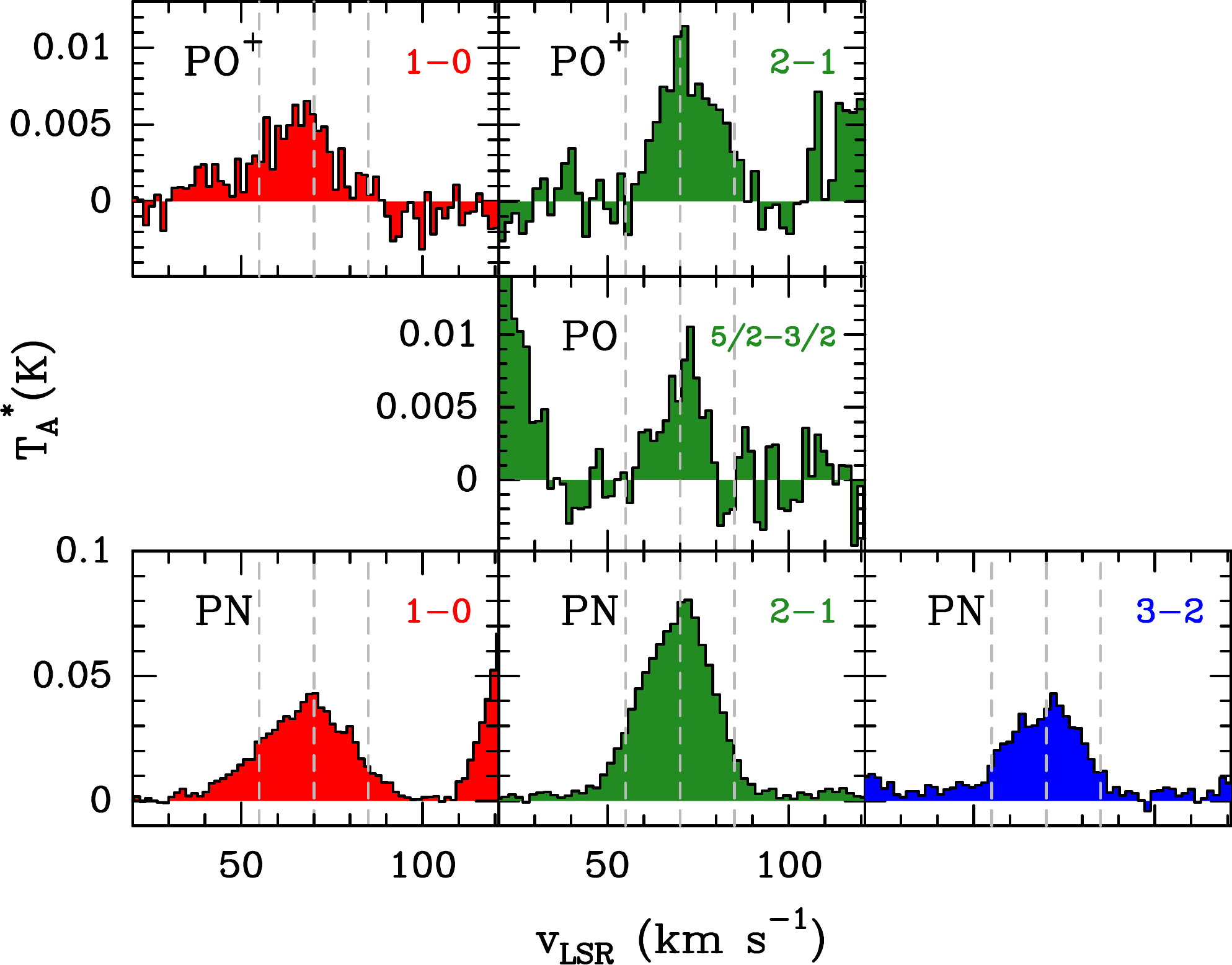}
\end{center}
\vskip-3mm
\caption{Spectral profiles of the molecular transitions of the P-bearing species detected towards G+0.693 (see Table \ref{tab:transitions}): PO$^{+}$ (upper panels), PO (middle panel), and PN (lower panels). The left/middle/right column shows transitions with $E_{\rm up}$=2.3/6.8$-$8.4/13.5 K, indicated with red/green/blue histograms, respectively. The gray vertical dashed lines indicate the velocities 70 km s$^{-1}$ and 70$\pm$15 km s$^{-1}$.}
\label{fig:profiles}
\end{figure}

\begin{figure}[]
\begin{center}
\includegraphics[width=5.8cm]{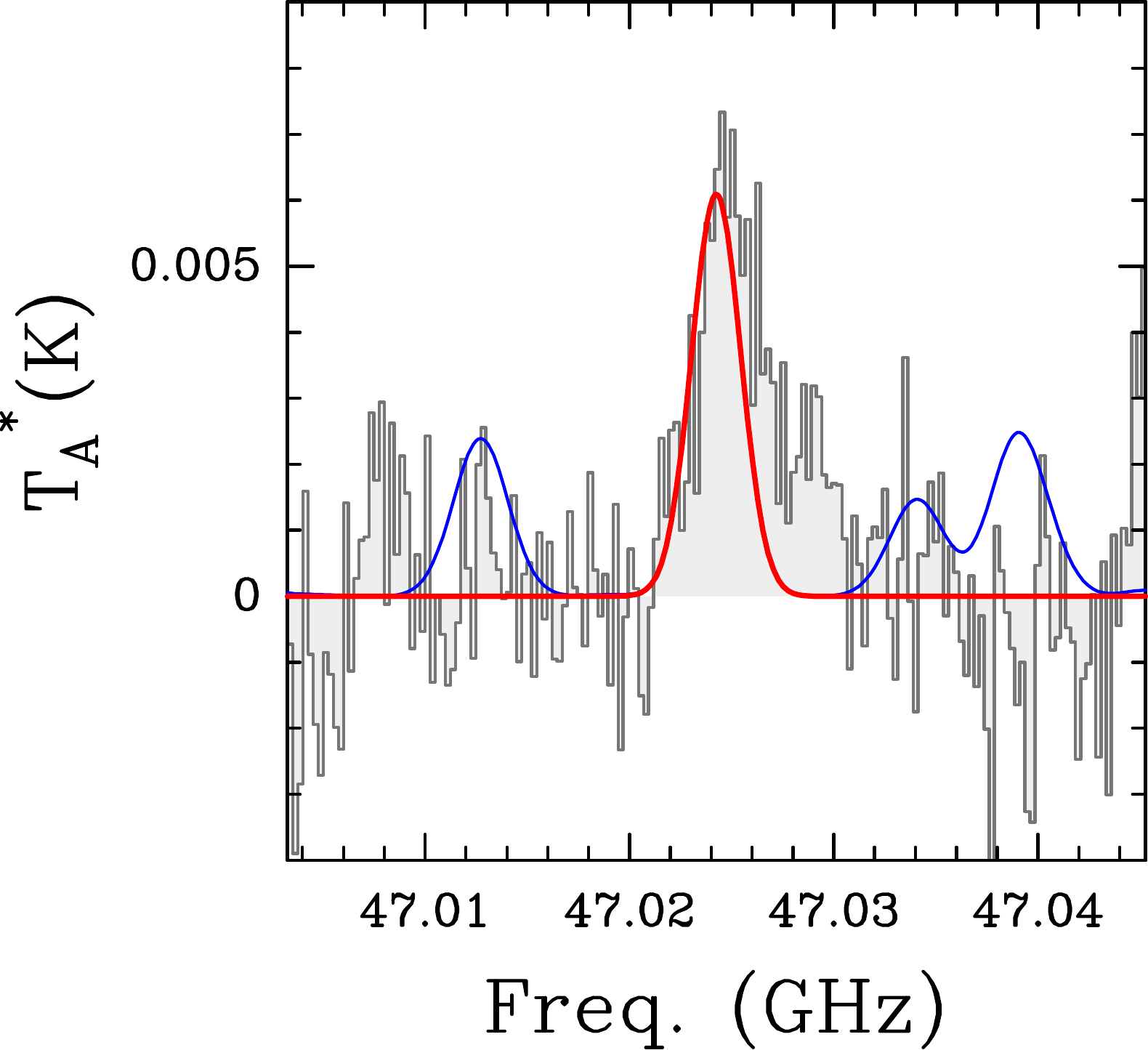}
\hspace{3mm}
\includegraphics[width=5.325cm]{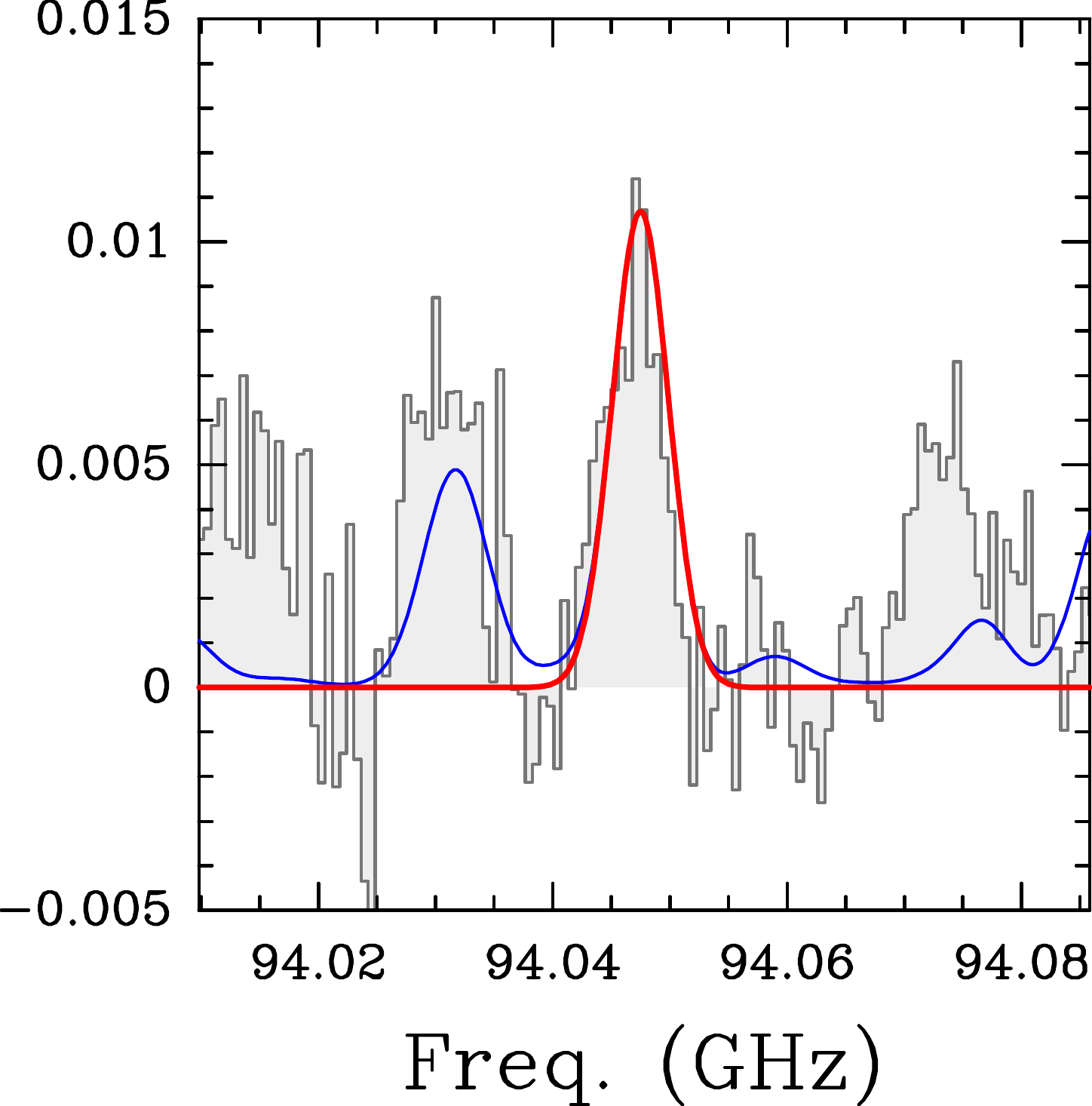}
\hspace{3mm}
\includegraphics[width=5cm]{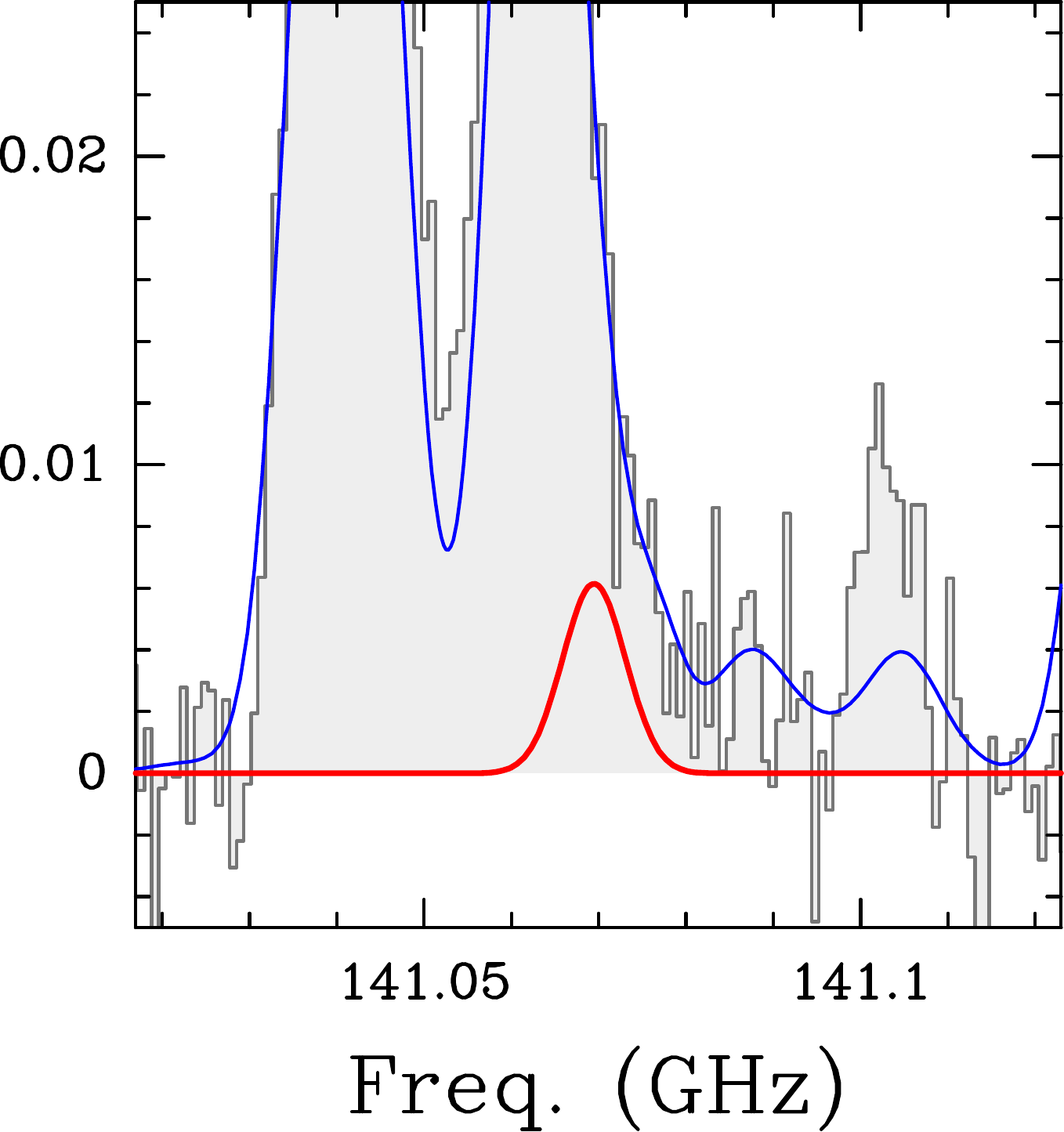}
\end{center}
\caption{PO$^+$ transitions detected towards G+0.693 (see Table \ref{tab:transitions}). The gray histograms show the observed spectra obtained with the Yebes 40m telescope ($J=$1$-$0 transition; leftmost panel), and IRAM 30m telescope ($J=$2$-$1 and $J=$3$-$2, transitions; middle and right panel). The red curve corresponds to the best LTE fit derived with MADCUBA, and the blue curve shows the total contribution considering all the molecular species identified, including PO$^{+}$.}
\label{fig:po+}
\end{figure}

\vskip2mm
\subsection{PO$^+$ detection}


We have used the molecular spectroscopy from the laboratory work by \citet{petrmichl1991}.
The $J=$1$-$0 and $J=$2$-$1 transitions of PO$^+$ at 47.024 GHz and 94.047 GHz (Table \ref{tab:transitions}), respectively, have been detected in our survey, as shown in the upper panels of Figure \ref{fig:profiles}.
This is the first detection of PO$^+$ reported in the ISM.
We note that the laboratory work by \citet{petrmichl1991} measured PO$^+$ transitions in the range 140–470 GHz, and thus the two unblended transitions detected in the present work are actually the very first observation of these two lines.
The $J=$3$-$2 and $J=$4$-$3 transitions were also covered by the survey. The former transition, at 141.070 GHz, is heavily blended with bright emission from H$^{13}$CCCN(16$-$15) (see right panel of Figure \ref{fig:po+}).
The $J=$4$-$3 transition at 235.107 GHz (not shown here) is also strongly blended with an unidentified species. We note that the expected line intensity of this higher-energy PO$^+$ transition (Table \ref{tab:transitions}), according to the best LTE fit (see below), is $<$0.25 mK, lower than the noise level of the observed spectra.

The upper panels of Figure \ref{fig:profiles} show that the spectral profiles of the $J=$1$-$0 and $J=$2$-$1 transitions are not completely identical, with the 1$-$0 transition showing a more prominent wing at blue-shifted velocities. 
This is consistent with the observed profiles other P-bearing molecule like PN. The lower panels of Figure \ref{fig:profiles} show that the lower the energy level of the PN transition, the more prominent the wing at blue-shifted velocities is, as also observed in PO$^+$. This suggests that there are two different velocity components, one peaking at $\sim$70 km s$^{-1}$, and another one at more blue-shifted velocities (producing the wing of the 1$-$0 transitions), with different excitation conditions. These two velocity components have also been identified in several other molecular species (\citealt{colzi2022}).
%

For simplicity, to perform the fit of PO$^{+}$, we have considered only the $\sim$70 km s$^{-1}$ velocity component, which dominates the emission of the two transitions. To run AUTOFIT we have considered the $J=$1$-$0, $J=$2$-$1, and $J=$3$-$2 transitions, taking also the contribution of all the other species into account. We have fixed the velocity to $v_{\rm LSR}$=70 km s$^{-1}$, the linewidth to $FWHM$=18 km s$^{-1}$, and the $T_{\rm ex}$ to 4.5 K, which is the value derived when fitting the transitions of PN shown in the lower panels of Figure \ref{fig:profiles} (Rivilla et al., in preparation). The result of the fit of PO$^+$ is shown in Figure \ref{fig:po+}.
We note that the JPL entry used to perform the fit considers a dipole moment of $\mu$=3.44 Debye, calculated by \citet{peterson1990}.
In Appendix \ref{app:po+dipole} we present higher level calculations of the PO$^+$ dipole moment, which gives a value of $\mu$=3.13 Debye. Therefore, to calculate the final molecular column density we have multiplied the value derived using the JPL entry by a factor of (3.44 / 3.13046)$^2$=1.21. The derived value of the PO$^{+}$ column density is (6.0$\pm$0.7)$\times$10$^{11}$ cm$^{-2}$ (Table \ref{tab:parameters}). This translates into a molecular abundance with respect to molecular hydrogen of (4.5$\pm$1.1)$\times$10$^{-12}$, using the value of $N$(H$_2$)=1.35$\times$10$^{23}$ cm$^{-2}$ from \citet{martin_tracing_2008}.
%
%
%
\vspace{5mm}
\subsection{PO}
\begin{figure}[]
\begin{center}
\includegraphics[width=13cm]{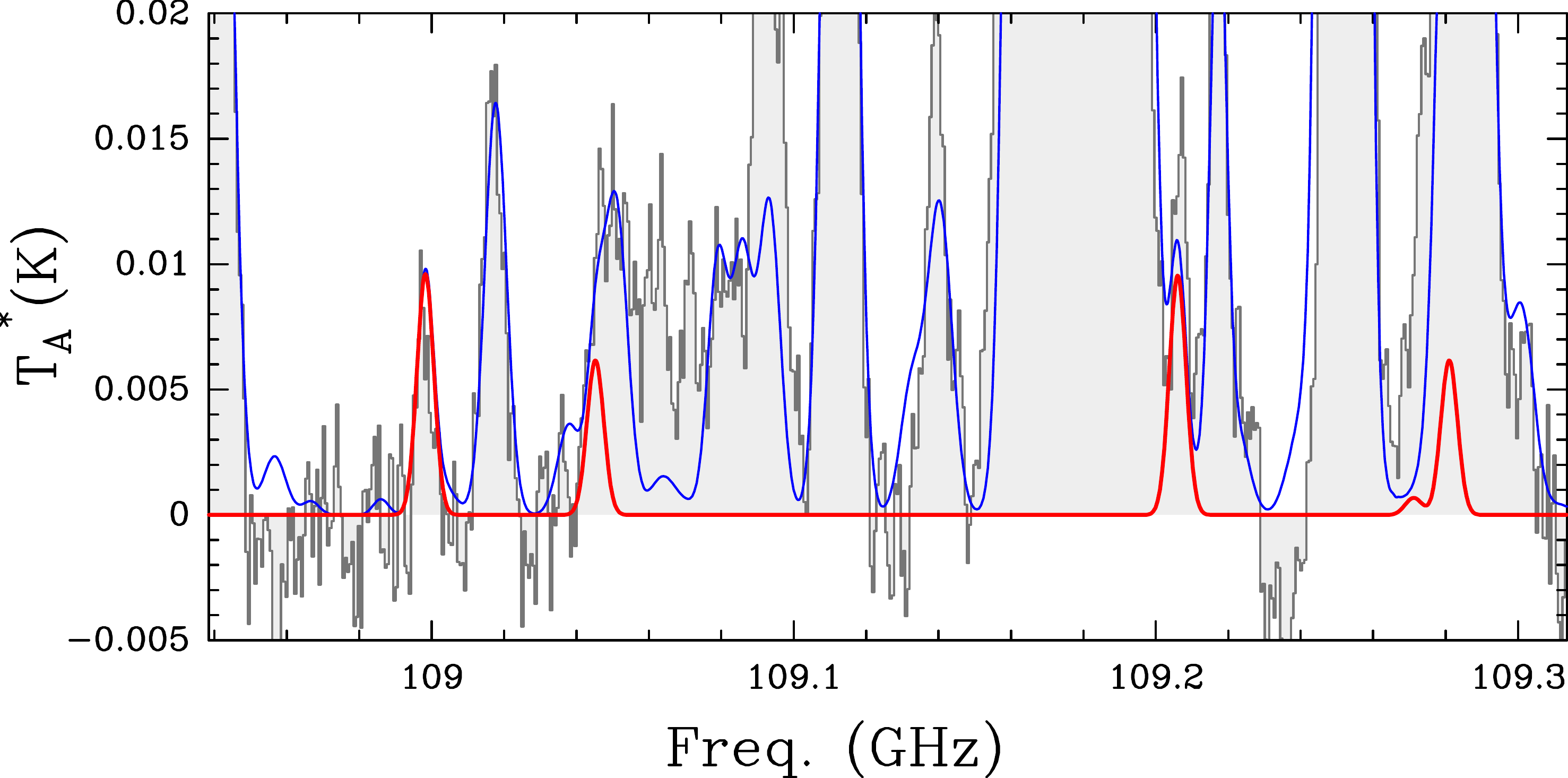}
\end{center}
\caption{PO transitions detected towards G+0.693 (see Table \ref{tab:transitions}). The gray histogram shows the observed spectra obtained with the IRAM 30m telescope. The red curve corresponds to the best LTE fit derived with MADCUBA, and the blue curve shows the total contribution considering all the molecular species identified, including PO.}
\label{fig:po}
\end{figure}

To study the relative abundance of PO$^+$ with respect to its neutral counterpart, we have also analysed the molecular emission of PO. This molecule was already detected towards G+0.693 by \citet{rivilla2018} using a previous less sensitive spectral survey carried out with the IRAM 30m (\citealt{zeng2018}). We repeat here the analysis using the data from the new deeper survey. The $J=$5/2$-$3/2, $\Omega$=1/2 quadruplet of PO (with $E_{\rm up}$=8.4 K,  Table \ref{tab:transitions}) is displayed in Figure \ref{fig:po}. 
The $F$=3$-$2 $l$=e transition is completely unblended (see also the middle panel of Figure \ref{fig:profiles}), while the $F$=3$-$2 $l$=f is slightly blended with a very weak line of ethanolamine (NH$_2$CH$_2$CH$_2$OH, \citealt{rivilla2021a}), and an unidentified species. The other two transitions appear blended with other species already identified in G+0.693, which are indicated in the last column of Table \ref{tab:transitions}.
The $J=$7/2$-$5/2, $\Omega$=1/2 quadruplet of PO at 152 GHz (with $E_{\rm up}$=15.7 K), which is also covered by our survey, is expected to be too weak to be detected, with predicted line intensities of $\leq$4 mK, according to the LTE fit of the $J=$5/2$-$3/2, $\Omega$=1/2 transitions (see below), lower than the noise of the data at that frequency ($\sim$5 mK).

To perform the fit, we have used the $J=$5/2$-$3/2, $\Omega$=1/2 $F$=3$-$2 $l$=e unblended transition. We fixed the $T_{\rm ex}$ to 4.5 K, as for PO$^{+}$, leaving $N$, $FWHM$ and $v_{\rm LSR}$ as free parameters. The results are shown in Table \ref{tab:parameters}. We obtained $v_{\rm LSR}$=69.5$\pm$1.4, and $FWHM$=15$\pm$3 km s$^{-1}$, which are consistent to the fixed values used for PO$^{+}$. The derived column density is (0.5$\pm$0.1)$\times$10$^{13}$ cm$^{-2}$, a factor of 1.6 lower than that reported by \citet{rivilla2018} previously. 
Using this value, we obtained a molecular column density ratio PO$^+$/PO=0.12$\pm$0.03.

\vspace{5mm}

\subsection{SO$^+$, S$^{18}$O and $^{34}$SO}

\begin{figure}[h!]
\begin{center}
\includegraphics[width=5.35cm]{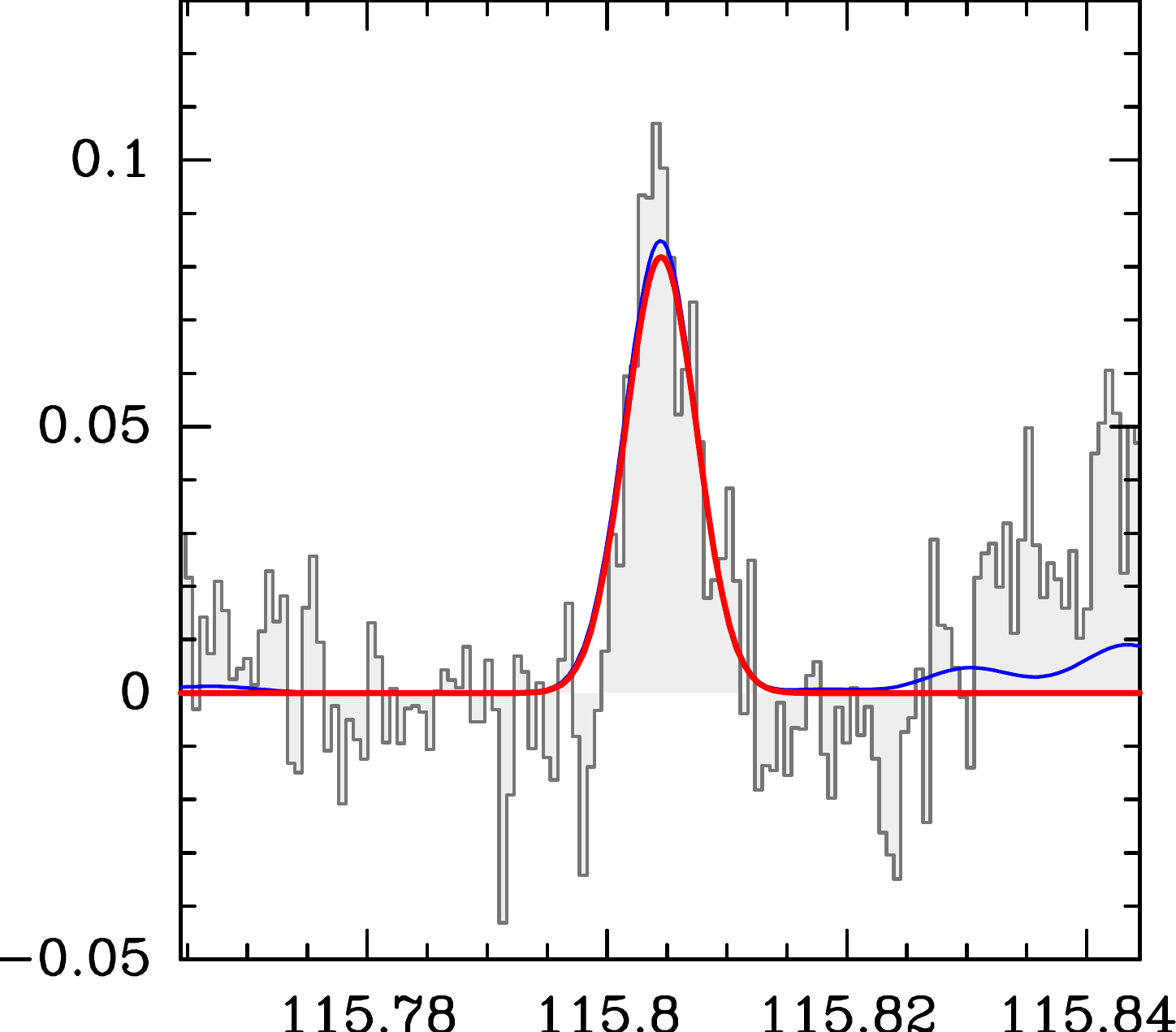}
\hspace{3mm}
\includegraphics[width=5cm]{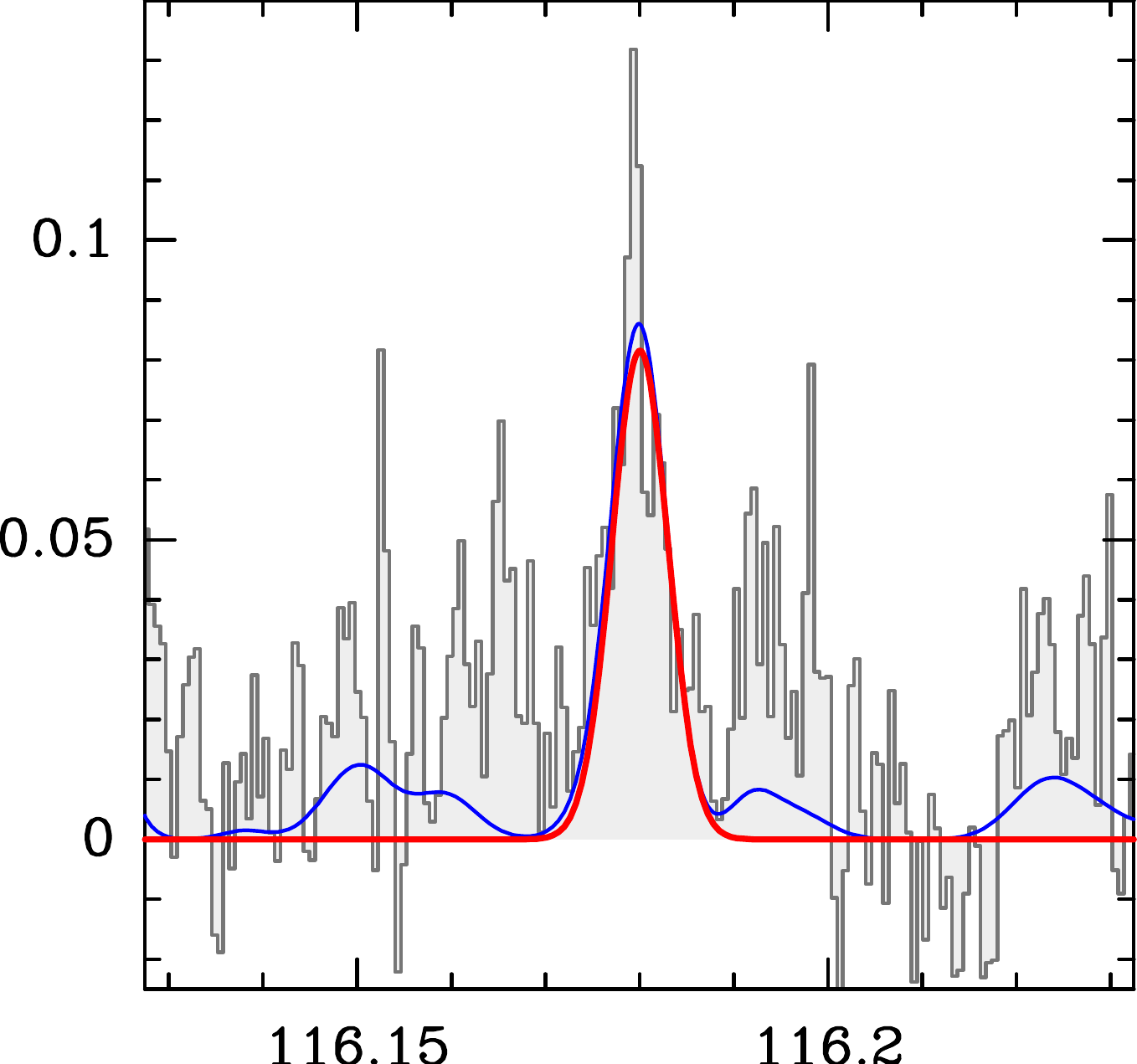}
\hspace{3mm}
\includegraphics[width=5cm]{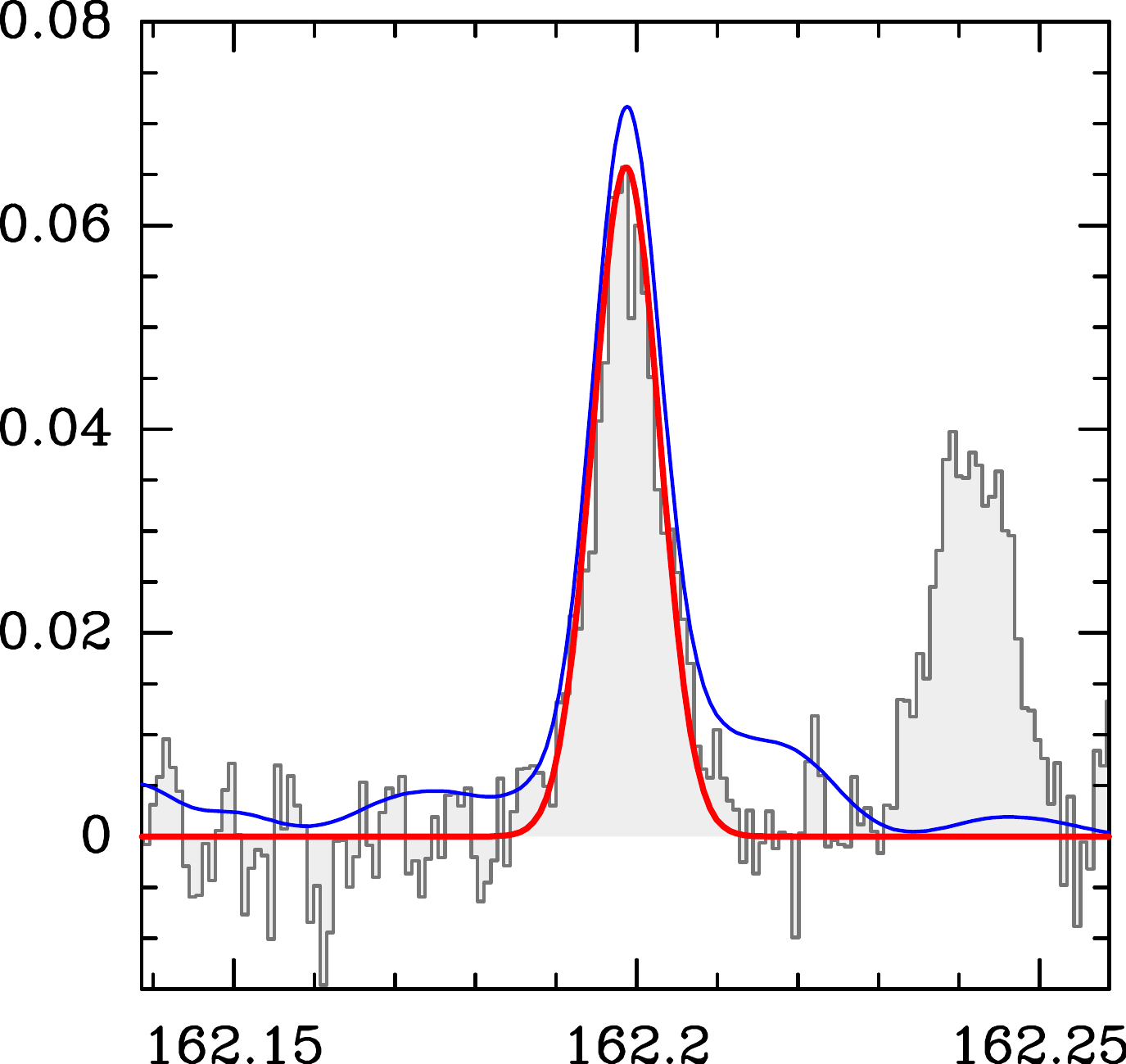}
\vskip5mm
\hspace{-9mm}
\includegraphics[width=5.8cm]{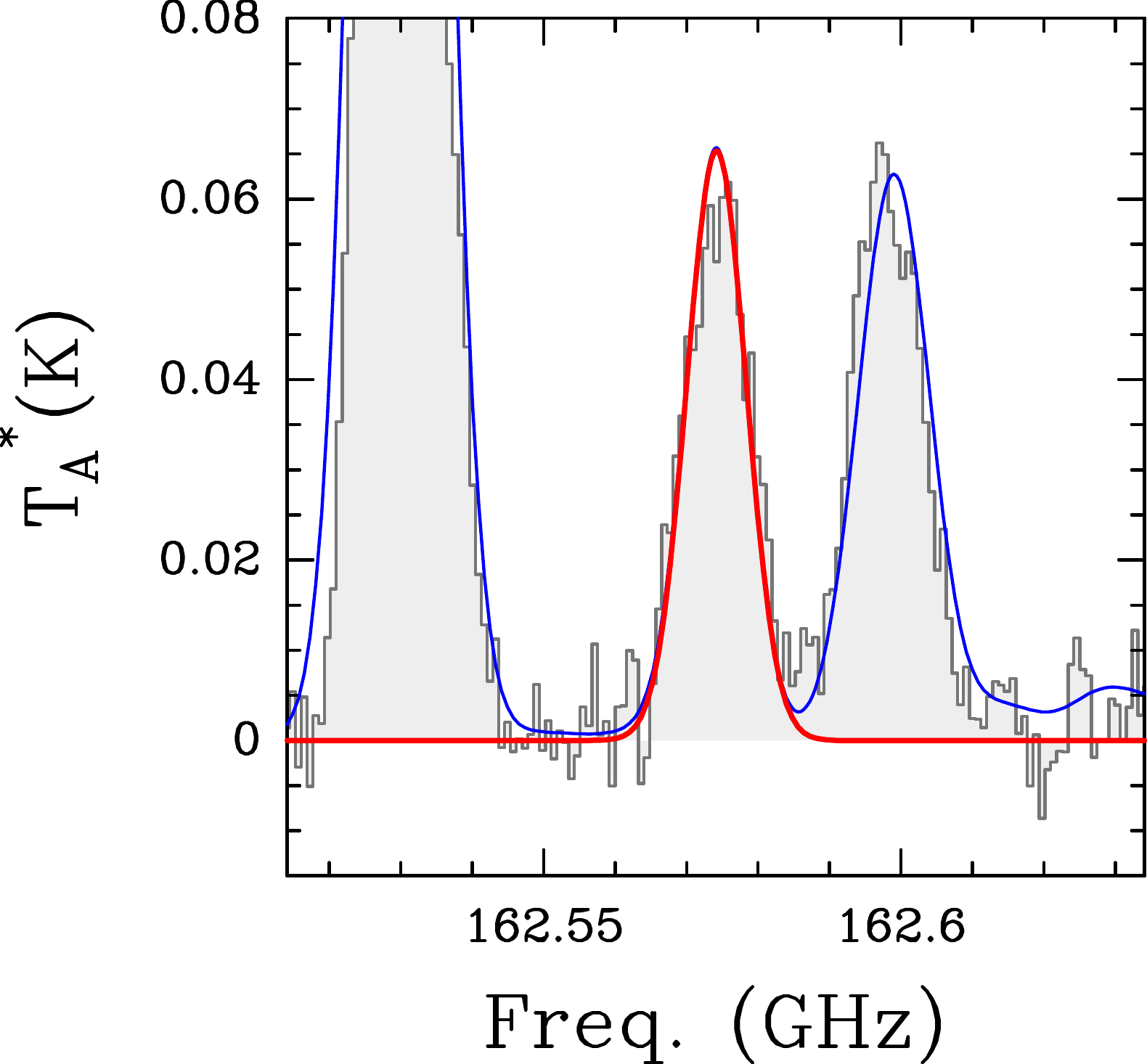}
\hspace{6mm}
\includegraphics[width=5cm]{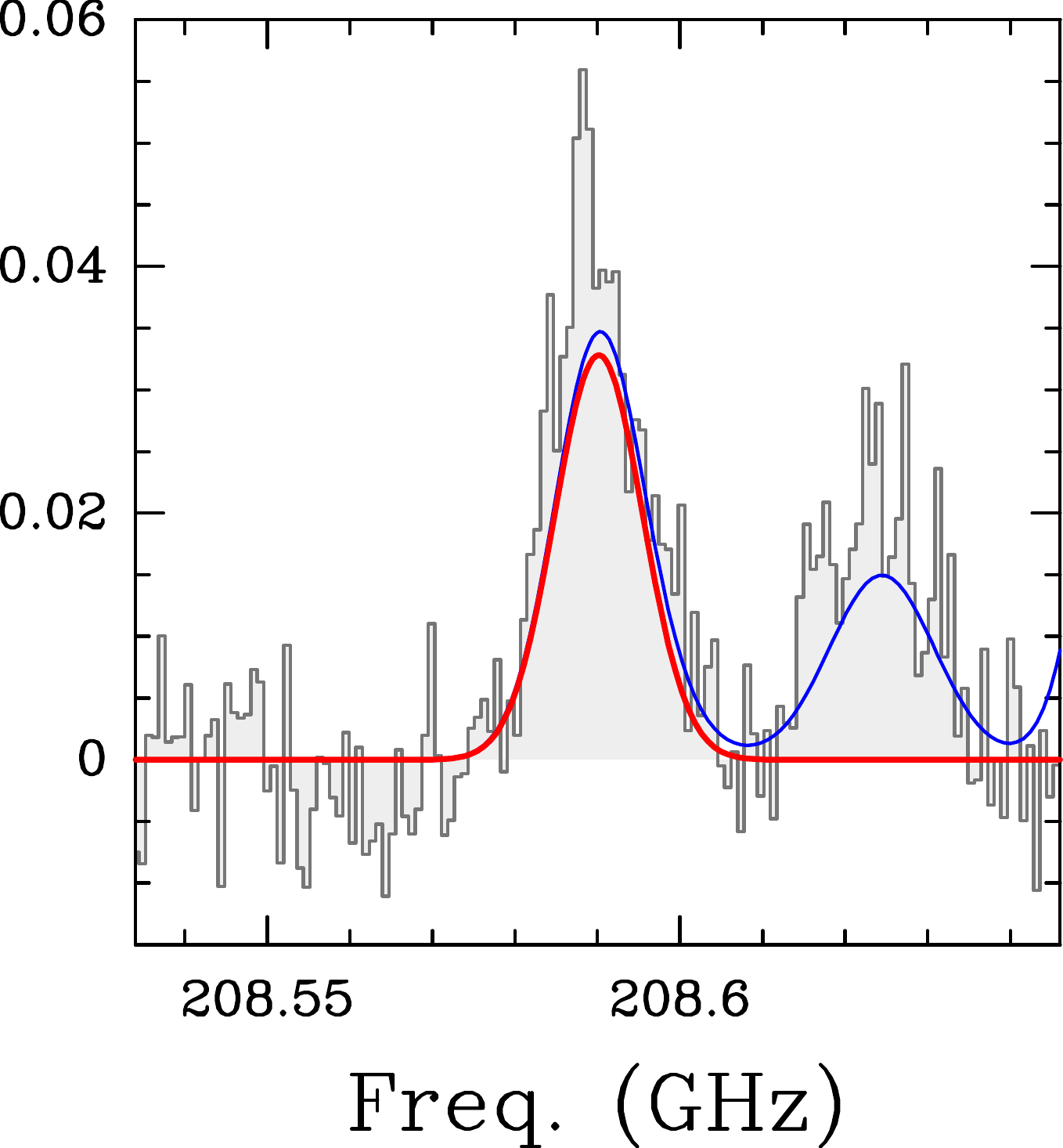}
\hspace{3mm}
\includegraphics[width=5cm]{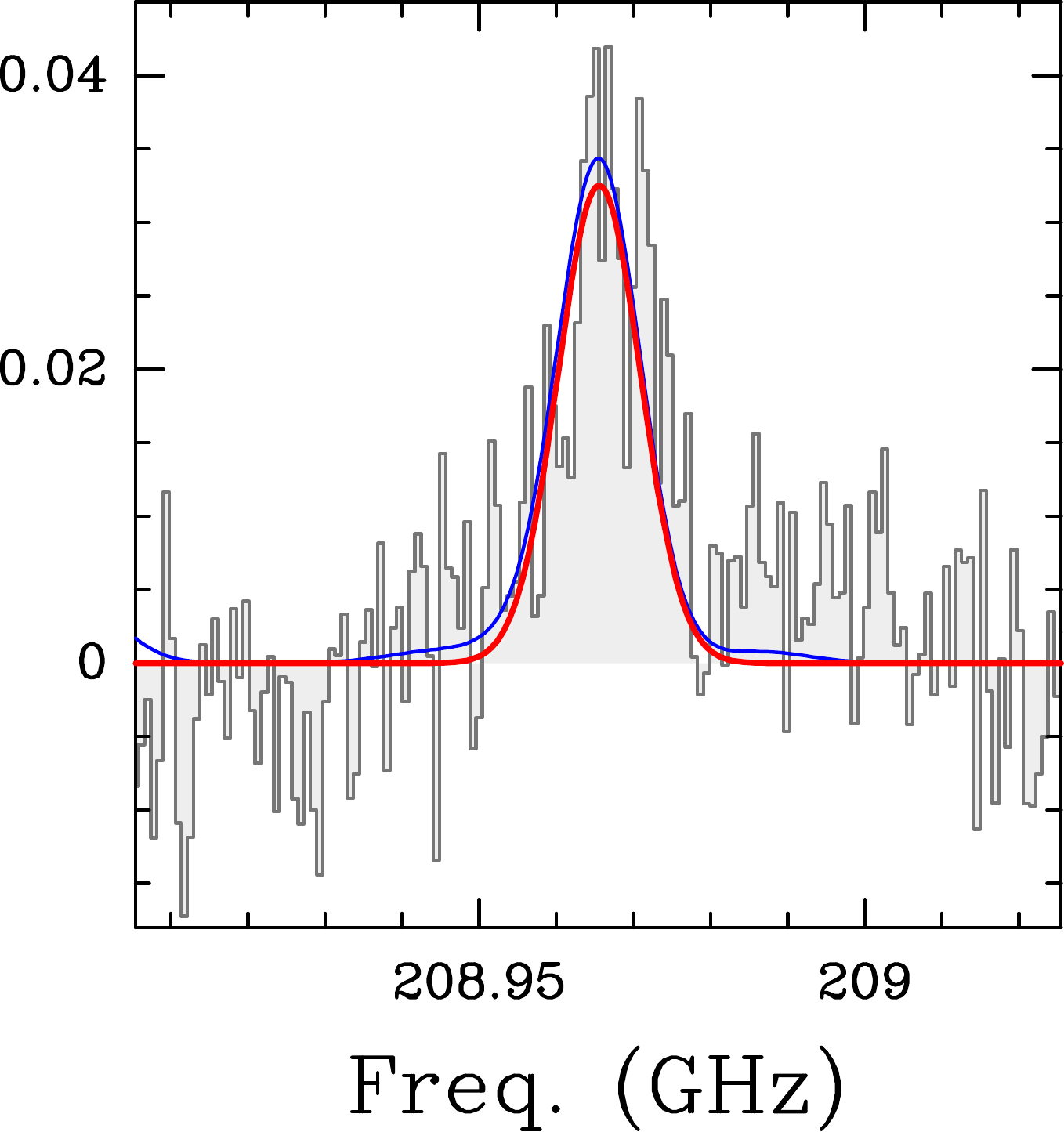}
\end{center}
\vskip-3mm
\caption{SO$^+$ transitions  detected towards G+0.693 (see Table \ref{tab:transitions}). The gray histogram shows the observed spectra obtained with the IRAM 30m telescope. The red curve corresponds to the best LTE fit derived with MADCUBA, and the blue curve shows the total contribution considering all the molecular species identified, including SO$^{+}$.}
\label{fig:so+}
\end{figure}

\begin{figure}[]
\begin{center}
\includegraphics[width=4cm]{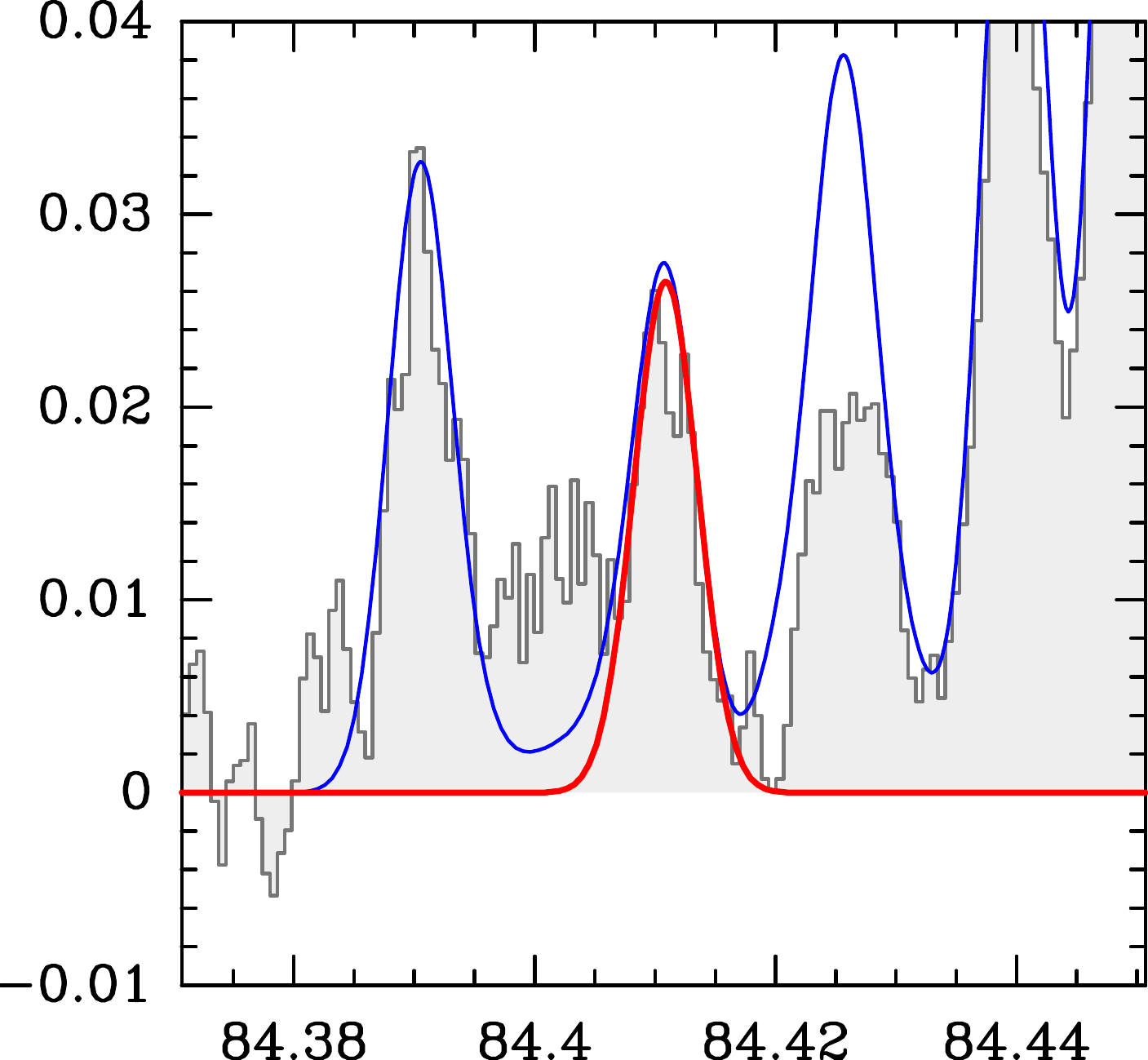}
\hspace{3mm}
\includegraphics[width=3.75cm]{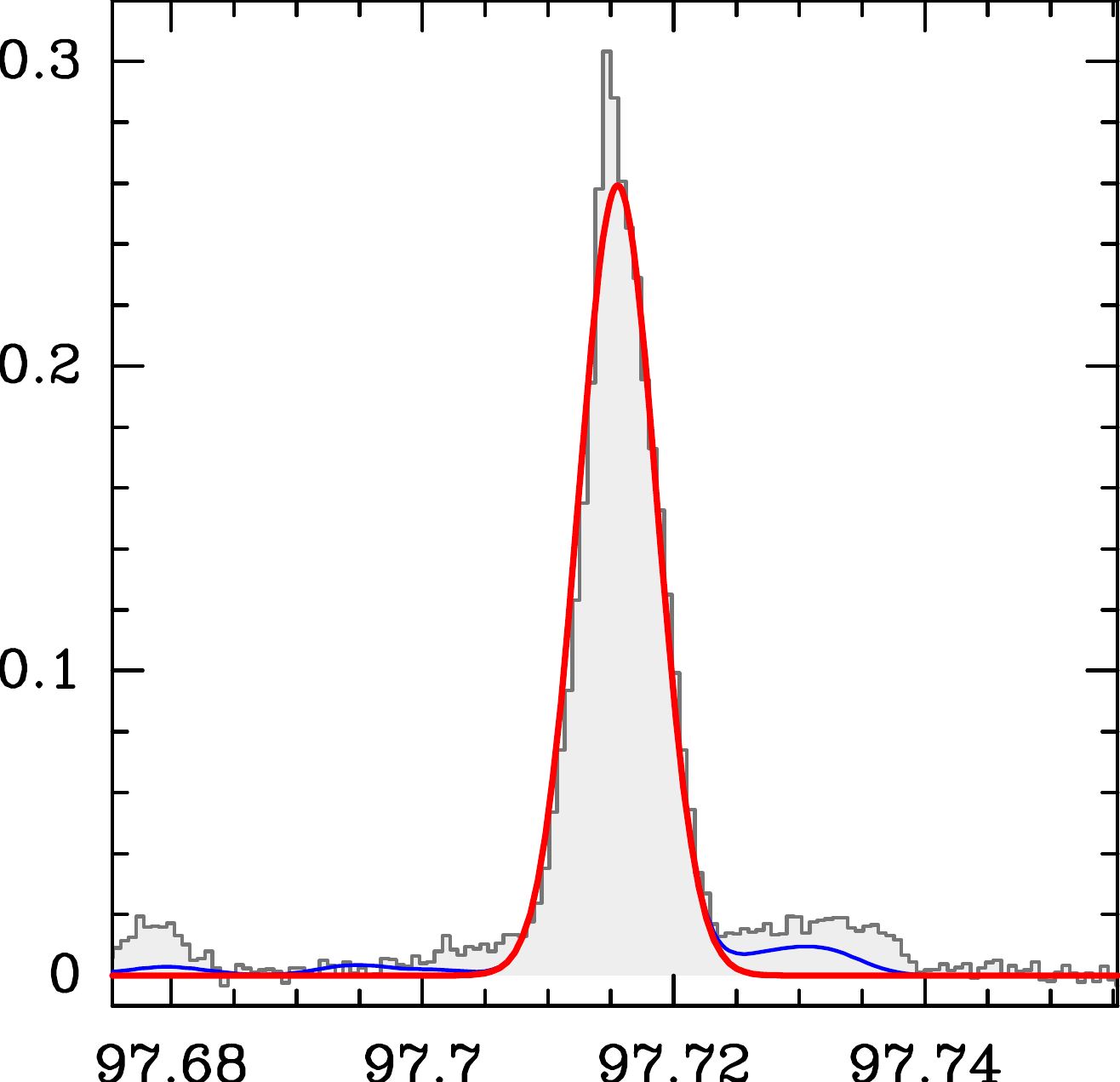}
\hspace{3mm}
\includegraphics[width=3.85cm]{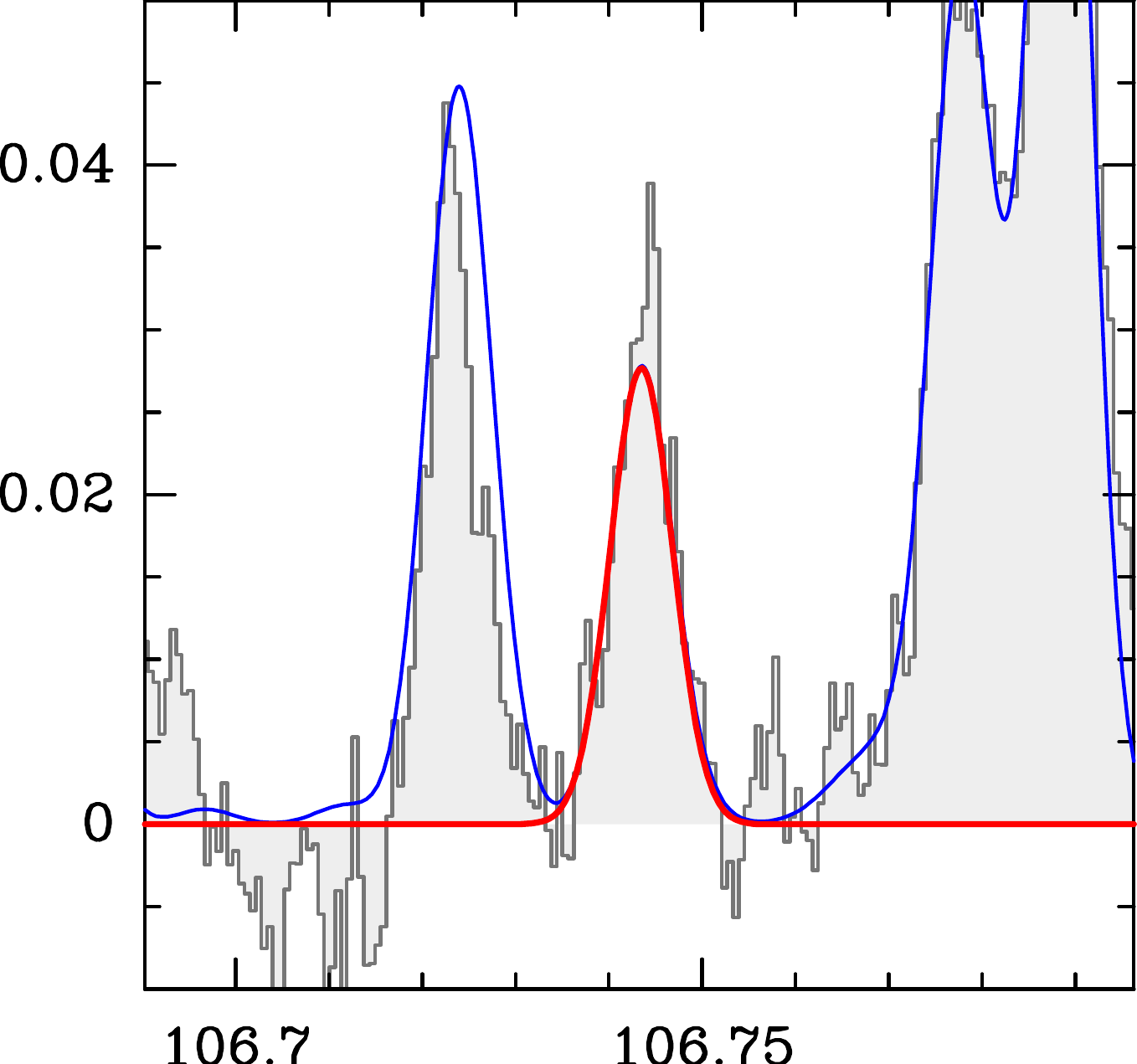}
\hspace{3mm}
\includegraphics[width=4cm]{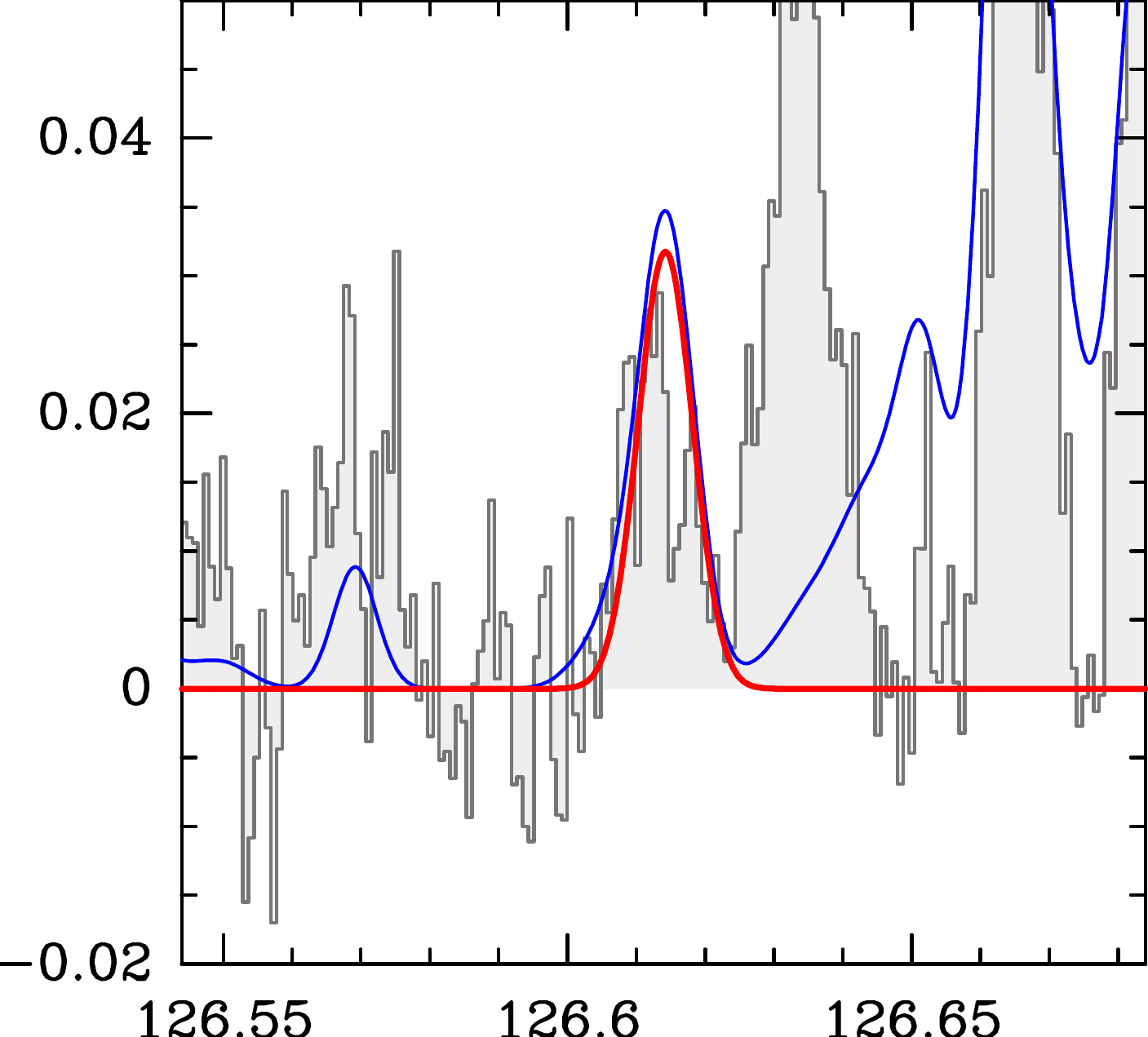}
\vskip5mm
\hspace{-14mm}
\includegraphics[width=4.5cm]{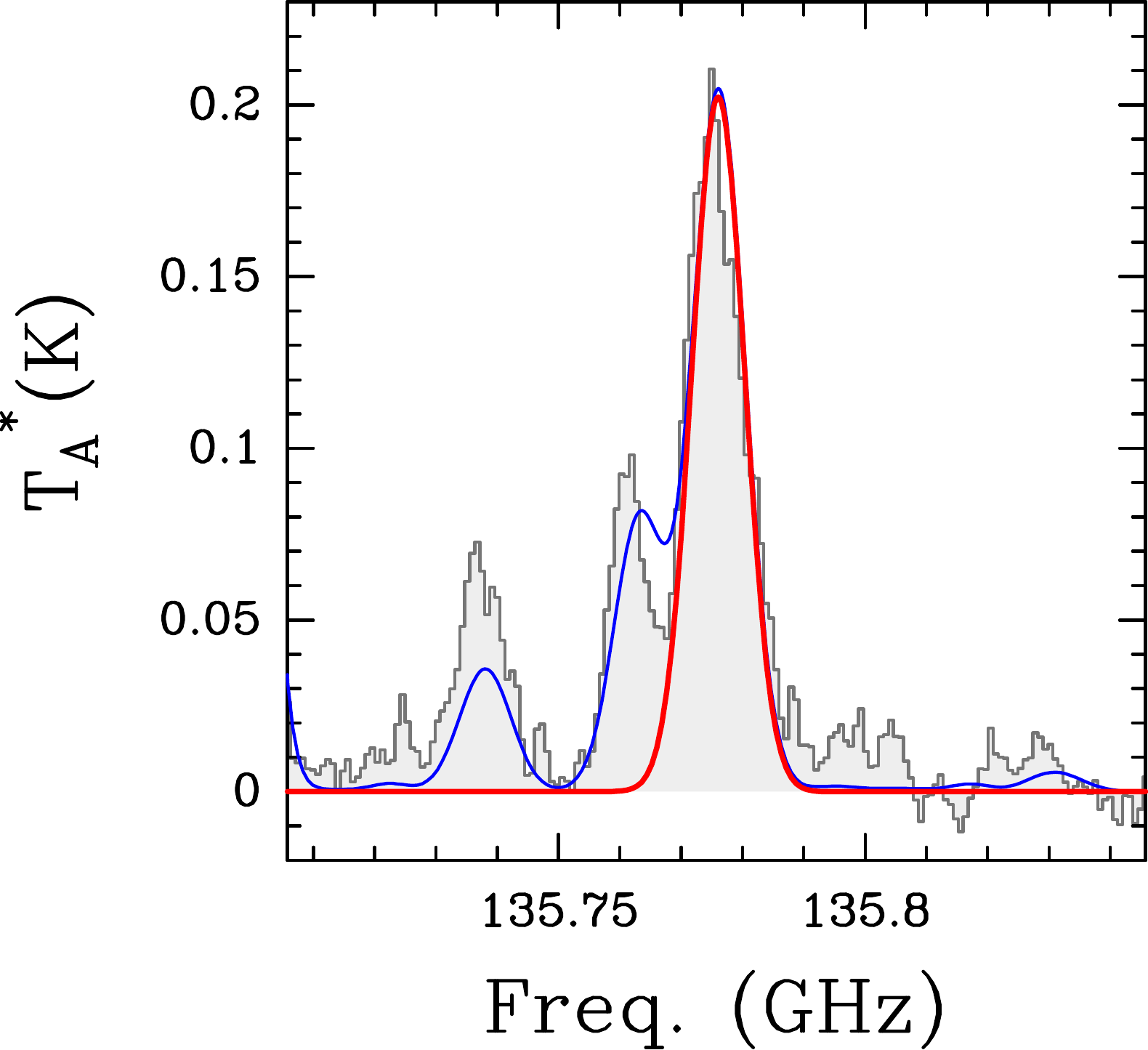}
\hspace{3mm}
\includegraphics[width=3.85cm]{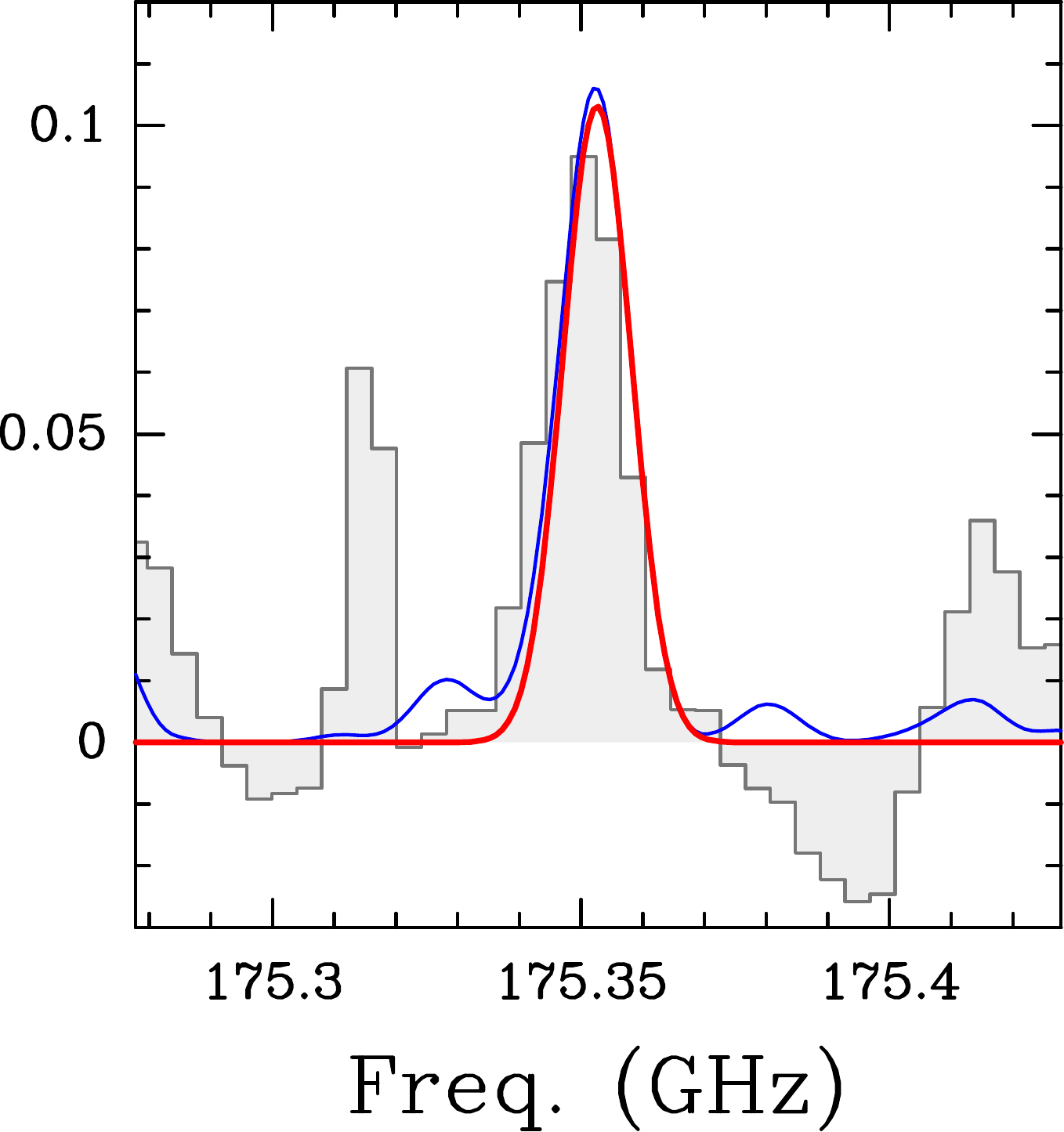}
\hspace{3mm}
\includegraphics[width=3.85cm]{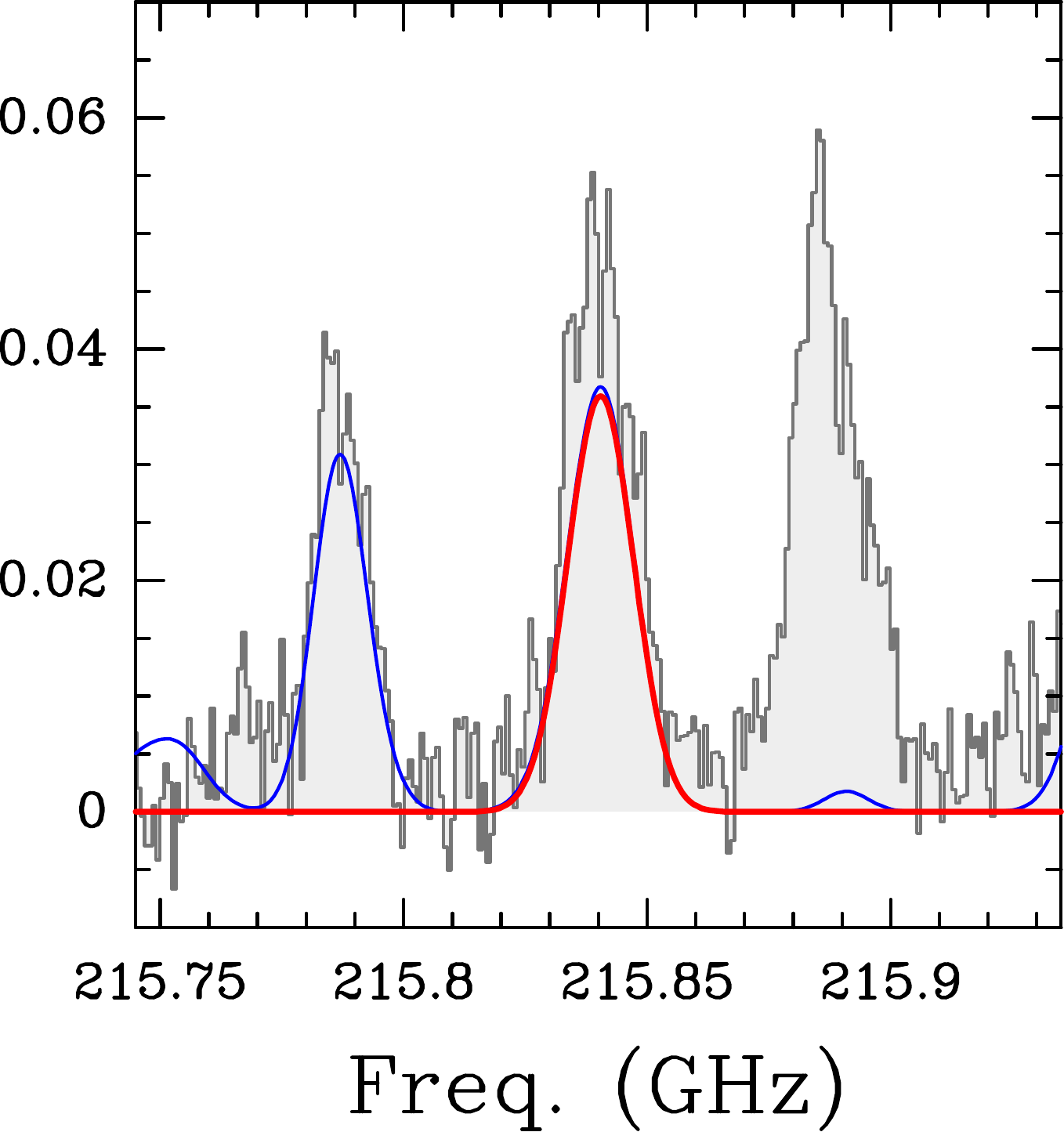}
\end{center}
\caption{$^{34}$SO transitions detected towards G+0.693 (see Table \ref{tab:transitions}). The gray histogram shows the observed spectra obtained with the IRAM 30m telescope. The red curve corresponds to the best LTE fit derived with MADCUBA, and the blue curve shows the total contribution considering all the molecular species identified, including $^{34}$SO.}
\label{fig:34so}
\end{figure}

\begin{figure}[]
\begin{center}
\includegraphics[width=8cm]{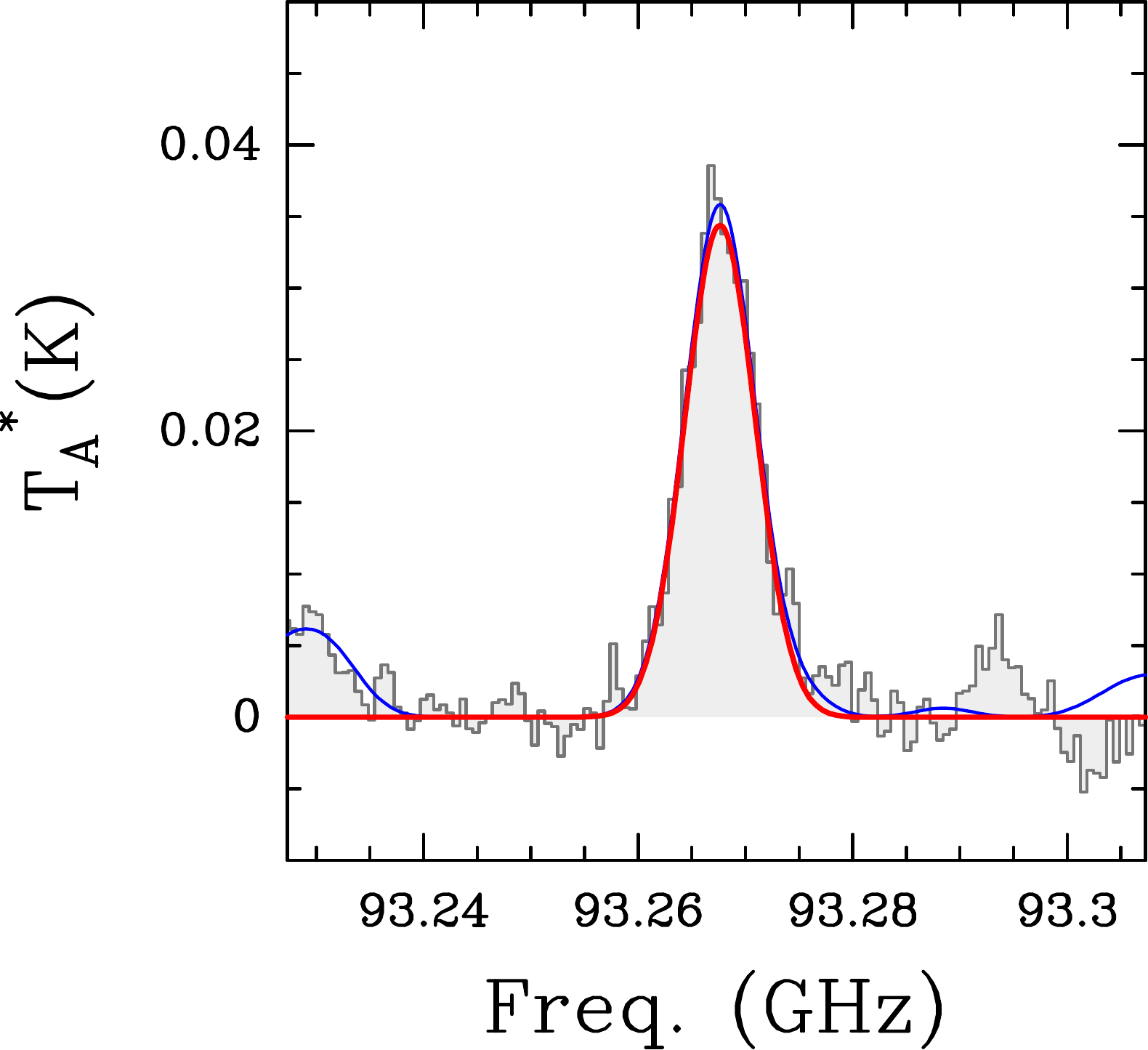}
\end{center}
\vskip-3mm
\caption{S$^{18}$O transition detected towards G+0.693 (see Table \ref{tab:transitions}). The gray histogram shows the observed spectra obtained with the IRAM 30m telescope. The red curve corresponds to the best LTE fit derived with MADCUBA, and the blue curve shows the total contribution considering all the molecular species identified, including S$^{18}$O.}
\label{fig:s18o}
\end{figure}

We show in Figures \ref{fig:so+}, \ref{fig:34so}, and \ref{fig:s18o} the molecular transitions of SO$^+$, $^{34}$SO, and S$^{18}$O, respectively, detected toward G+0.693. 
We note that the JPL entry used for SO$^{+}$ (see Table \ref{tab:spectroscopy} in Appendix \ref{app:spectroscopy}) assumes a dipole moment of $\mu$=1, while theoretical {\it ab initio} calculations reported in \citet{turner1992} from a private communication from Peterson $\&$ Woods derived $\mu$=2.3$\pm$0.2 Debye. As for PO$^+$, in Appendix \ref{app:so+dipole} we have performed new higher level calculations, which result in a dipole moment of $\mu$=2.016. Therefore, we have corrected the column density derived by AUTOFIT by the factor (1 / 2.016)$^2$. 
The derived physical parameters from the SO$^+$ emission are shown in Table \ref{tab:parameters}. We obtained $T_{\rm ex}$=8.0$\pm$0.3 K, $v_{\rm LSR}$=68.8$\pm$0.3 km s$^{-1}$, $FWHM$=18.0$\pm$0.7 km s$^{-1}$, and $N$=(1.34$\pm$0.07)$\times$10$^{13}$ cm$^{-2}$.

Since the SO transitions detected towards G+0.693 are optically thick, we have analysed its optically thin isotopologues to better derive its column density. For S$^{18}$O only a single transition is detected (Figure \ref{fig:s18o}), so we have fixed $T_{\rm ex}$ to the value derived for $^{34}$SO, 6.9$\pm$0.1 K. The physical parameters derived by AUTOFIT are shown in Table \ref{tab:parameters}.
Using the derived column densities for both isotopologues, and assuming the isotopic ratios $^{32}$S/$^{34}$S=22 and $^{16}$O/$^{18}$O=250 in the CMZ (\citealt{wilson_abundances_1994}), we have derived that the column density of SO is (142$\pm$4)$\times$10$^{13}$ cm$^{-2}$ and (300.6$\pm$0.7)$\times$10$^{13}$ cm$^{-2}$, from $^{34}$SO and S$^{18}$O, respectively. Since the most optically thin isotopologue is S$^{18}$O, we use its value hereafter to compute relative molecular abundances ratios. The ratio SO$^+$/SO=0.0045$\pm$0.0003, a factor of $\sim$30 lower than the PO$^+$/PO ratio. Note that even using the lower limit of (142$\pm$4)$\times$10$^{13}$ cm$^{-2}$, the ratio SO$^+$/SO would still be a factor of 15 lower than that measured for the PO$^+$/PO ratio

\vspace{3mm}
\subsection{NO$^+$ and NO}
\begin{figure}[h!]
\begin{center}
\includegraphics[width=6cm]{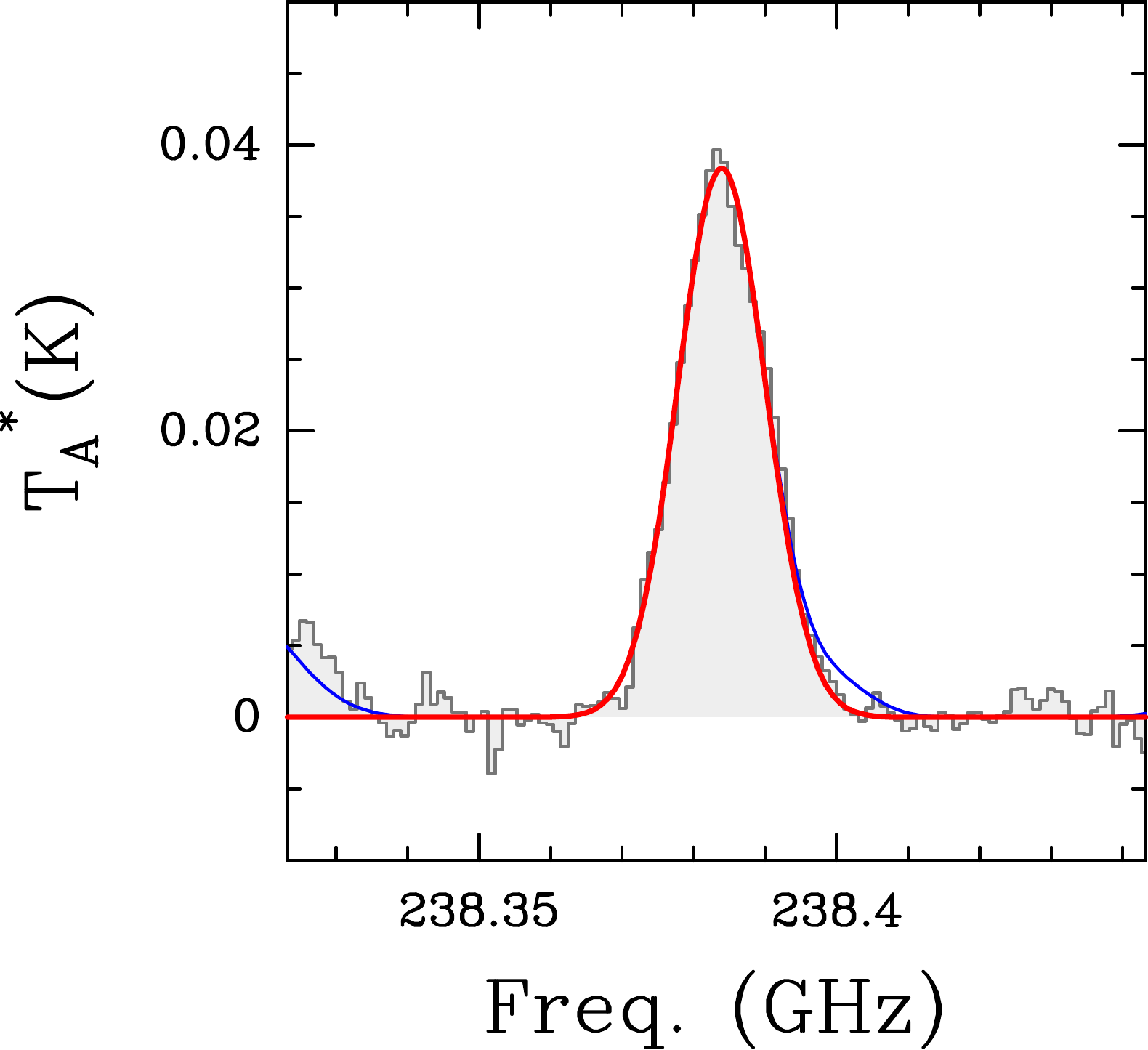}
\end{center}
\vskip-3mm
\caption{NO$^{+}$ transition detected towards G+0.693 (see Table \ref{tab:transitions}). The gray histogram shows the observed spectra obtained with the APEX telescope. The red curve corresponds to the best LTE fit derived with MADCUBA, and the blue curve shows the total contribution considering all the molecular species identified, including NO$^+$.}
\label{fig:no+}
\end{figure}

We show in Figure \ref{fig:no+} the detection of the $N=$2$-$1 transition of NO$^{+}$, which is the only transition of this species that falls in the frequency range covered by our survey. This is the second detection of this species towards the ISM, after the one reported towards the cold dense core Barnard 1$-$b by \citet{cernicharo2014}, using the same transition. Since only a single transition is detected, we fixed the $T_{\rm ex}$ to the value derived from NO, which is 11.2 K (see below). The result of the fit is shown in Table \ref{tab:parameters}. We obtained a column density of (3.89$\pm$0.05)$\times$10$^{13}$ cm$^{-2}$.

NO was already reported towards G+0.693 in \citet{zeng2020}, who detected the $J=$3/2$-$1/2 transitions (see Table \ref{tab:transitions}) using a previous IRAM 30m survey (\citealt{zeng2018}). Using data from the new survey, we have also detected the higher energy $J=$5/2$-$3/2 transitions with APEX (Table \ref{tab:transitions}), which allow us to constrain its excitation temperature more accurately.
We have obtained $T_{\rm ex}$=11.2$\pm$0.2 K, and a column density of (1.58$\pm$0.03)$\times$10$^{16}$ cm$^{-2}$, a factor of $\sim$2 lower than that reported by \citet{zeng2020}.
The best LTE fit slightly underestimates ($\sim$10$\%$) the line intensities of the transitions at 150 GHz. This  small discrepancy is within the typical uncertainties in the calibration of the IRAM 30m data, and therefore the LTE fit reproduces reasonably well both sets of lines, within the calibration uncertainties.
The derived ratio NO$^+$/NO is 0.00245$\pm$0.00005, very similar to that derived in the cold core Barnard 1-b of $\sim$0.002 (\citealt{cernicharo2014}). The NO$^+$/NO ratio found in G+0.693 is also similar to the SO$^+$/SO ratio (which is a factor of 1.7 higher), and a factor of $\sim$50 lower than the PO$^+$/PO ratio.



\begin{figure}[h]
\begin{center}
\includegraphics[width=12cm]{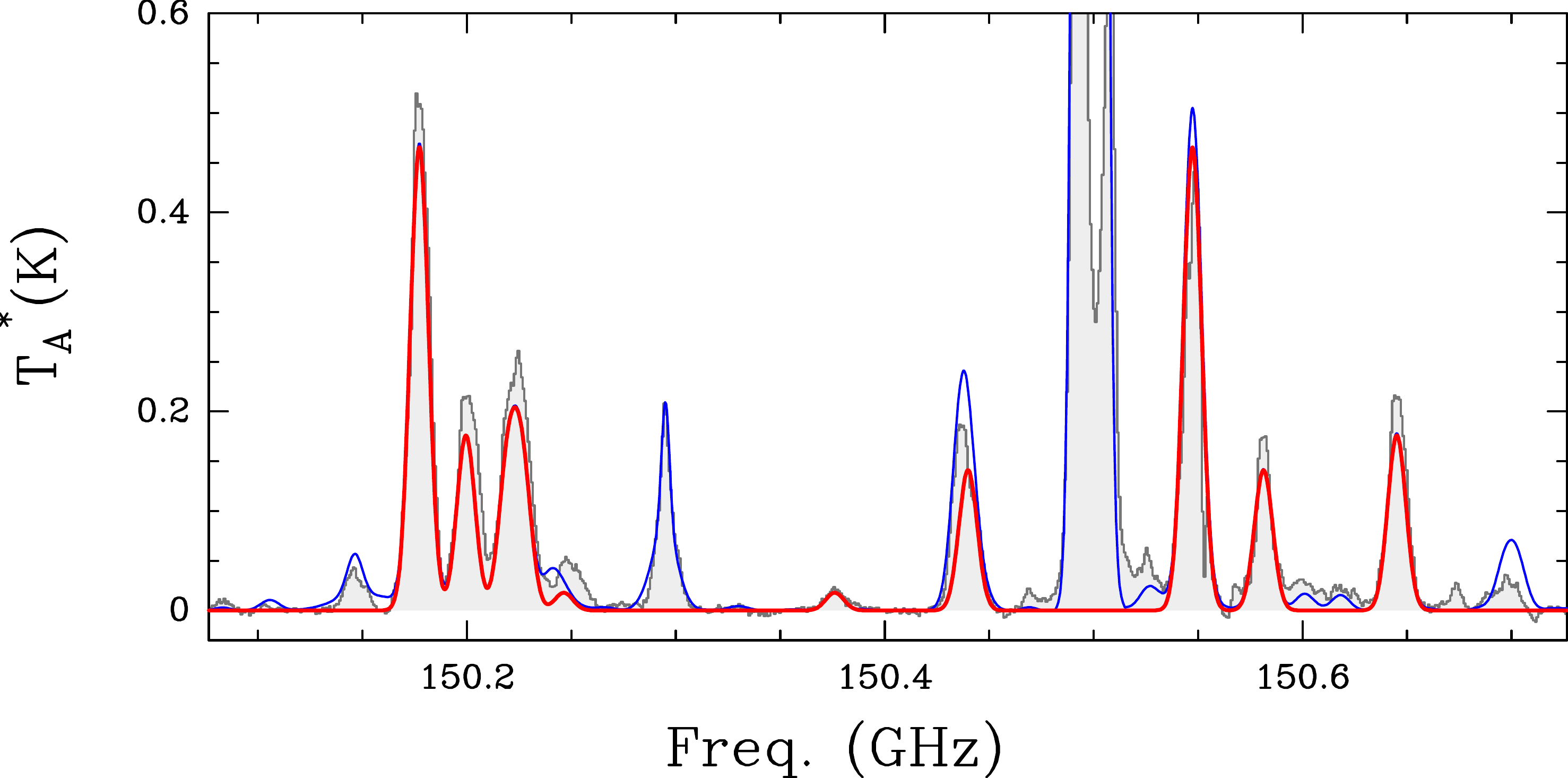}
%
%
\includegraphics[width=12cm]{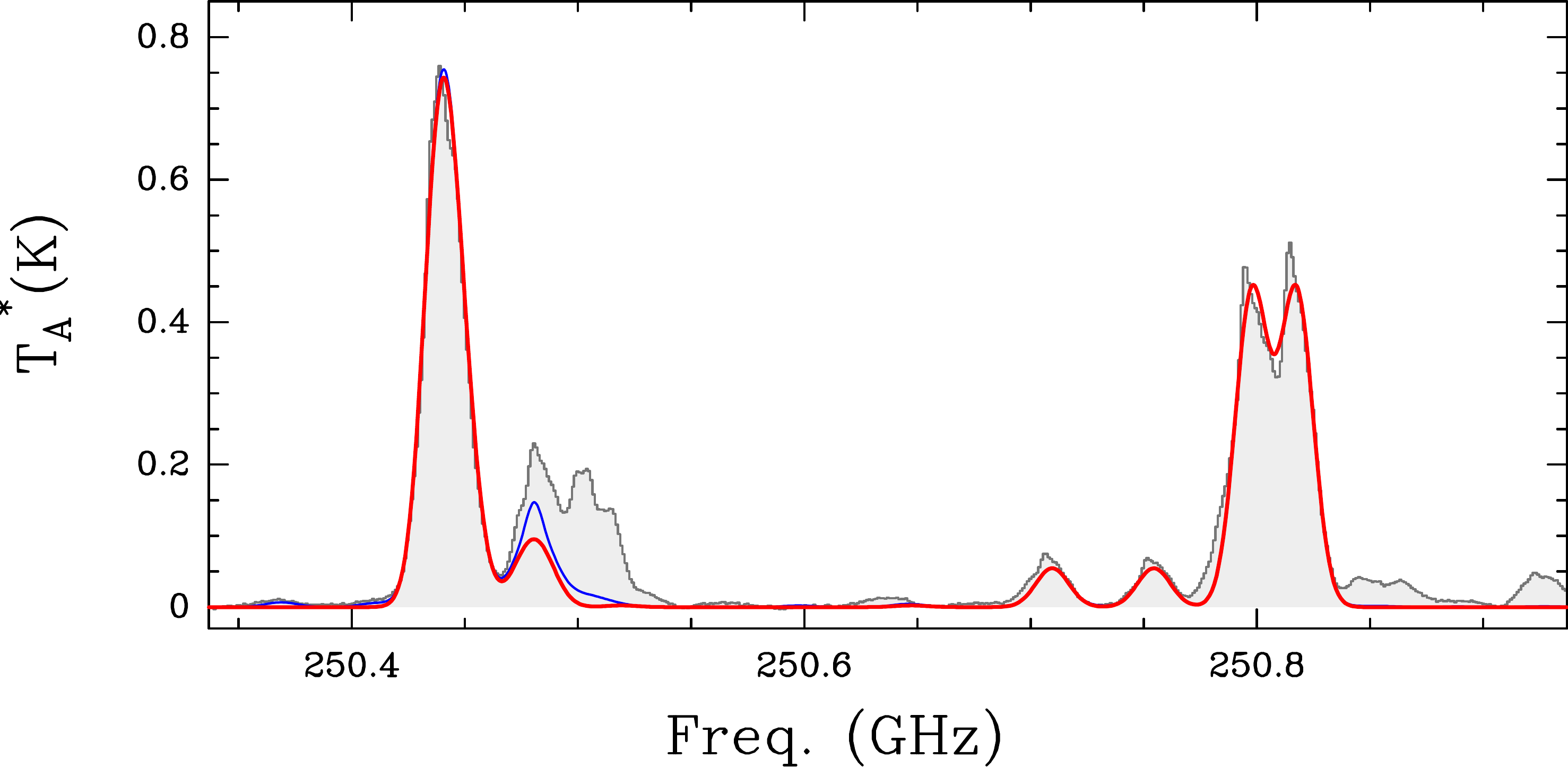}
\end{center}
\caption{NO transitions detected towards G+0.693. The gray histograms show the observed spectra obtained with the IRAM 30m telescope (upper panel) and the APEX telescope (lower panel). The unblended transitions used to perform the fit with MADCUBA are listed in Table \ref{tab:transitions}.  The red curve corresponds to the best LTE fit derived with MADCUBA, and the blue curve shows the total contribution considering all the molecular species identified including NO.}
\label{fig:no}
\end{figure}

\clearpage

\section{Discussion}

\subsection{Chemistry of PO$^+$, NO$^+$ and SO$^+$}
\label{sec:discussion}

We discuss in this section the chemistry of PO$^+$, so far poorly explored, comparing it with those of NO$^+$ and SO$^+$ proposed previously in the literature. Then, in section \ref{sec:model} we present the results of a new detailed chemical modelling focused on the formation of PO$^+$.

\begin{figure}[]
\begin{center}
\includegraphics[width=15cm]{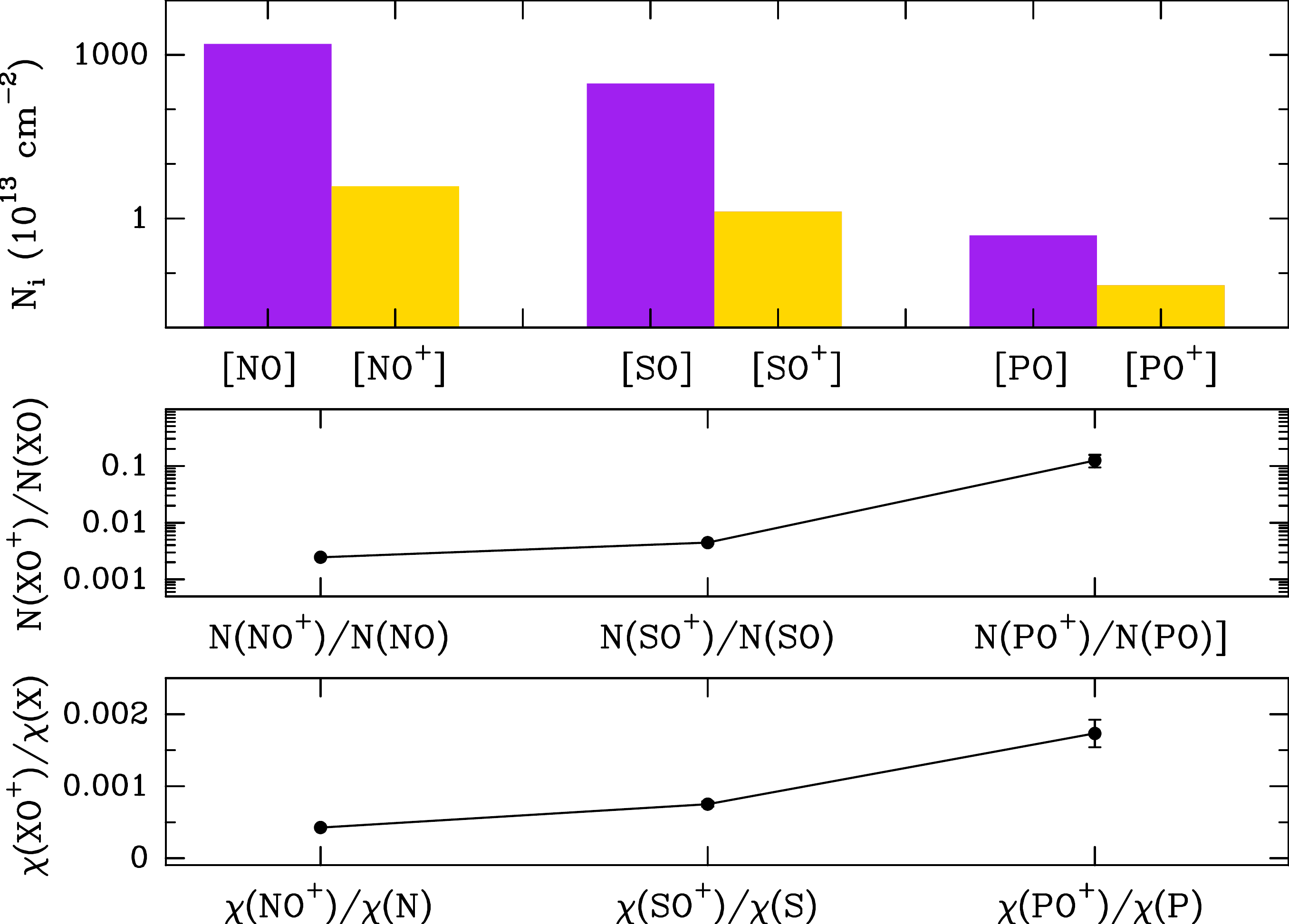}
\vskip3mm
\hspace{-1mm}
\includegraphics[width=14.9cm]{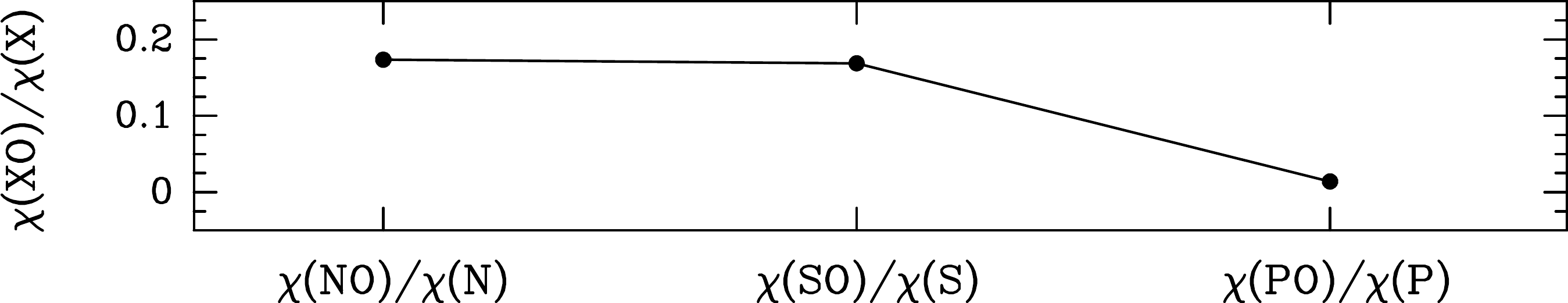}
\end{center}
\caption{{\it Upper panel}: molecular column densities of the N, S and P monoxides (purple) and ion monoxides (yellow) derived in G+0.693.
{\it Middle upper panel:} Molecular column density ratios of N, S and P monoxides with respect their associated ions in G+0.693.
{\it Middle Lower panel}: Molecular abundances of NO$^+$, SO$^+$, and PO$^+$  normalised by the cosmic abundances of N, S and P with respect to H.  Cosmic abundances of N, S and P are from \citealt{asplund2009}
{\it Lower panel:}  Molecular abundances of NO, SO, and PO normalised by the cosmic abundances of N, S and P with respect to H.}
\label{fig:ratios}
\end{figure}

The upper panel of Figure \ref{fig:ratios} shows the molecular column densities of NO$^+$, SO$^+$, and PO$^+$, and of their neutral counterparts derived in this work towards G+0.693. 
The two lowermost panels of Figure \ref{fig:ratios} show the molecular abundances of the ion monoxides (middle lower panel) and the monoxides (lower panel), normalised by the cosmic abundances of N, S and P with respect to H (from \citealt{asplund2009}). 
After the normalisation by cosmic abundances, PO$^{+}$ is the most abundant ion; however, PO is the least abundant monoxide (two lowermost panels of Figure \ref{fig:ratios}). These two facts together produce the observed enhancement of the $N$(PO$^{+}$)/$N$(PO) ratio (middle upper panel of Figure \ref{fig:ratios}), which is 
0.12$\pm$0.03, significantly higher than the $N$(SO$^{+}$)/$N$(SO) ratio (0.0045$\pm$0.0003), and the $N$(NO$^{+}$)/$N$(NO) ratio (0.00245$\pm$0.00005). 

%
 
\begin{figure}[]
\begin{center}
\includegraphics[width=18cm]{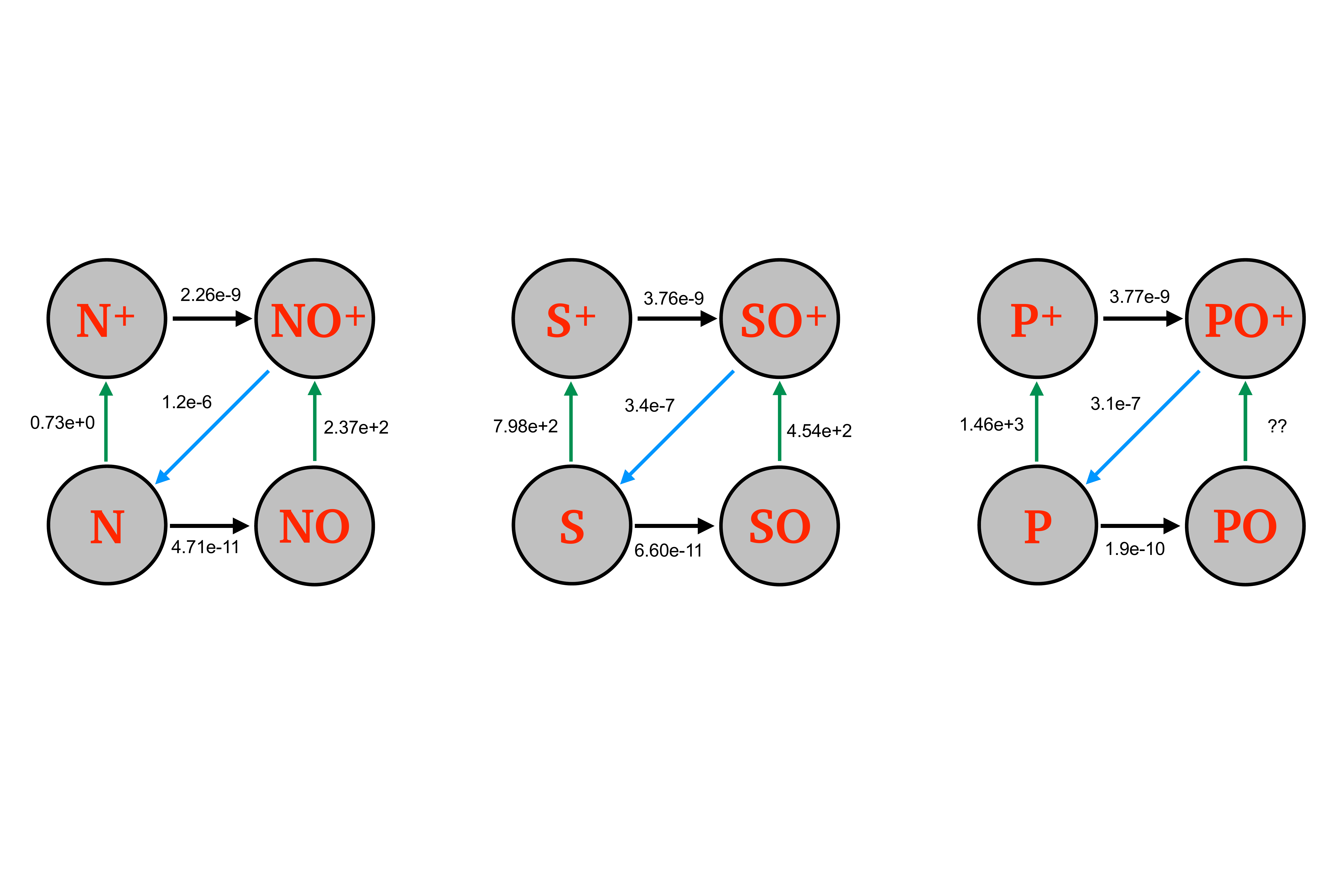}
\end{center}
\vskip-40mm
\caption{Chemical network of nitrogen, phosphorus and sulfur monoxides and ion monoxides. The green arrows indicate cosmic-ray ionisation reactions, the black arrows denote ion$-$molecule reactions with OH, and the blue arrows correspond to recombination with electron. The numbers indicate the values of $\alpha$ (where the reaction constant is $k$=$\alpha\times\zeta$
) for the cosmic-ray ionisation reactions, and the values of the $k$ for the ion$-$molecule and electron recombination reactions (calculated at $T$=100 K.)}
\label{fig:network}
\end{figure}

We discuss here the possible formations routes of these three ions, which are summarised in Figure \ref{fig:network}. We have included the rates ($k$) of the chemical reactions, obtained from KIDA (Kinetic Database for Astrochemistry, \citealt{wakelam2012}), and the calculations of \citet{garcia_de_la_concepcion2021}. The values, indicated with numbers above each arrow, denote the $\alpha$ parameter (where the reaction constant is $k$ = $\alpha\times\zeta$, with $\zeta$ being the cosmic-ray ionisation rate) for the cosmic-ray ionisation reactions, and the values of $k$ for the ion$-$molecule and electron recombination reactions calculated at $T$=100 K, which is the average gas kinetic temperature of the CMZ (e.g. \citealt{huettemeister_kinetic_1993,zeng2018}). 

Molecular ions are expected to be formed through gas-phase chemistry (e.g., \citealt{herbst_leung1986}). For the case of SO$^+$, several works (\citealt{neufeld1989,herbst_leung1989,turner1996}) have shown that the dominant formation route of SO$^+$ is the ion-molecule reaction S$^+$ +  OH $\rightarrow$  SO$^+$ + H.
%
%
\citet{neufeld1989} argued that the SO$^+$ abundance can be strongly enhanced in shocked regions due to the release of S from dust mantles, and its subsequent ionisation by cosmic rays, followed by the previous reaction. 
This chemical pathway has been supported by the detection of SO$^+$ in the shocked molecular clump associated with the supernova remnant IC 443G (\citealt{turner1992}), and in the protostellar shock L1157$-$B1 (\citealt{podio2014}). The detection of SO$^+$ towards G+0.693, where large-scale shocks are also present (e.g. \citealt{requena-torres_largest_2008,martin_tracing_2008,zeng2020}) further confirms the primary shock origin of SO$^+$. Moreover, regions with enhanced cosmic-ray ionisation rate, such as the CMZ in general (\citealt{goto2014}), and G+0.693 in particular (\citealt{zeng2018}), or the L1157$-$B1 shock (\citealt{podio2014}), favors the formation of S$^+$ through cosmic-ray ionisation reactions (Figure \ref{fig:network}), which also increase the formation efficiency of SO$^+$. 

Analogously, the same chemical route can be applied for N and P (left and right panels of Figure \ref{fig:network}). In this scenario, NO$^{+}$ and PO$^{+}$ can be formed through N$^+$ + OH $\rightarrow$  NO$^+$ + H, and P$^+$ + OH $\rightarrow$ PO$^+$ + H.
%
%
%
The reaction rates of S$^+$, N$^+$, and P$^+$ with the OH radical, calculated by \citet{woon2009}, are similar regardless of the temperature (see e.g. the values at 100 K in Figure \ref{fig:network}). However, the observations indicate that the abundance of PO$^{+}$ after normalisation by the cosmic abundance of P is higher than those of NO$^{+}$ and SO$^{+}$ (lower middle panel of Figure \ref{fig:ratios}).
This might be due to a higher cosmic-ray ionisation of atomic P compared with N and S giving P$^+$, N$^+$, and S$^+$, respectively (Figure \ref{fig:ratios}).
Indeed, the quantum calculations by \citet{heays2017} indicate that the reaction rates of P $\rightarrow$ P$^+$ is higher than that of S$\rightarrow$ S$^+$ (by a factor of 1.8), and N$\rightarrow$ N$^+$ (by a factor of $\sim$2000), as shown in Figure \ref{fig:network}. 

Another complementary chemical route for the formation of the monoxide ions (NO$^{+}$, SO$^{+}$ and PO$^{+}$) is the cosmic-ray ionisation of their neutral counterparts (NO, SO, and PO; Figure \ref{fig:network}). The latter can be produced by gas-phase neutral reactions between atomic N, S, and P with the OH radical (\citealt{pineau1990,turner1996,cernicharo2014,garcia_de_la_concepcion2021}). We note that they can also be formed with reactions with O$_2$ (KIDA and \citealt{garcia_de_la_concepcion2021}), although the reaction rates at least one order of magnitude lower. 
Previous observations have shown that the abundances of NO, SO, and PO are significantly enhanced in shocked regions (\citealt{bachiller1997,codella2018,rivilla2018,rivilla2020a}), which make them abundant possible progenitors for their associated ions. 
For the case of S, \citet{turner1992} found in the molecular clump IC 443C that SO$^{+}$ and SO do not coexist spatially, which seems to rule out this pathway, at least in this particular source. However, this might not be the case in other regions such as G+0.693, where both species have been detected, nor can be extrapolated to the cases of P and N.

As indicated in Figure \ref{fig:network}, the rates of the reactions with OH to form the oxides are higher for P than for S and N by a factor of 3$-$4. However, the lower panel of Figure \ref{fig:ratios} shows that the observed $\chi$(PO)/$\chi$(P) is lower than $\chi$(NO)/$\chi$(N) and $\chi$(SO)/$\chi$(S). This might indicate that PO is more efficiently destroyed in the gas phase than SO and NO. In this sense, a possible destruction route would be the cosmic-ray ionisation of PO to form PO$^+$ (Figure \ref{fig:network}). If this reaction were highly efficient, the abundance of PO would decrease, and the PO$^+/$PO ratio would be significantly enhanced, in agreement with observations. Unfortunately, this reaction is not included in the available chemical databases such as KIDA  (\citealt{wakelam2012}) or UMIST (\citealt{mcelroy2013}), so its rate is not known.
One can argue that the ionisation rates of PO should be higher that those of SO and NO, as occurs for the atomic species (see values in Figure \ref{fig:network}).
P is the less electronegative atom, thus the -I inductive effect produced by the O atom is more noticeable in PO than in SO or NO, as can be inferred from the dipole moments of the three oxides: $\mu$(PO)=1.88 D $>$ $\mu$(SO)=1.535 D $>$ $\mu$(NO)=0.159 D (Appendix \ref{app:spectroscopy}).
As a consequence, the valence electrons on the P atom of PO are more weakly retained than in the S of SO, and much less than in the N of NO. 
Furthermore, the highest occupied molecular orbital in PO is half-filled, being easier to lose an electron in comparison to the closed-shell orbital of SO. 
In section \ref{sec:model} we have included the cosmic-ray ionisation of PO in a chemical model to evaluate its role in the formation of  PO$^+$.


Finally, the main destruction mechanism of the monoxide ions is the dissociative recombination reaction with electrons (blue arrows in Figure \ref{fig:network}), which produces back the atomic species.
NO$^{+}$ is more rapidly destroyed by electrons than SO$^{+}$ and PO$^{+}$. This would further decrease the abundance of NO$^{+}$, which might explain why $\chi$(NO$^{+}$)/$\chi$(N) and $N$(NO$^{+}$)/$N$(NO) exhibits the lowest ratios (see Figure \ref{fig:ratios}).



Regarding the still poorly constrained chemistry of P in the ISM, the relatively high gas-phase abundance of PO$^{+}$ with respect to PO, $N$(PO$^{+}$)/$N$(PO)=0.12$\pm$0.03, stresses the more predominant role that P-bearing ions like PO$^+$ and P$^+$ can play in the chemical network of P, compared to N and S chemistries. In the next section we studied in more detail the formation of PO$^{+}$ in G+0.693 using the chemical models of P chemistry presented in \citet{Jimenez-Serra2018}.

\vskip3mm

\subsection{Chemical modelling of PO$^+$}
\label{sec:model}

To explain the abundances of PO$^+$ measured in G+0.693, we have used the chemical code UCLCHEM\footnote{https://uclchem.github.io/} (\citealt{holdship2017}) to simulate the P chemistry toward this source. It has been proposed that the chemistry of this source is characterised by low-velocity shocks (which explain the line widths of the molecular emission of 20 km s$^{-1}$) and by an enhanced cosmic-ray ionisation rate with respect to its standard value ($\zeta_0$=1.3$\times$10$^{-13}$ s$^{-1}$). UCLCHEM runs in three phases: Phase 0 considers the chemistry of a translucent cloud with $n$(H)=10$^{3}$ cm$^{-3}$ and $T_{\rm kin}$=20 K for 10$^{6}$ yrs. Phase 1 simulates the collapse of a molecular cloud from $n$(H)=10$^{3}$ cm$^{-3}$ to $n$(H)=2$\times$10$^{4}$ cm$^{-3}$ at a constant temperature of $T_{\rm kin}$=10 K. At this stage, most of atomic P is locked into solid PH$_3$. In Phase 2, we simulate the passage of a low-velocity C-type shock with v$_s$=20 km s$^{-1}$ and initial gas density of $n$(H)=10$^{4}$ cm$^{-3}$ using the parametric approximation for the physical structure of the C-type shocks of \citet{jimenez-serra2008}. The assumed shock velocity of v$_s$=20 km $^{-1}$ is consistent with the observed line widths of the molecular line emission (\citealt{requena-torres_organic_2006,zeng2018}), and with the gas densities measured toward G+0.693 (\citealt{zeng2020}). The maximum temperature reached within the shock is 900 K (see Table 4 in \citealt{jimenez-serra2008}). For the P chemistry network, we use the one built by \citet{Jimenez-Serra2018}, which has recently been updated with the rates of the reactions P + OH $\rightarrow$ PO + H and P + H$_2$O $\rightarrow$ PO + H$_2$ (\citealt{garcia_de_la_concepcion2021}). The initial elemental abundances are as in \citet{Jimenez-Serra2018}.
In this work we have implemented several updates:

\begin{enumerate}
    \item The reaction rate of the ionisation of P with a cosmic-ray photon (CR-photon),  P + CR-photon $\rightarrow$ P$^+$ + e$^{-}$, was obtained from the quantum chemical calculations by \citet{heays2017};
    \item The reaction rate of  P$^+$ + OH $\rightarrow$ PO$^+$ + H was obtained from the calculations by \citet{woon2009};
    \item We have introduced for the first time the formation route of PO$^{+}$ proposed in this work: PO + CR-photon $\rightarrow$ PO$^+$ + e$^{-}$. To our knowledge, there are neither laboratory experiments nor theoretical calculations carried out for this reaction. Therefore, we have assumed the rate of the analogous reaction with SO, from \citet{heays2017}. As discussed in Section \ref{sec:discussion}, the ionisation rate of PO might be higher than that of SO, so the value assumed here should be considered as a lower limit.
\end{enumerate}

\begin{figure*}
\begin{center}
\includegraphics[width=17.5cm]{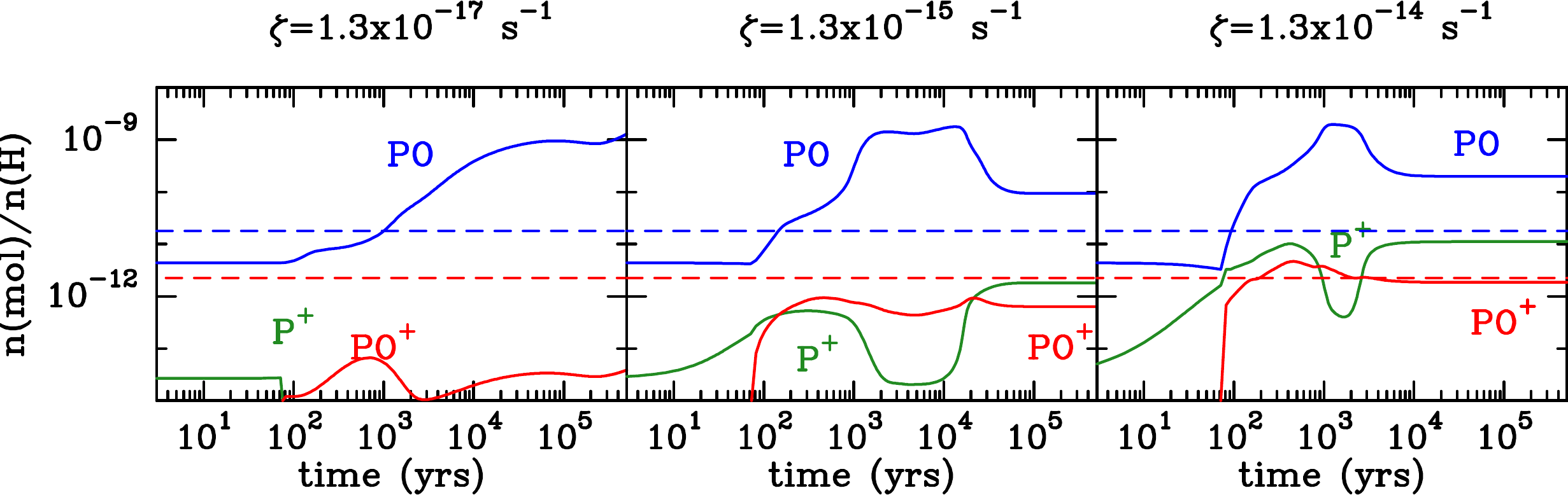}
\end{center}
\caption{Results of the chemical model: evolution of the abundances of PO$^+$ (red), PO (blue) and P$^+$ (green) as a function of time across a C-type shock with a pre-shock density of $n$(H)=2$\times$10$^4$ cm$^{-3}$, and
shock speed of v$_s$=20 km s$^{-1}$. We also consider that the shocked gas is affected by cosmic-ray ionisation. We compare the results using the standard Galactic value $\zeta$=1.3$\times$10$^{-17}$ s$^{-1}$ (left panel), with those of enhanced cosmic-ray ionisation rates of $\zeta$=1.3$\times$10$^{-15}$ s$^{-1}$ (middle panel), and $\zeta$=1.3$\times$10$^{-14}$ s$^{-1}$ (right panel), to simulate the extreme conditions in the Galactic center. The dashed horizontal lines denote the molecular abundances of PO$^+$ and PO derived from the observations of G+0.693, assuming $n$(H)=2$\times n$(H$_2$).}
\label{fig:model}
\end{figure*}

The chemistry of PO$^+$ is explored for three different values of the cosmic-ray ionisation rate $\zeta$=1.3$\times$10$^{-17}$ s$^{-1}$, 1.3$\times$10$^{-15}$ s$^{-1}$, and 1.3$\times$10$^{-14}$ s$^{-1}$, i.e. 1, 100 and 1000 times the standard value. The results of Phase 2 of the models for the abundances of PO$^+$, PO and P$^+$ are shown in Figure \ref{fig:model} (solid curves), compared to the observed values of PO$^+$ and PO (dashed horizontal lines). It is clear that an enhanced cosmic-ray ionisation rate of at least factors of 100$-$1000 is needed in order to obtain PO$^+$ abundances close to the observed value $\sim$10$^{-12}$. 
In our models, most of the P contained on dust grains is in the form of PH$_3$, which is initially released into the gas phase by the sputtering of the icy mantles
once their saturation time-scale is reached\footnote{The saturation time-scale is defined as the time at which 90\% of the content of the icy mantles is sputtered off grains (see \citealt{Jimenez-Serra2018}).}. PH$_3$ is rapidly converted into P due to the endothermic destruction reactions PH$_3$ + H $\rightarrow$ PH$_2$ + H$_2$, PH$_2$ + H $\rightarrow$ PH + H$_2$ and PH + H $\rightarrow$ P + H$_2$,
and photo-dissociation by secondary ultraviolet photons 
(see \citealt{Jimenez-Serra2018}). Atomic P is then ionised efficiently with enhanced values of $\zeta$, as can be seen from the higher abundance of P$^+$ in the models with $\zeta$=1.3$\times$10$^{-15}$ s$^{-1}$ and 1.3$\times$10$^{-14}$ s$^{-1}$ (Figure \ref{fig:model}), via the reaction P + CR-photon $\rightarrow$ P$^+$ + e$^-$. This is consistent with the conclusion drawn in Section \ref{sec:discussion}, of P being more easily ionised by cosmic rays than other elements such as N or S. 
In the model with $\zeta$=1.3$\times$10$^{-14}$ s$^{-1}$, which reproduces better the observed abundance of PO$^+$, the production of PO$^+$ is dominated by the reaction P$^+$ + O$_2$ $\rightarrow$ PO$^+$ + O at early post-shock times (60-600 yr), when O$_2$ is very abundant due to its release to the gas phase from the icy mantles after grain sputtering.
Later, the production of PO$^{+}$ is dominated by the reaction PO + CR-photon $\rightarrow$ PO$^+$ until 2500 yr. After that time, PO$^+$ is mainly formed by  P$^+$ + OH $\rightarrow$ PO$^+$ + H, with a contribution of the P$^+$ + O$_2$ pathway.
We stress that we have assumed a reaction rate for the cosmic-ray ionisation of PO, and that theoretical and experimental studies of this reaction would be helpful to evaluate its possible contribution to the formation of PO$^{+}$.



\section{Summary and conclusions}

We report the first detection of PO$^+$ in the interstellar medium, towards the molecular cloud G+0.693-0.027. We have detected the $J$=1$-$0 and $J$=2$-$1 transitions with the Yebes 40m and the IRAM 30m telescopes, respectively, which appear free of contamination from other species. The LTE analysis performed derives a column density of $N$=(6.0$\pm$0.7)$\times$10$^{11}$ cm$^{-2}$, and an abundance with respect to molecular hydrogen of 4.5$\times$10$^{-12}$.
The abundance of PO$^{+}$ normalised by the cosmic abundance of P is larger than those of NO$^{+}$ and SO$^{+}$, normalised by N and S, by factors of 3.6 and 2.3, respectively. As well, the $N$(PO$^{+}$)/$N$(PO) ratio is 0.12$\pm$0.03, more than one order of magnitude higher than those of $N$(SO$^{+}$)/$N$(SO) (0.0045$\pm$0.0003) and $N$(NO$^{+}$)/$N$(NO) (0.00264$\pm$0.00005). These results indicate that P is more efficiently ionised in the ISM than N and S.

We have performed a detailed chemical model that includes the effects of a C-type shock and high cosmic-ray ionisation rates ($\zeta$), to reproduce the physical conditions of G+0.693. The results show that the abundance of PO$^+$ is enhanced in shocked regions with high values of $\zeta$. Most of the P contained on dust grains, which is in the form of PH$_3$, is released into the gas phase by the sputtering of the icy mantles produced by the shock.
PH$_3$ is rapidly converted into atomic P, which is then ionised efficiently by cosmic rays. Later, PO$^+$ is formed by the reactions of P$^+$ with O$_2$ at early post-shock times, and by cosmic-ray ionisation of PO and by the reaction of P$^+$ with OH afterwards. 
Values of the cosmic-ray ionisation rates of $\zeta$=10$^{-15}-$10$^{-14}$ s$^{-1}$ are needed to obtain PO$^+$ abundances close to the observed value of several 10$^{-12}$. The relatively high gas-phase abundance of PO$^{+}$ with respect to PO, $N$(PO$^{+}$)/$N$(PO)=0.12$\pm$0.03, stresses the predominant role that P-bearing ions like PO$^+$ and P$^+$ can play in the chemical network of P.

\clearpage

\begin{table}
\centering
\tabcolsep 4pt
\caption{List of targeted transitions of PO$^+$, SO$^+$, and NO$^+$  ions, and their associated neutrals PO, S$^{18}$O, $^{34}$SO, and NO. We provide the transitions frequencies, quantum numbers, the base 10 logarithm of Einstein coefficients (log $A_{\rm ul}$), the upper state degeneracy ($g_{\rm u}$), and the upper energy level (E$_{\rm u}$). The last column states if the transitions are unblended or blended with another molecular species.}
\vspace{2mm}
\begin{tabular}{l c c c  c c l}
\hline
Molecule & Frequency & Transition    & log  $A_{\rm ul}$  & $g_{\rm u}$ & E$_{\rm u}$ &   Blending$^{(a)}$ \\
& (GHz) &   &  (s$^{-1}$) & & (K) &   \\
\hline
PO$^+$& 47.024252    &  $J=$1$-$0   & -5.32112  &  3  & 2.3  &  unblended  \\ 
PO$^+$ & 94.047798     &   $J=$2$-$1    & -4.33884  &  5  & 6.8  &  unblended   \\ 
PO$^+$ & 141.069934    &   $J=$3$-$2   & -3.78062	 & 7   & 13.5  &  H$^{13}$CCCN  \\ 
PO$^+$ & 235.107153    &   $J=$5$-$4   & -3.08956	 & 11   & 33.9  & Un.    \\ 
\hline
PO   & 108.99844500  &  $J=$5/2$-$3/2, $\Omega=$1/2, $F$=3$-$2 $l=$e   	 &  -4.67120  & 7  &   8.4 & unblended \\ 
PO  & 109.04539600 &  $J=$5/2$-$3/2, $\Omega=$1/2, $F$=2$-$1 $l=$e  & -4.71676	 &  5     &  8.4 & $s-$C$_2$H$_5$CHO \\ 
PO  & 109.20620000  & $J=$5/2$-$3/2, $\Omega=$1/2, $F$=3$-$2 $l=$f   & -4.66892	     &  7 &   8.4 & NH$_2$CH$_2$CH$_2$OH + Un. \\ 
PO  & 109.28118900 &  $J=$5/2$-$3/2, $\Omega=$1/2, $F$=2$-$1 $l=$f   	 &  -4.71433  & 5  &  8.4 & aGg$-$(CH$_2$OH)$_2$ \\ 
\hline
SO$^+$& 115.804405   &  $J=$5/2$-$3/2, $\Omega=$1/2, $l=$e   & 	-5.14768  &  6  & 8.9  &  unblended  \\ 
SO$^+$& 116.179947   & $J=$5/2$-$3/2, $\Omega=$1/2, $l=$f   & 	-5.14351  &  6  & 8.9  &  unblended  \\ 
SO$^+$& 162.198598   &  $J=$7/2$-$5/2, $\Omega=$1/2, $l=$e    & 	-4.67885  &  8  & 16.7  &  unblended  \\ 
SO$^+$& 162.574058   & $J=$7/2$-$5/2, $\Omega=$1/2, $l=$f   & 	-4.67575  &  8  & 16.7  &  unblended  \\ 
SO$^+$& 208.590016   &  $J=$9/2$-$7/2, $\Omega=$1/2, $l=$e    & 	-4.33524  & 10   & 26.7  &  unblended  \\ 
SO$^+$& 208.965420    &  $J=$9/2$-$7/2, $\Omega=$1/2, $l=$f   & -4.33294  &  10 &  26.8  &  unblended  \\ 
\hline
S$^{18}$O    & 93.267270   & $N_{J}$ = 2$_3-$1$_2$	 & -5.02955   & 7  &   8.7 & unblended  \\ 
\hline
$^{34}$SO    & 84.4106900  & $N_{J}$ = 2$_2-$1$_1$	 &  -5.30558  &  5 & 19.2  & unblended  \\ 
$^{34}$SO      & 97.715317   & $N_{J}$ = 3$_2-$2$_1$ 	 & -4.96948   &  7 & 9.1 &  unblended  \\ 
$^{34}$SO    & 106.743244  &  $N_{J}$ = 2$_1-$3$_2$ 	 &  -4.99699  & 5  &  20.9 & unblended   \\ 
$^{34}$SO      & 126.613930   &  $N_{J}$ = 3$_3-$2$_2$ 	 &  	-4.67350  & 7  & 25.3 &  unblended   \\ 
$^{34}$SO    & 135.775728   &  $N_{J}$ = 4$_3-$3$_2$  & -4.52226   & 9  & 15.6  & unblended  \\ 
$^{34}$SO      & 175.352766   & $N_{J}$ = 5$_4-$4$_3$ 	 &  -4.17742  &  11 & 24.0  &  unblended   \\ 
$^{34}$SO    &  215.839920  &  $N_{J}$ = 6$_5-$5$_4$ 	 &  -3.89897  &  13 & 34.4  & unblended  \\
\hline 
NO$^{+}$    &  238.3831574  & $N$=2$-$1	 & -5.0635   & 15  & 17.1  & unblended \\
\hline
NO    & 150.1764800  & $J=$3/2$-$1/2, $\Omega=$1/2$^+$, $F=$5/2$-$3/2	 & -6.4802   & 6  & 7.2  & unblended \\ 
NO    & 150.1987600  & $J=$3/2$-$1/2, $\Omega=$1/2$^+$, $F=$3/2$-$1/2	 & -6.7355   & 4  & 7.2  & Un.   \\ 
NO    & 150.2187300  & $J=$3/2$-$1/2, $\Omega=$1/2$^+$, $F=$3/2$-$3/2	 & -6.8323   & 4  & 7.2  & autoblended \\ 
NO    & 150.2256600  & $J=$3/2$-$1/2, $\Omega=$1/2$^+$, $F=$1/2$-$1/2	 & -6.5313   & 2  & 7.2  & autoblended \\ 
NO    & 150.2456356  & $J=$3/2$-$1/2, $\Omega=$1/2$^+$, $F=$1/2$-$3/2	         &    -7.4343       & 2  &  7.2     & HOCO$^+$ \\ 
NO    & 150.3752987  & $J=$3/2$-$1/2, $\Omega=$1/2$^-$, $F=$1/2$-$3/2	 &  -7.4319  & 2  & 7.2  & unblended  \\ 
NO    & 150.4391200  & $J=$3/2$-$1/2, $\Omega=$1/2$^-$, $F=$3/2$-$3/2	 & -6.8297   & 4  & 7.2  & c$-$C$_3$H$_2$ \\ 
NO    & 150.5465200  & $J=$3/2$-$1/2, $\Omega=$1/2$^-$, $F=$5/2$-$3/2	 & -6.4771   & 6  & 7.2  & unblended \\ 
NO    & 150.5805600  & $J=$3/2$-$1/2, $\Omega=$1/2$^-$, $F=$1/2$-$1/2	 & -6.5282   & 2  & 7.2  & unblended \\ 
NO    & 150.6443400  & $J=$3/2$-$1/2, $\Omega=$1/2$^-$, $F=$3/2$-$1/2	 & -6.7321   & 4  & 7.2  & unblended    \\ 
NO    & 250.4368480  & $J=$5/2$-$3/2, $\Omega=$1/2$^+$, $F=$7/2$-$5/2	 &  -5.7349  &  8  &  19.2 & autoblended \\ 
NO    & 250.4406590  & $J=$5/2$-$3/2, $\Omega=$1/2$^+$, $F=$5/2$-$3/2	 &  -5.8105  &  6  &  19.2 & autoblended \\ 
NO    & 250.4485300  & $J=$5/2$-$3/2, $\Omega=$1/2$^+$, $F=$3/2$-$1/2	 &  -5.8597  &  4  &  19.2 & autoblended \\ 
NO    & 250.4754140 & $J=$5/2$-$3/2, $\Omega=$1/2$^+$, $F=$3/2$-$3/2	 &  -6.3546  &  4  & 19.2 & autoblended + NS$^+$\\ 
NO    & 250.4829390 & $J=$5/2$-$3/2, $\Omega=$1/2$^+$, $F=$5/2$-$5/2	 &  	-6.5307  & 6   & 19.2 & autoblended + NS$^+$ \\ 
NO    & 250.5177030 & $J=$5/2$-$3/2, $\Omega=$1/2$^+$, $F=$3/2$-$5/2	 & -7.7347   &  4  & 19.2 & Un. \\ 
NO    & 250.6458000 & $J=$5/2$-$3/2, $\Omega=$1/2$^-$, $F=$3/2$-$5/2 &  -7.7332  & 4   & 19.3 & Un. \\ 
NO    & 250.7082450  & $J=$5/2$-$3/2, $\Omega=$1/2$^-$, $F=$5/2$-$5/2	 &  -6.5290  &  6  &  19.3 & unblended \\ 
NO    & 250.7531400  & $J=$5/2$-$3/2, $\Omega=$1/2$^-$, $F=$3/2$-$3/2	 &  -6.3529  &  4  &  19.3 & unblended \\ 
NO    & 250.7964360  & $J=$5/2$-$3/2, $\Omega=$1/2$^-$, $F=$7/2$-$5/2	 &  -5.7330  &  8  &  19.3 & autoblended \\ 
NO    & 250.8155940  & $J=$5/2$-$3/2, $\Omega=$1/2$^-$, $F=$5/2$-$3/2	 &  -5.8087  &  6  &  19.3 & autoblended \\ 
NO    & 250.8169540  & $J=$5/2$-$3/2, $\Omega=$1/2$^-$, $F=$3/2$-$1/2	 &  -5.8579  &  4  &  19.3 & autoblended \\ 
\hline
\end{tabular}
\label{tab:transitions}
{\\ (a) We indicate if the transition is blended with an already identified species, with an unidentified species (Un.), or autoblended (with a transition close in frequency of the same molecule).}
\end{table}

\begin{table}
\centering
\tabcolsep 5pt
\caption{Derived physical parameters of the molecules towards G+0.693 analysed in this work.}
\begin{tabular}{ l   c c c c c  }
\hline
 Molecule  &$N$   &  $T_{\rm ex}$ & v$_{\rm LSR}$ & $FWHM$  & Abundance$^a$   \\
 &  ($\times$10$^{13}$ cm$^{-2}$) & (K) & (km s$^{-1}$) & (km s$^{-1}$) & ($\times$10$^{-10}$)   \\
\hline
PO$^{+}$  & 0.060$\pm$0.007  & 4.5  & 70  & 18 & 0.045   \\ 
PO   & 0.49$\pm$0.09  & 4.5  & 69.5$\pm$1.4  & 15$\pm$3 & 0.36    \\
\hline
SO$^+$   & 1.34$\pm$0.07  & 8.0$\pm$0.3  & 68.8$\pm$0.3  & 18.0$\pm$0.7 & 1.0   \\
$^{34}$SO  &  6.5$\pm$0.2  & 6.9$\pm$0.1  & 68.3$\pm$0.2  & 22.1$\pm$0.4 & 4.8   \\
S$^{18}$O  &  1.20$\pm$0.03  & 6.9  & 67.9$\pm$0.3  & 24.8$\pm$0.7 & 0.89   \\
\hline
NO$^+$    &  3.89$\pm$0.05   & 11.2  & 68.1$\pm$0.2 & 18.1$\pm$0.3 & 3.2  \\
NO    & 1580$\pm$30   & 11.2$\pm$0.2  & 67.4$\pm$0.1 & 20.4$\pm$0.3 &  1170  \\
\hline 
\end{tabular}
\label{tab:parameters}
\vspace{0mm}
{\\ (a) We adopted $N_{\rm H_2}$=1.35$\times$10$^{23}$ cm$^{-2}$, from \citet{martin_tracing_2008}.
}
\end{table}

\clearpage

\section*{Conflict of Interest Statement}

The authors declare that the research was conducted in the absence of any commercial or financial relationships that could be construed as a potential conflict of interest.




\section*{Funding}
V.M.R., L.C. and A.L.-G. acknowledge support from the Comunidad de Madrid through the Atracción de Talento Investigador Modalidad 1 (Doctores con experiencia) Grant (COOL:Cosmic Origins of Life; 2019-T1/TIC-15379).
I.J.-S. and J.M.-P. have received partial support from the Spanish State Research Agency (AEI) through project number PID2019-105552RB-C41.
J.G.d.l.C. acknowledges the Spanish State Research Agency (AEI) through project number MDM-2017-0737 Unidad de Excelencia “María de Maeztu”—Centro de Astrobiología and the Spanish State Research Agency (AEI) for partial financial support through Project No. PID2019-105552RB-C41. Computational assistance was provided by the Supercomputer facilities of LUSITANIA founded by Cénits and Computaex Foundation.

\section*{Acknowledgments}

We thank the two reviewers for their careful reading and helpful comments that have contributed to improve the original version of the manuscript.
We are very grateful to the IRAM 30m, Yebes 40m and APEX telescope staff for their precious help during the different observing runs. IRAM is supported by the National Institute for Universe Sciences and Astronomy/National Center for Scientific Research (France), Max Planck Society for the Advancement of Science (Germany), and the National Geographic Institute (IGN) (Spain). The 40m radio telescope at Yebes Observatory is operated by the IGN, Ministerio de Transportes, Movilidad y Agenda Urbana. This publication is based on data acquired with the Atacama Pathfinder Experiment (APEX) under programmes ID O-0108.F-9308A-2021 and E-0108.C-0306A-2021. APEX is a collaboration between the
Max-Planck-Institut fur Radioastronomie, the European Southern Observatory,
and the Onsala Space Observatory.


\section*{Data Availability Statement}

The data underlying this article will be shared on reasonable request to the corresponding author.


 \bibliographystyle{frontiersinSCNS_ENG_HUMS} 
\bibliography{PO+}

\begin{thebibliography}{99}
\providecommand{\natexlab}[1]{#1}
\expandafter\ifx\csname urlstyle\endcsname\relax
  \providecommand{\doi}[1]{doi:\discretionary{}{}{}#1}\else
  \providecommand{\doi}{doi:\discretionary{}{}{}\begingroup
  \urlstyle{rm}\Url}\fi
\providecommand{\selectlanguage}[1]{\relax}
\providecommand{\bibAnnoteFile}[1]{%
  \IfFileExists{#1}{\begin{quotation}\noindent\textsc{Key:} #1\\
  \textsc{Annotation:}\ \input{#1}\end{quotation}}{}}
\providecommand{\bibAnnote}[2]{%
  \begin{quotation}\noindent\textsc{Key:} #1\\
  \textsc{Annotation:}\ #2\end{quotation}}

\bibitem[{Adler et~al.(2007)Adler, Knizia, and Werner}]{adler2007}
Adler, T.~B., Knizia, G., and Werner, H.-J. (2007).
\newblock {A simple and efficient CCSD(T)-F12 approximation}.
\newblock \emph{The Journal of Chemical Physics} 127, 221106.
\newblock \doi{10.1063/1.2817618}
\bibAnnoteFile{adler2007}

\bibitem[{{Ag{\'u}ndez} et~al.(2014){Ag{\'u}ndez}, {Cernicharo}, {Decin},
  {Encrenaz}, and {Teyssier}}]{Agundez2014}
{Ag{\'u}ndez}, M., {Cernicharo}, J., {Decin}, L., {Encrenaz}, P., and
  {Teyssier}, D. (2014).
\newblock {Confirmation of Circumstellar Phosphine}.
\newblock \emph{\apjl} 790, L27.
\newblock \doi{10.1088/2041-8205/790/2/L27}
\bibAnnoteFile{Agundez2014}

\bibitem[{{Ag{\'u}ndez} et~al.(2007){Ag{\'u}ndez}, {Cernicharo}, and
  {Gu{\'e}lin}}]{Agundez2007}
{Ag{\'u}ndez}, M., {Cernicharo}, J., and {Gu{\'e}lin}, M. (2007).
\newblock {Discovery of Phosphaethyne (HCP) in Space: Phosphorus Chemistry in
  Circumstellar Envelopes}.
\newblock \emph{\apjl} 662, L91--L94.
\newblock \doi{10.1086/519561}
\bibAnnoteFile{Agundez2007}

\bibitem[{{Altwegg} et~al.(2016){Altwegg}, {Balsiger}, {Bar-Nun}, {Berthelier},
  {Bieler}, {Bochsler} et~al.}]{altwegg2016}
{Altwegg}, K., {Balsiger}, H., {Bar-Nun}, A., {Berthelier}, J.~J., {Bieler},
  A., {Bochsler}, P., et~al. (2016).
\newblock {Prebiotic chemicals--amino acid and phosphorus--in the coma of comet
  67P/Churyumov-Gerasimenko}.
\newblock \emph{Science Advances} 2, e1600285--e1600285.
\newblock \doi{10.1126/sciadv.1600285}
\bibAnnoteFile{altwegg2016}

\bibitem[{{Amano} et~al.(1991){Amano}, {Amano}, and {Warner}}]{amano1991}
{Amano}, T., {Amano}, T., and {Warner}, H.~E. (1991).
\newblock {The microwave spectrum of SO $^{+}$}.
\newblock \emph{Journal of Molecular Spectroscopy} 146, 519--523.
\newblock \doi{10.1016/0022-2852(91)90023-4}
\bibAnnoteFile{amano1991}

\bibitem[{Andersson et~al.(1990)Andersson, Malmqvist, Roos, Sadlej, and
  Wolinski}]{CASPT2a}
Andersson, K., Malmqvist, P.~A., Roos, B.~O., Sadlej, A.~J., and Wolinski, K.
  (1990).
\newblock {Second-order perturbation theory with a CASSCF reference function}.
\newblock \emph{The Journal of Physical Chemistry} 94, 5483--5488.
\newblock \doi{10.1021/j100377a012}
\bibAnnoteFile{CASPT2a}

\bibitem[{{Asplund} et~al.(2009){Asplund}, {Grevesse}, {Sauval}, and
  {Scott}}]{asplund2009}
{Asplund}, M., {Grevesse}, N., {Sauval}, A.~J., and {Scott}, P. (2009).
\newblock {The Chemical Composition of the Sun}.
\newblock \emph{\araa} 47, 481--522.
\newblock \doi{10.1146/annurev.astro.46.060407.145222}
\bibAnnoteFile{asplund2009}

\bibitem[{{Bachiller} and {P{\'e}rez Guti{\'e}rrez}(1997)}]{bachiller1997}
{Bachiller}, R. and {P{\'e}rez Guti{\'e}rrez}, M. (1997).
\newblock {Shock Chemistry in the Young Bipolar Outflow L1157}.
\newblock \emph{\apjl} 487, L93--L96.
\newblock \doi{10.1086/310877}
\bibAnnoteFile{bachiller1997}

\bibitem[{Bailleux et~al.(2002)Bailleux, Bogey, Demuynck, Liu, and
  Walters}]{BAILLEUX2002}
Bailleux, S., Bogey, M., Demuynck, C., Liu, Y., and Walters, A. (2002).
\newblock {Millimeter-Wave Spectroscopy of PO in Excited Vibrational States up
  to v=7}.
\newblock \emph{Journal of Molecular Spectroscopy} 216, 465--471.
\newblock \doi{https://doi.org/10.1006/jmsp.2002.8665}
\bibAnnoteFile{BAILLEUX2002}

\bibitem[{Ben~Houria et~al.(2006)Ben~Houria, Ben~Lakhdar, and
  Hochlaf}]{Houria2006}
Ben~Houria, A., Ben~Lakhdar, Z., and Hochlaf, M. (2006).
\newblock {Spectroscopic and spin-orbit calculations on the SO$^+$ radical
  cation}.
\newblock \emph{The Journal of Chemical Physics} 124, 054313.
\newblock \doi{10.1063/1.2163872}
\bibAnnoteFile{Houria2006}

\bibitem[{{Bergner} et~al.(2019){Bergner}, {{\"O}berg}, {Walker}, {Guzm{\'a}n},
  {Rice}, and {Bergin}}]{Bergner2019}
{Bergner}, J.~B., {{\"O}berg}, K.~I., {Walker}, S., {Guzm{\'a}n}, V.~V.,
  {Rice}, T.~S., and {Bergin}, E.~A. (2019).
\newblock {Detection of Phosphorus-bearing Molecules toward a Solar-type
  Protostar}.
\newblock \emph{\apjl} 884, L36.
\newblock \doi{10.3847/2041-8213/ab48f9}
\bibAnnoteFile{Bergner2019}

\bibitem[{{Bernal} et~al.(2021){Bernal}, {Koelemay}, and {Ziurys}}]{Bernal2021}
{Bernal}, J.~J., {Koelemay}, L.~A., and {Ziurys}, L.~M. (2021).
\newblock {Detection of PO in Orion-KL: Phosphorus Chemistry in the Plateau
  Outflow}.
\newblock \emph{\apj} 906, 55.
\newblock \doi{10.3847/1538-4357/abc87b}
\bibAnnoteFile{Bernal2021}

\bibitem[{Bowman et~al.(1982)Bowman, Herbst, and De~Lucia}]{bowman1982}
Bowman, W.~C., Herbst, E., and De~Lucia, F.~C. (1982).
\newblock {Millimeter and submillimeter spectrum of NO$^+$}.
\newblock \emph{The Journal of Chemical Physics} 77, 4261--4262.
\newblock \doi{10.1063/1.444307}
\bibAnnoteFile{bowman1982}

\bibitem[{{Bregman} et~al.(1975){Bregman}, {Lester}, and {Rank}}]{Bregman1975}
{Bregman}, J.~D., {Lester}, D.~F., and {Rank}, D.~M. (1975).
\newblock {Observations of the {\ensuremath{\nu}}$_{2}$ band of PH$_{3}$ in the
  atmosphere of Saturn.}
\newblock \emph{\apjl} 202, L55.
\newblock \doi{10.1086/181979}
\bibAnnoteFile{Bregman1975}

\bibitem[{{Cernicharo} et~al.(2014){Cernicharo}, {Bailleux}, {Alekseev},
  {Fuente}, {Roueff}, {Gerin} et~al.}]{cernicharo2014}
{Cernicharo}, J., {Bailleux}, S., {Alekseev}, E., {Fuente}, A., {Roueff}, E.,
  {Gerin}, M., et~al. (2014).
\newblock {Tentative Detection of the Nitrosylium Ion in Space}.
\newblock \emph{\apj} 795, 40.
\newblock \doi{10.1088/0004-637X/795/1/40}
\bibAnnoteFile{cernicharo2014}

\bibitem[{{Chantzos} et~al.(2020){Chantzos}, {Rivilla}, {Vasyunin}, {Redaelli},
  {Bizzocchi}, {Fontani} et~al.}]{chantzos2020}
{Chantzos}, J., {Rivilla}, V.~M., {Vasyunin}, A., {Redaelli}, E., {Bizzocchi},
  L., {Fontani}, F., et~al. (2020).
\newblock {The first steps of interstellar phosphorus chemistry}.
\newblock \emph{\aap} 633, A54.
\newblock \doi{10.1051/0004-6361/201936531}
\bibAnnoteFile{chantzos2020}

\bibitem[{{Codella} et~al.(2018){Codella}, {Viti}, {Lefloch}, {Holdship},
  {Bachiller}, {Bianchi} et~al.}]{codella2018}
{Codella}, C., {Viti}, S., {Lefloch}, B., {Holdship}, J., {Bachiller}, R.,
  {Bianchi}, E., et~al. (2018).
\newblock {Nitrogen oxide in protostellar envelopes and shocks: the ASAI
  survey}.
\newblock \emph{\mnras} 474, 5694--5703.
\newblock \doi{10.1093/mnras/stx3196}
\bibAnnoteFile{codella2018}

\bibitem[{{Colzi} et~al.(2022){Colzi}, {Mart{\'\i}n-Pintado}, {Rivilla},
  {Jim{\'e}nez-Serra}, {Zeng}, {Rodr{\'\i}guez-Almeida} et~al.}]{colzi2022}
{Colzi}, L., {Mart{\'\i}n-Pintado}, J., {Rivilla}, V.~M., {Jim{\'e}nez-Serra},
  I., {Zeng}, S., {Rodr{\'\i}guez-Almeida}, L.~F., et~al. (2022).
\newblock {Deuterium Fractionation as a Multiphase Component Tracer in the
  Galactic Center}.
\newblock \emph{\apjl} 926, L22.
\newblock \doi{10.3847/2041-8213/ac52ac}
\bibAnnoteFile{colzi2022}

\bibitem[{Dyke et~al.(1982)Dyke, Morris, and Ridha}]{dyke1982}
Dyke, J.~M., Morris, A., and Ridha, A. (1982).
\newblock {Study of the ground state of PO$^+$ using photoelectron
  spectroscopy}.
\newblock \emph{J. Chem. Soc.{,} Faraday Trans. 2} 78, 2077--2082.
\newblock \doi{10.1039/F29827802077}
\bibAnnoteFile{dyke1982}

\bibitem[{{Endres} et~al.(2016){Endres}, {Schlemmer}, {Schilke}, {Stutzki}, and
  {M{\"u}ller}}]{endres2016}
{Endres}, C.~P., {Schlemmer}, S., {Schilke}, P., {Stutzki}, J., and
  {M{\"u}ller}, H. S.~P. (2016).
\newblock {The Cologne Database for Molecular Spectroscopy, CDMS, in the
  Virtual Atomic and Molecular Data Centre, VAMDC}.
\newblock \emph{Journal of Molecular Spectroscopy} 327, 95--104.
\newblock \doi{10.1016/j.jms.2016.03.005}
\bibAnnoteFile{endres2016}

\bibitem[{{Fagerbakke} et~al.(1996){Fagerbakke}, {Heldal}, and
  {Norland}}]{fagerbakke1996}
{Fagerbakke}, K., {Heldal}, M., and {Norland}, S. (1996).
\newblock {Content of carbon, nitrogen, oxygen, sulfur and phosphorus in native
  aquatic and cultured bacteria}.
\newblock \emph{Aquatic Microbial Ecology} 10.
\newblock \doi{10.3354/ame010015}
\bibAnnoteFile{fagerbakke1996}

\bibitem[{Feyereisen et~al.(1993)Feyereisen, Fitzgerald, and
  Komornicki}]{feyereisen1993}
Feyereisen, M., Fitzgerald, G., and Komornicki, A. (1993).
\newblock {Use of approximate integrals in ab initio theory. An application in
  MP2 energy calculations}.
\newblock \emph{Chemical Physics Letters} 208, 359--363.
\newblock \doi{https://doi.org/10.1016/0009-2614(93)87156-W}
\bibAnnoteFile{feyereisen1993}

\bibitem[{{Fontani} et~al.(2016){Fontani}, {Rivilla}, {Caselli}, {Vasyunin},
  and {Palau}}]{Fontani2016}
{Fontani}, F., {Rivilla}, V.~M., {Caselli}, P., {Vasyunin}, A., and {Palau}, A.
  (2016).
\newblock {Phosphorus-bearing Molecules in Massive Dense Cores}.
\newblock \emph{\apjl} 822, L30.
\newblock \doi{10.3847/2041-8205/822/2/L30}
\bibAnnoteFile{Fontani2016}

\bibitem[{{Fontani} et~al.(2019){Fontani}, {Rivilla}, {van der Tak}, {Mininni},
  {Beltr{\'a}n}, and {Caselli}}]{Fontani2019}
{Fontani}, F., {Rivilla}, V.~M., {van der Tak}, F.~F.~S., {Mininni}, C.,
  {Beltr{\'a}n}, M.~T., and {Caselli}, P. (2019).
\newblock {Origin of the PN molecule in star-forming regions: the enlarged
  sample}.
\newblock \emph{\mnras} 489, 4530--4542.
\newblock \doi{10.1093/mnras/stz2446}
\bibAnnoteFile{Fontani2019}

\bibitem[{{Garc{\'\i}a de la Concepci{\'o}n} et~al.(2021){Garc{\'\i}a de la
  Concepci{\'o}n}, {Puzzarini}, {Barone}, {Jim{\'e}nez-Serra}, and
  {Roncero}}]{garcia_de_la_concepcion2021}
{Garc{\'\i}a de la Concepci{\'o}n}, J., {Puzzarini}, C., {Barone}, V.,
  {Jim{\'e}nez-Serra}, I., and {Roncero}, O. (2021).
\newblock {Formation of phosphorus monoxide (PO) in the interstellar medium:
  insights from quantum-chemical and kinetic calculations}.
\newblock \emph{arXiv e-prints} , arXiv:2108.08530
\bibAnnoteFile{garcia_de_la_concepcion2021}

\bibitem[{{Goto} et~al.(2014){Goto}, {Geballe}, {Indriolo}, {Yusef-Zadeh},
  {Usuda}, {Henning} et~al.}]{goto2014}
{Goto}, M., {Geballe}, T.~R., {Indriolo}, N., {Yusef-Zadeh}, F., {Usuda}, T.,
  {Henning}, T., et~al. (2014).
\newblock {Infrared H\_3\^+ and CO Studies of the Galactic Core: GCIRS 3 and
  GCIRS 1W}.
\newblock \emph{\apj} 786, 96.
\newblock \doi{10.1088/0004-637X/786/2/96}
\bibAnnoteFile{goto2014}

\bibitem[{{Guelin} et~al.(1990){Guelin}, {Cernicharo}, {Paubert}, and
  {Turner}}]{Guelin1990}
{Guelin}, M., {Cernicharo}, J., {Paubert}, G., and {Turner}, B.~E. (1990).
\newblock {Free CP in IRC +10216.}
\newblock \emph{\aap} 230, L9--L11
\bibAnnoteFile{Guelin1990}

\bibitem[{{Haasler} et~al.(2021){Haasler}, {Rivilla}, {Mart{\'\i}n},
  {Holdship}, {Viti}, {Harada} et~al.}]{haasler2021}
{Haasler}, D., {Rivilla}, V.~M., {Mart{\'\i}n}, S., {Holdship}, J., {Viti}, S.,
  {Harada}, N., et~al. (2021).
\newblock {First extragalactic detection of a phosphorus-bearing molecule with
  ALCHEMI: phosphorus nitride (PN)}.
\newblock \emph{arXiv e-prints} , arXiv:2112.04849
\bibAnnoteFile{haasler2021}

\bibitem[{{Halfen} et~al.(2008){Halfen}, {Clouthier}, and
  {Ziurys}}]{Halfen2008}
{Halfen}, D.~T., {Clouthier}, D.~J., and {Ziurys}, L.~M. (2008).
\newblock {Detection of the CCP Radical (X$^{2}${\ensuremath{\Pi}}$_{r}$) in
  IRC +10216: A New Interstellar Phosphorus-containing Species}.
\newblock \emph{\apjl} 677, L101.
\newblock \doi{10.1086/588024}
\bibAnnoteFile{Halfen2008}

\bibitem[{Heays et~al.(2017)Heays, Bosman, and van Dishoeck}]{heays2017}
Heays, A.~N., Bosman, A.~D., and van Dishoeck, E.~F. (2017).
\newblock {Photodissociation and photoionisation of atoms and molecules of
  astrophysical interest}.
\newblock \emph{\aap} 602, A105.
\newblock \doi{10.1051/0004-6361/201628742}
\bibAnnoteFile{heays2017}

\bibitem[{{Herbst} and {Leung}(1986)}]{herbst_leung1986}
{Herbst}, E. and {Leung}, C.~M. (1986).
\newblock {Effects of Large Rate Coefficients for Ion-Polar Neutral Reactions
  on Chemical Models of Dense Interstellar Clouds}.
\newblock \emph{\apj} 310, 378.
\newblock \doi{10.1086/164691}
\bibAnnoteFile{herbst_leung1986}

\bibitem[{{Herbst} and {Leung}(1989)}]{herbst_leung1989}
{Herbst}, E. and {Leung}, C.~M. (1989).
\newblock {Gas Phase Production of Complex Hydrocarbons, Cyanopolyynes, and
  Related Compounds in Dense Interstellar Clouds}.
\newblock \emph{\apjs} 69, 271.
\newblock \doi{10.1086/191314}
\bibAnnoteFile{herbst_leung1989}

\bibitem[{{Holdship} et~al.(2017){Holdship}, {Viti}, {Jim{\'e}nez-Serra},
  {Makrymallis}, and {Priestley}}]{holdship2017}
{Holdship}, J., {Viti}, S., {Jim{\'e}nez-Serra}, I., {Makrymallis}, A., and
  {Priestley}, F. (2017).
\newblock {UCLCHEM: A Gas-grain Chemical Code for Clouds, Cores, and C-Shocks}.
\newblock \emph{\aj} 154, 38.
\newblock \doi{10.3847/1538-3881/aa773f}
\bibAnnoteFile{holdship2017}

\bibitem[{H\"uettemeister et~al.(1993)H\"uettemeister, Wilson, Bania, and
  Mart\'in-Pintado}]{huettemeister_kinetic_1993}
H\"uettemeister, S., Wilson, T.~L., Bania, T.~M., and Mart\'in-Pintado, J.
  (1993).
\newblock Kinetic temperatures in {Galactic} {Center} molecular clouds.
\newblock \emph{Astronomy and Astrophysics} 280, 255--267
\bibAnnoteFile{huettemeister_kinetic_1993}

\bibitem[{{Jim{\'e}nez-Serra} et~al.(2008){Jim{\'e}nez-Serra}, {Caselli},
  {Mart{\'\i}n-Pintado}, and {Hartquist}}]{jimenez-serra2008}
{Jim{\'e}nez-Serra}, I., {Caselli}, P., {Mart{\'\i}n-Pintado}, J., and
  {Hartquist}, T.~W. (2008).
\newblock {Parametrization of C-shocks. Evolution of the sputtering of grains}.
\newblock \emph{\aap} 482, 549--559.
\newblock \doi{10.1051/0004-6361:20078054}
\bibAnnoteFile{jimenez-serra2008}

\bibitem[{{Jim{\'e}nez-Serra} et~al.(2020){Jim{\'e}nez-Serra},
  {Mart{\'\i}n-Pintado}, {Rivilla}, {Rodr{\'\i}guez-Almeida}, {Alonso Alonso},
  {Zeng} et~al.}]{jimenez-serra2020}
{Jim{\'e}nez-Serra}, I., {Mart{\'\i}n-Pintado}, J., {Rivilla}, V.~M.,
  {Rodr{\'\i}guez-Almeida}, L., {Alonso Alonso}, E.~R., {Zeng}, S., et~al.
  (2020).
\newblock {Toward the RNA-World in the Interstellar
  Medium{\textemdash}Detection of Urea and Search of 2-Amino-oxazole and Simple
  Sugars}.
\newblock \emph{Astrobiology} 20, 1048--1066.
\newblock \doi{10.1089/ast.2019.2125}
\bibAnnoteFile{jimenez-serra2020}

\bibitem[{{Jim{\'e}nez-Serra} et~al.(2018){Jim{\'e}nez-Serra}, {Viti},
  {Qu{\'e}nard}, and {Holdship}}]{Jimenez-Serra2018}
{Jim{\'e}nez-Serra}, I., {Viti}, S., {Qu{\'e}nard}, D., and {Holdship}, J.
  (2018).
\newblock {The Chemistry of Phosphorus-bearing Molecules under Energetic
  Phenomena}.
\newblock \emph{\apj} 862, 128.
\newblock \doi{10.3847/1538-4357/aacdf2}
\bibAnnoteFile{Jimenez-Serra2018}

\bibitem[{Jura and York(1978)}]{jura_observations_1978}
Jura, M. and York, D.~G. (1978).
\newblock Observations of interstellar chlorine and phosphorus.
\newblock \emph{The Astrophysical Journal} 219, 861--869.
\newblock \doi{10.1086/155847}
\bibAnnoteFile{jura_observations_1978}

\bibitem[{Kanata et~al.(1988)Kanata, Yamamoto, and Saito}]{KANATA1988}
Kanata, H., Yamamoto, S., and Saito, S. (1988).
\newblock {The dipole moment of the PO radical determined by microwave
  spectroscopy}.
\newblock \emph{Journal of Molecular Spectroscopy} 131, 89--95.
\newblock \doi{https://doi.org/10.1016/0022-2852(88)90109-9}
\bibAnnoteFile{KANATA1988}

\bibitem[{Kawaguchi et~al.(1983)Kawaguchi, Saito, and Hirota}]{Kawaguchi1983}
Kawaguchi, K., Saito, S., and Hirota, E. (1983).
\newblock Far‐infrared laser magnetic resonance detection and microwave
  spectroscopy of the po radical.
\newblock \emph{The Journal of Chemical Physics} 79, 629--634.
\newblock \doi{10.1063/1.445810}
\bibAnnoteFile{Kawaguchi1983}

\bibitem[{Kendall et~al.(1992)Kendall, Dunning, and Harrison}]{dunningTZ}
Kendall, R.~A., Dunning, T.~H., and Harrison, R.~J. (1992).
\newblock {Electron affinities of the first‐row atoms revisited. Systematic
  basis sets and wave functions}.
\newblock \emph{The Journal of Chemical Physics} 96, 6796--6806.
\newblock \doi{10.1063/1.462569}
\bibAnnoteFile{dunningTZ}

\bibitem[{Knizia et~al.(2009)Knizia, Adler, and Werner}]{Knizia2009}
Knizia, G., Adler, T.~B., and Werner, H.-J. (2009).
\newblock {Simplified CCSD(T)-F12 methods: Theory and benchmarks}.
\newblock \emph{The Journal of Chemical Physics} 130, 054104.
\newblock \doi{10.1063/1.3054300}
\bibAnnoteFile{Knizia2009}

\bibitem[{{Latter} et~al.(1993){Latter}, {Walker}, and {Maloney}}]{latter1993}
{Latter}, W.~B., {Walker}, C.~K., and {Maloney}, P.~R. (1993).
\newblock {Detection of the Carbon Monoxide Ion (CO$^+$) in the Interstellar
  Medium and a Planetary Nebula}.
\newblock \emph{\apjl} 419, L97.
\newblock \doi{10.1086/187146}
\bibAnnoteFile{latter1993}

\bibitem[{Lee and Taylor(1989)}]{T1}
Lee, T.~J. and Taylor, P.~R. (1989).
\newblock A diagnostic for determining the quality of single-reference electron
  correlation methods.
\newblock \emph{International Journal of Quantum Chemistry} 36, 199--207.
\newblock \doi{https://doi.org/10.1002/qua.560360824}
\bibAnnoteFile{T1}

\bibitem[{{Lefloch} et~al.(2016){Lefloch}, {Vastel}, {Viti}, {Jimenez-Serra},
  {Codella}, {Podio} et~al.}]{lefloch2016}
{Lefloch}, B., {Vastel}, C., {Viti}, S., {Jimenez-Serra}, I., {Codella}, C.,
  {Podio}, L., et~al. (2016).
\newblock {Phosphorus-bearing molecules in solar-type star-forming regions:
  first PO detection}.
\newblock \emph{\mnras} 462, 3937--3944.
\newblock \doi{10.1093/mnras/stw1918}
\bibAnnoteFile{lefloch2016}

\bibitem[{{Lovas} et~al.(1992){Lovas}, {Suenram}, {Ogata}, and
  {Yamamoto}}]{lovas1992}
{Lovas}, F.~J., {Suenram}, R.~D., {Ogata}, T., and {Yamamoto}, S. (1992).
\newblock {Microwave Spectra and Electric Dipole Moments for Low- J Levels of
  Interstellar Radicals: SO, C 2S, C 3S, c-HC 3, CH 2CC, and c-C 3H 2}.
\newblock \emph{\apj} 399, 325.
\newblock \doi{10.1086/171928}
\bibAnnoteFile{lovas1992}

\bibitem[{{Mancini} et~al.(2020){Mancini}, {Rosi}, {Balucani}, {Skouteris},
  {Codella}, and {Ceccarelli}}]{mancini2020}
{Mancini}, L., {Rosi}, M., {Balucani}, N., {Skouteris}, D., {Codella}, C., and
  {Ceccarelli}, C. (2020).
\newblock {Probing the Chemistry of P-Bearing Molecules in Interstellar
  Environments and other Extraterrestrial Environments}.
\newblock In \emph{European Planetary Science Congress}. EPSC2020--643
\bibAnnoteFile{mancini2020}

\bibitem[{{Mart{\'\i}n} et~al.(2019){Mart{\'\i}n}, {Mart{\'\i}n-Pintado},
  {Blanco-S{\'a}nchez}, {Rivilla}, {Rodr{\'\i}guez-Franco}, and
  {Rico-Villas}}]{martin2019}
{Mart{\'\i}n}, S., {Mart{\'\i}n-Pintado}, J., {Blanco-S{\'a}nchez}, C.,
  {Rivilla}, V.~M., {Rodr{\'\i}guez-Franco}, A., and {Rico-Villas}, F. (2019).
\newblock {Spectral Line Identification and Modelling (SLIM) in the MAdrid Data
  CUBe Analysis (MADCUBA) package. Interactive software for data cube
  analysis}.
\newblock \emph{\aap} 631, A159.
\newblock \doi{10.1051/0004-6361/201936144}
\bibAnnoteFile{martin2019}

\bibitem[{Mart\'in et~al.(2008)Mart\'in, Requena-Torres, Mart\'in-Pintado, and
  Mauersberger}]{martin_tracing_2008}
Mart\'in, S., Requena-Torres, M.~A., Mart\'in-Pintado, J., and Mauersberger, R.
  (2008).
\newblock Tracing shocks and photodissociation in the {Galactic} center region.
\newblock \emph{The Astrophysical Journal} 678, 245--254.
\newblock \doi{10.1086/533409}.
\newblock ArXiv: 0801.3614
\bibAnnoteFile{martin_tracing_2008}

\bibitem[{{McElroy} et~al.(2013){McElroy}, {Walsh}, {Markwick}, {Cordiner},
  {Smith}, and {Millar}}]{mcelroy2013}
{McElroy}, D., {Walsh}, C., {Markwick}, A.~J., {Cordiner}, M.~A., {Smith}, K.,
  and {Millar}, T.~J. (2013).
\newblock {The UMIST database for astrochemistry 2012}.
\newblock \emph{Astronomy and Astrophysics} 550, A36.
\newblock \doi{10.1051/0004-6361/201220465}
\bibAnnoteFile{mcelroy2013}

\bibitem[{{Mininni} et~al.(2018){Mininni}, {Fontani}, {Rivilla}, {Beltr{\'a}n},
  {Caselli}, and {Vasyunin}}]{mininni2018}
{Mininni}, C., {Fontani}, F., {Rivilla}, V.~M., {Beltr{\'a}n}, M.~T.,
  {Caselli}, P., and {Vasyunin}, A. (2018).
\newblock {On the origin of phosphorus nitride in star-forming regions}.
\newblock \emph{\mnras} 476, L39--L44.
\newblock \doi{10.1093/mnrasl/sly026}
\bibAnnoteFile{mininni2018}

\bibitem[{Moussaoui et~al.(2003)Moussaoui, Ouamerali, and
  Maré}]{Moussaoui2003}
Moussaoui, Y., Ouamerali, O., and Maré, G. R.~D. (2003).
\newblock {Properties of the phosphorus oxide radical, PO, its cation and anion
  in their ground electronic states: comparison of theoretical and experimental
  data}.
\newblock \emph{International Reviews in Physical Chemistry} 22, 641--675.
\newblock \doi{10.1080/01442350310001617011}
\bibAnnoteFile{Moussaoui2003}

\bibitem[{Müller et~al.(2015)Müller, Kobayashi, Takahashi, Tomaru, and
  Matsushima}]{MULLER2015}
Müller, H.~S., Kobayashi, K., Takahashi, K., Tomaru, K., and Matsushima, F.
  (2015).
\newblock {Terahertz spectroscopy of N$^{18}$O and isotopic invariant fit of
  several nitric oxide isotopologs}.
\newblock \emph{Journal of Molecular Spectroscopy} 310, 92--98.
\newblock \doi{https://doi.org/10.1016/j.jms.2014.12.002}.
\newblock Spectroscopy of Radicals and Ions in Memory of Marilyn Jacox
\bibAnnoteFile{MULLER2015}

\bibitem[{{Neufeld} and {Dalgarno}(1989)}]{neufeld1989}
{Neufeld}, D.~A. and {Dalgarno}, A. (1989).
\newblock {Fast Molecular Shocks. I. Re-formation of Molecules behind a
  Dissociative Shock}.
\newblock \emph{\apj} 340, 869.
\newblock \doi{10.1086/167441}
\bibAnnoteFile{neufeld1989}

\bibitem[{{Neumann}(1970)}]{neumann1970}
{Neumann}, R.~M. (1970).
\newblock {High-Precision Radiofrequency Spectrum of $^{14}$N $^{16}$O}.
\newblock \emph{\apj} 161, 779.
\newblock \doi{10.1086/150578}
\bibAnnoteFile{neumann1970}

\bibitem[{{Padovani} et~al.(2009){Padovani}, {Galli}, and
  {Glassgold}}]{Padovani2009}
{Padovani}, M., {Galli}, D., and {Glassgold}, A.~E. (2009).
\newblock {Cosmic-ray ionization of molecular clouds}.
\newblock \emph{\aap} 501, 619--631.
\newblock \doi{10.1051/0004-6361/200911794}
\bibAnnoteFile{Padovani2009}

\bibitem[{{Pasek} and {Lauretta}(2005)}]{pasek2005}
{Pasek}, M.~A. and {Lauretta}, D.~S. (2005).
\newblock {Aqueous Corrosion of Phosphide Minerals from Iron Meteorites: A
  Highly Reactive Source of Prebiotic Phosphorus on the Surface of the Early
  Earth}.
\newblock \emph{Astrobiology} 5, 515--535.
\newblock \doi{10.1089/ast.2005.5.515}
\bibAnnoteFile{pasek2005}

\bibitem[{Peterson and Woods(1990)}]{peterson1990}
Peterson, K.~A. and Woods, R.~C. (1990).
\newblock Configuration interaction potential energy and dipole moment
  functions for thirteen 22 electron diatomics.
\newblock \emph{The Journal of Chemical Physics} 92, 6061--6068.
\newblock \doi{10.1063/1.458378}
\bibAnnoteFile{peterson1990}

\bibitem[{Peterson et~al.(1994)Peterson, Woon, and Dunning}]{dunning5Z}
Peterson, K.~A., Woon, D.~E., and Dunning, T.~H. (1994).
\newblock {Benchmark calculations with correlated molecular wave functions. IV.
  The classical barrier height of the H+H$_2$→H$_2$+H reaction}.
\newblock \emph{The Journal of Chemical Physics} 100, 7410--7415.
\newblock \doi{10.1063/1.466884}
\bibAnnoteFile{dunning5Z}

\bibitem[{Petrmichl et~al.(1991)Petrmichl, Peterson, and Woods}]{petrmichl1991}
Petrmichl, R.~H., Peterson, K.~A., and Woods, R.~C. (1991).
\newblock {The microwave spectrum of PO$^+$: Comparison to SiF$^+$}.
\newblock \emph{The Journal of Chemical Physics} 94, 3504--3510.
\newblock \doi{10.1063/1.459771}
\bibAnnoteFile{petrmichl1991}

\bibitem[{{Pickett} et~al.(1998){Pickett}, {Poynter}, {Cohen}, {Delitsky},
  {Pearson}, and {M{\"u}ller}}]{pickett1998}
{Pickett}, H.~M., {Poynter}, R.~L., {Cohen}, E.~A., {Delitsky}, M.~L.,
  {Pearson}, J.~C., and {M{\"u}ller}, H.~S.~P. (1998).
\newblock {Submillimeter, millimeter and microwave spectral line catalog.}
\newblock \emph{\jqsrt} 60, 883--890.
\newblock \doi{10.1016/S0022-4073(98)00091-0}
\bibAnnoteFile{pickett1998}

\bibitem[{{Pineau des Forets} et~al.(1990){Pineau des Forets}, {Roueff}, and
  {Flower}}]{pineau1990}
{Pineau des Forets}, G., {Roueff}, E., and {Flower}, D.~R. (1990).
\newblock {The formation of nitrogen-bearing species in dark interstellar
  clouds.}
\newblock \emph{\mnras} 244, 668--674
\bibAnnoteFile{pineau1990}

\bibitem[{{Podio} et~al.(2014){Podio}, {Lefloch}, {Ceccarelli}, {Codella}, and
  {Bachiller}}]{podio2014}
{Podio}, L., {Lefloch}, B., {Ceccarelli}, C., {Codella}, C., and {Bachiller},
  R. (2014).
\newblock {Molecular ions in the protostellar shock L1157-B1}.
\newblock \emph{\aap} 565, A64.
\newblock \doi{10.1051/0004-6361/201322928}
\bibAnnoteFile{podio2014}

\bibitem[{Polák and Fišer(2004)}]{POLAK2004}
Polák, R. and Fišer, J. (2004).
\newblock {A comparative icMRCI study of some NO$^+$, NO and NO$^-$ electronic
  ground state properties}.
\newblock \emph{Chemical Physics} 303, 73--83.
\newblock \doi{https://doi.org/10.1016/j.chemphys.2004.04.027}
\bibAnnoteFile{POLAK2004}

\bibitem[{Powell and Lide(1964)}]{powell1964}
Powell, F.~X. and Lide, D.~R. (1964).
\newblock Microwave spectrum of the so radical.
\newblock \emph{The Journal of Chemical Physics} 41, 1413--1419.
\newblock \doi{10.1063/1.1726082}
\bibAnnoteFile{powell1964}

\bibitem[{Requena-Torres et~al.(2008)Requena-Torres, Mart\'in-Pintado,
  Mart\'in, and Morris}]{requena-torres_largest_2008}
Requena-Torres, M.~A., Mart\'in-Pintado, J., Mart\'in, S., and Morris, M.~R.
  (2008).
\newblock The largest oxigen bearing organic molecule repository.
\newblock \emph{The Astrophysical Journal} 672, 352--360.
\newblock \doi{10.1086/523627}.
\newblock ArXiv: 0709.0542
\bibAnnoteFile{requena-torres_largest_2008}

\bibitem[{Requena-Torres et~al.(2006)Requena-Torres, Mart\'in-Pintado,
  Rodr\'iguez-Franco, Mart\'in, Rodr\'iguez-Fern\'andez, and
  de~Vicente}]{requena-torres_organic_2006}
Requena-Torres, M.~A., Mart\'in-Pintado, J., Rodr\'iguez-Franco, A., Mart\'in,
  S., Rodr\'iguez-Fern\'andez, N.~J., and de~Vicente, P. (2006).
\newblock Organic {Molecules} in the {Galactic} {Center}. {Hot} {Core}
  {Chemistry} without {Hot} {Cores}.
\newblock \emph{Astronomy \& Astrophysics} 455, 971--985.
\newblock \doi{10.1051/0004-6361:20065190}.
\newblock ArXiv: astro-ph/0605031
\bibAnnoteFile{requena-torres_organic_2006}

\bibitem[{{Ridgway} et~al.(1976){Ridgway}, {Wallace}, and
  {Smith}}]{ridgway1976}
{Ridgway}, S.~T., {Wallace}, L., and {Smith}, G.~R. (1976).
\newblock {The 800 - 1200 inverse centimeter absorption spectrum of Jupiter.}
\newblock \emph{\apj} 207, 1002--1006.
\newblock \doi{10.1086/154570}
\bibAnnoteFile{ridgway1976}

\bibitem[{Rienstra-Kiracofe et~al.(2000)Rienstra-Kiracofe, Allen, and
  Schaefer}]{T12}
Rienstra-Kiracofe, J.~C., Allen, W.~D., and Schaefer, H.~F. (2000).
\newblock {The C$_2$H$_5$ + O$_2$ Reaction Mechanism: High-Level ab Initio
  Characterizations}.
\newblock \emph{The Journal of Physical Chemistry A} 104, 9823--9840.
\newblock \doi{10.1021/jp001041k}
\bibAnnoteFile{T12}

\bibitem[{{Rivilla} et~al.(2019){Rivilla}, {Beltr{\'a}n}, {Vasyunin},
  {Caselli}, {Viti}, {Fontani} et~al.}]{rivilla2019a}
{Rivilla}, V.~M., {Beltr{\'a}n}, M.~T., {Vasyunin}, A., {Caselli}, P., {Viti},
  S., {Fontani}, F., et~al. (2019).
\newblock {First ALMA maps of HCO, an important precursor of complex organic
  molecules, towards IRAS 16293-2422}.
\newblock \emph{\mnras} 483, 806--823.
\newblock \doi{10.1093/mnras/sty3078}
\bibAnnoteFile{rivilla2019a}

\bibitem[{{Rivilla} et~al.(2020{\natexlab{a}}){Rivilla}, {Drozdovskaya},
  {Altwegg}, {Caselli}, {Beltr{\'a}n}, {Fontani} et~al.}]{rivilla2020a}
{Rivilla}, V.~M., {Drozdovskaya}, M.~N., {Altwegg}, K., {Caselli}, P.,
  {Beltr{\'a}n}, M.~T., {Fontani}, F., et~al. (2020{\natexlab{a}}).
\newblock {ALMA and ROSINA detections of phosphorus-bearing molecules: the
  interstellar thread between star-forming regions and comets}.
\newblock \emph{\mnras} 492, 1180--1198.
\newblock \doi{10.1093/mnras/stz3336}
\bibAnnoteFile{rivilla2020a}

\bibitem[{Rivilla et~al.(2016)Rivilla, Fontani, Beltr\'an, Vasyunin, Caselli,
  Mart\'in-Pintado et~al.}]{rivilla_first_2016}
Rivilla, V.~M., Fontani, F., Beltr\'an, M.~T., Vasyunin, A., Caselli, P.,
  Mart\'in-Pintado, J., et~al. (2016).
\newblock The {First} {Detections} of the {Key} {Prebiotic} {Molecule} {PO} in
  {Star}-forming {Regions}.
\newblock \emph{The Astrophysical Journal} 826, 161.
\newblock \doi{10.3847/0004-637X/826/2/161}
\bibAnnoteFile{rivilla_first_2016}

\bibitem[{{Rivilla} et~al.(2016){Rivilla}, {Fontani}, {Beltr{\'a}n},
  {Vasyunin}, {Caselli}, {Mart{\'\i}n-Pintado} et~al.}]{Rivilla2016}
{Rivilla}, V.~M., {Fontani}, F., {Beltr{\'a}n}, M.~T., {Vasyunin}, A.,
  {Caselli}, P., {Mart{\'\i}n-Pintado}, J., et~al. (2016).
\newblock {The First Detections of the Key Prebiotic Molecule PO in
  Star-forming Regions}.
\newblock \emph{\apj} 826, 161.
\newblock \doi{10.3847/0004-637X/826/2/161}
\bibAnnoteFile{Rivilla2016}

\bibitem[{{Rivilla} et~al.(2021b){Rivilla}, {Jim{\'e}nez-Serra}, {Garc{\'\i}a
  de la Concepci{\'o}n}, {Mart{\'\i}n-Pintado}, {Colzi},
  {Rodr{\'\i}guez-Almeida} et~al.}]{rivilla2021b}
{Rivilla}, V.~M., {Jim{\'e}nez-Serra}, I., {Garc{\'\i}a de la Concepci{\'o}n},
  J., {Mart{\'\i}n-Pintado}, J., {Colzi}, L., {Rodr{\'\i}guez-Almeida}, L.~F.,
  et~al. (2021b).
\newblock {Detection of the cyanomidyl radical (HNCN): a new interstellar
  species with the NCN backbone}.
\newblock \emph{\mnras} 506, L79--L84.
\newblock \doi{10.1093/mnrasl/slab074}
\bibAnnoteFile{rivilla2021b}

\bibitem[{Rivilla et~al.(2021a)Rivilla, Jim{\'e}nez-Serra, Mart{\'\i}n-Pintado,
  Briones, Rodr{\'\i}guez-Almeida, Rico-Villas et~al.}]{rivilla2021a}
Rivilla, V.~M., Jim{\'e}nez-Serra, I., Mart{\'\i}n-Pintado, J., Briones, C.,
  Rodr{\'\i}guez-Almeida, L.~F., Rico-Villas, F., et~al. (2021a).
\newblock Discovery in space of ethanolamine, the simplest phospholipid head
  group.
\newblock \emph{PNAS} 118.
\newblock \doi{10.1073/pnas.2101314118}
\bibAnnoteFile{rivilla2021a}

\bibitem[{{Rivilla} et~al.(2018){Rivilla}, {Jim{\'e}nez-Serra}, {Zeng},
  {Mart{\'\i}n}, {Mart{\'\i}n-Pintado}, {Armijos-Abenda{\~n}o}
  et~al.}]{rivilla2018}
{Rivilla}, V.~M., {Jim{\'e}nez-Serra}, I., {Zeng}, S., {Mart{\'\i}n}, S.,
  {Mart{\'\i}n-Pintado}, J., {Armijos-Abenda{\~n}o}, J., et~al. (2018).
\newblock {Phosphorus-bearing molecules in the Galactic Center}.
\newblock \emph{\mnras} 475, L30--L34.
\newblock \doi{10.1093/mnrasl/slx208}
\bibAnnoteFile{rivilla2018}

\bibitem[{{Rivilla} et~al.(2020{\natexlab{b}}){Rivilla}, {Mart{\'\i}n-Pintado},
  {Jim{\'e}nez-Serra}, {Mart{\'\i}n}, {Rodr{\'\i}guez-Almeida},
  {Requena-Torres} et~al.}]{rivilla2020b}
{Rivilla}, V.~M., {Mart{\'\i}n-Pintado}, J., {Jim{\'e}nez-Serra}, I.,
  {Mart{\'\i}n}, S., {Rodr{\'\i}guez-Almeida}, L.~F., {Requena-Torres}, M.~A.,
  et~al. (2020{\natexlab{b}}).
\newblock {Prebiotic Precursors of the Primordial RNA World in Space: Detection
  of NH$_{2}$OH}.
\newblock \emph{\apjl} 899, L28.
\newblock \doi{10.3847/2041-8213/abac55}
\bibAnnoteFile{rivilla2020b}

\bibitem[{Roca-Sanjuán et~al.(2012)Roca-Sanjuán, Aquilante, and
  Lindh}]{CASPT2b}
Roca-Sanjuán, D., Aquilante, F., and Lindh, R. (2012).
\newblock {Multiconfiguration second-order perturbation theory approach to
  strong electron correlation in chemistry and photochemistry}.
\newblock \emph{WIREs Computational Molecular Science} 2, 585--603.
\newblock \doi{https://doi.org/10.1002/wcms.97}
\bibAnnoteFile{CASPT2b}

\bibitem[{{Rodr{\'\i}guez-Almeida}
  et~al.(2021{\natexlab{a}}){Rodr{\'\i}guez-Almeida}, {Jim{\'e}nez-Serra},
  {Rivilla}, {Mart{\'\i}n-Pintado}, {Zeng}, {Tercero}
  et~al.}]{rodriguez-almeida2021}
{Rodr{\'\i}guez-Almeida}, L.~F., {Jim{\'e}nez-Serra}, I., {Rivilla}, V.~M.,
  {Mart{\'\i}n-Pintado}, J., {Zeng}, S., {Tercero}, B., et~al.
  (2021{\natexlab{a}}).
\newblock {Thiols in the Interstellar Medium: First Detection of HC(O)SH and
  Confirmation of C$_{2}$H$_{5}$SH}.
\newblock \emph{\apjl} 912, L11.
\newblock \doi{10.3847/2041-8213/abf7cb}
\bibAnnoteFile{rodriguez-almeida2021}

\bibitem[{{Rodr{\'\i}guez-Almeida}
  et~al.(2021{\natexlab{b}}){Rodr{\'\i}guez-Almeida}, {Rivilla},
  {Jim{\'e}nez-Serra}, {Melosso}, {Colzi}, {Zeng}
  et~al.}]{rodriguez-Almeida2021b}
{Rodr{\'\i}guez-Almeida}, L.~F., {Rivilla}, V.~M., {Jim{\'e}nez-Serra}, I.,
  {Melosso}, M., {Colzi}, L., {Zeng}, S., et~al. (2021{\natexlab{b}}).
\newblock {First detection of C$_{2}$H$_{5}$NCO in the ISM and search of other
  isocyanates towards the G+0.693-0.027 molecular cloud}.
\newblock \emph{\aap} 654, L1.
\newblock \doi{10.1051/0004-6361/202141989}
\bibAnnoteFile{rodriguez-Almeida2021b}

\bibitem[{Schwartz(2006)}]{schwartz2006}
Schwartz, A.~W. (2006).
\newblock {Phosphorus in prebiotic chemistry}.
\newblock \emph{Philosophical Transactions of the Royal Society B: Biological
  Sciences} 361, 1743--1749.
\newblock \doi{10.1098/rstb.2006.1901}
\bibAnnoteFile{schwartz2006}

\bibitem[{Tenenbaum et~al.(2007)Tenenbaum, Woolf, and
  Ziurys}]{tenenbaum_identification_2007}
Tenenbaum, E.~D., Woolf, N.~J., and Ziurys, L.~M. (2007).
\newblock Identification of {Phosphorus} {Monoxide} ({X}2{$\pi$r}) in {VY}
  {Canis} {Majoris}: {Detection} of the {First} {PO} {Bond} in {Space}.
\newblock \emph{The Astrophysical Journal Letters} 666, L29--L32.
\newblock \doi{10.1086/521361}
\bibAnnoteFile{tenenbaum_identification_2007}

\bibitem[{{Tercero} et~al.(2021){Tercero}, {L{\'o}pez-P{\'e}rez}, {Gallego},
  {Beltr{\'a}n}, {Garc{\'\i}a}, {Patino-Esteban} et~al.}]{tercero2021}
{Tercero}, F., {L{\'o}pez-P{\'e}rez}, J.~A., {Gallego}, J.~D., {Beltr{\'a}n},
  F., {Garc{\'\i}a}, O., {Patino-Esteban}, M., et~al. (2021).
\newblock {Yebes 40 m radio telescope and the broad band Nanocosmos receivers
  at 7 mm and 3 mm for line surveys}.
\newblock \emph{\aap} 645, A37.
\newblock \doi{10.1051/0004-6361/202038701}
\bibAnnoteFile{tercero2021}

\bibitem[{{Thorne} et~al.(1984){Thorne}, {Anicich}, {Prasad}, and
  {Huntress}}]{Thorne1987}
{Thorne}, L.~R., {Anicich}, V.~G., {Prasad}, S.~S., and {Huntress}, J., W.~T.
  (1984).
\newblock {The chemistry of phosphorus in dense interstellar clouds.}
\newblock \emph{\apj} 280, 139--143.
\newblock \doi{10.1086/161977}
\bibAnnoteFile{Thorne1987}

\bibitem[{Tiemann(1974)}]{tiemann1974}
Tiemann, E. (1974).
\newblock {Microwave Spectra of Molecules of Astrophysical Interest VIII.
  Sulfur Monoxide}.
\newblock \emph{Journal of Physical and Chemical Reference Data} 3, 259--268.
\newblock \doi{10.1063/1.3253141}
\bibAnnoteFile{tiemann1974}

\bibitem[{{Turner}(1992)}]{turner1992}
{Turner}, B.~E. (1992).
\newblock {Detection of Interstellar SO$^+$: A Diagnostic of Dissociative Shock
  Chemistry}.
\newblock \emph{\apjl} 396, L107.
\newblock \doi{10.1086/186528}
\bibAnnoteFile{turner1992}

\bibitem[{{Turner}(1996)}]{turner1996}
{Turner}, B.~E. (1996).
\newblock {The Physics and Chemistry of Small Translucent Molecular Clouds.
  VII. SO$^+$ and H 2S}.
\newblock \emph{\apj} 468, 694.
\newblock \doi{10.1086/177727}
\bibAnnoteFile{turner1996}

\bibitem[{Turner and Bally(1987)}]{turner_detection_1987}
Turner, B.~E. and Bally, J. (1987).
\newblock Detection of interstellar {PN} - {The} first identified phosphorus
  compound in the interstellar medium.
\newblock \emph{The Astrophysical Journal Letters} 321, L75--L79.
\newblock \doi{10.1086/185009}
\bibAnnoteFile{turner_detection_1987}

\bibitem[{{Wakelam} et~al.(2012){Wakelam}, {Herbst}, {Loison}, {Smith},
  {Chandrasekaran}, {Pavone} et~al.}]{wakelam2012}
{Wakelam}, V., {Herbst}, E., {Loison}, J.-C., {Smith}, I.~W.~M.,
  {Chandrasekaran}, V., {Pavone}, B., et~al. (2012).
\newblock {A KInetic Database for Astrochemistry (KIDA)}.
\newblock \emph{The Astrophysical Journal Supplement} 199, 21.
\newblock \doi{10.1088/0067-0049/199/1/21}
\bibAnnoteFile{wakelam2012}

\bibitem[{Wilson et~al.(1996)Wilson, {van Mourik}, and Dunning}]{dunning6Z}
Wilson, A.~K., {van Mourik}, T., and Dunning, T.~H. (1996).
\newblock {Gaussian basis sets for use in correlated molecular calculations.
  VI. Sextuple zeta correlation consistent basis sets for boron through neon}.
\newblock \emph{Journal of Molecular Structure: THEOCHEM} 388, 339--349.
\newblock \doi{https://doi.org/10.1016/S0166-1280(96)80048-0}
\bibAnnoteFile{dunning6Z}

\bibitem[{Wilson and Rood(1994)}]{wilson_abundances_1994}
Wilson, T.~L. and Rood, R. (1994).
\newblock Abundances in the {Interstellar} {Medium}.
\newblock \emph{Annual Review of Astronomy and Astrophysics} 32, 191--226.
\newblock \doi{10.1146/annurev.aa.32.090194.001203}
\bibAnnoteFile{wilson_abundances_1994}

\bibitem[{Woon and Dunning(1993)}]{dunningQZ}
Woon, D.~E. and Dunning, T.~H. (1993).
\newblock {Gaussian basis sets for use in correlated molecular calculations.
  III. The atoms aluminum through argon}.
\newblock \emph{The Journal of Chemical Physics} 98, 1358--1371.
\newblock \doi{10.1063/1.464303}
\bibAnnoteFile{dunningQZ}

\bibitem[{Woon and Herbst(2009)}]{woon2009}
Woon, D.~E. and Herbst, E. (2009).
\newblock {Quantum Chemical Predictions of the Properties of Known and
  Postulated Neutral Interstellar Molecules}.
\newblock \emph{Astrophysical Journal Supplement Series} 185, 273--288.
\newblock \doi{10.1088/0067-0049/185/2/273}
\bibAnnoteFile{woon2009}

\bibitem[{Xing et~al.(2012)Xing, Shi, Sun, and Zhu}]{wei2012}
Xing, W., Shi, D., Sun, J., and Zhu, Z. (2012).
\newblock {Investigation of Spectroscopic Properties and Spin-Orbit Splitting
  in the $X2\pi$ and $A2\pi$ Electronic States of the SO$^+$ Cation}.
\newblock \emph{International Journal of Molecular Sciences} 13, 8189--8209.
\newblock \doi{10.3390/ijms13078189}
\bibAnnoteFile{wei2012}

\bibitem[{{Zeng} et~al.(2018){Zeng}, {Jim{\'e}nez-Serra}, {Rivilla},
  {Mart{\'{\i}}n}, {Mart{\'{\i}}n-Pintado}, {Requena-Torres} et~al.}]{zeng2018}
{Zeng}, S., {Jim{\'e}nez-Serra}, I., {Rivilla}, V.~M., {Mart{\'{\i}}n}, S.,
  {Mart{\'{\i}}n-Pintado}, J., {Requena-Torres}, M.~A., et~al. (2018).
\newblock {Complex organic molecules in the Galactic Centre: the N-bearing
  family}.
\newblock \emph{Monthly Notices of the Royal Astronomical Society} 478,
  2962--2975.
\newblock \doi{10.1093/mnras/sty1174}
\bibAnnoteFile{zeng2018}

\bibitem[{{Zeng} et~al.(2021){Zeng}, {Jim{\'e}nez-Serra}, {Rivilla},
  {Mart{\'\i}n-Pintado}, {Rodr{\'\i}guez-Almeida}, {Tercero} et~al.}]{zeng2021}
{Zeng}, S., {Jim{\'e}nez-Serra}, I., {Rivilla}, V.~M., {Mart{\'\i}n-Pintado},
  J., {Rodr{\'\i}guez-Almeida}, L.~F., {Tercero}, B., et~al. (2021).
\newblock {Probing the Chemical Complexity of Amines in the ISM: Detection of
  Vinylamine (C$_{2}$H$_{3}$NH$_{2}$) and Tentative Detection of Ethylamine
  (C$_{2}$H$_{5}$NH$_{2}$)}.
\newblock \emph{\apjl} 920, L27.
\newblock \doi{10.3847/2041-8213/ac2c7e}
\bibAnnoteFile{zeng2021}

\bibitem[{{Zeng} et~al.(2020){Zeng}, {Zhang}, {Jim{\'e}nez-Serra}, {Tercero},
  {Lu}, {Mart{\'\i}n-Pintado} et~al.}]{zeng2020}
{Zeng}, S., {Zhang}, Q., {Jim{\'e}nez-Serra}, I., {Tercero}, B., {Lu}, X.,
  {Mart{\'\i}n-Pintado}, J., et~al. (2020).
\newblock {Cloud-cloud collision as drivers of the chemical complexity in
  Galactic Centre molecular clouds}.
\newblock \emph{MNRAS} 497, 4896--4909.
\newblock \doi{10.1093/mnras/staa2187}
\bibAnnoteFile{zeng2020}

\bibitem[{Zhang and Shi(2021)}]{Zhang2021}
Zhang, M. and Shi, D. (2021).
\newblock {Transition properties of X1$\Sigma$+, A1$\Sigma$-, B1$\Delta$,
  C1$\Pi$, a3$\Sigma$+, b3$\Delta$, c3$\Pi$, and d3$\Sigma$- states of PO$^+$}.
\newblock \emph{Journal of Quantitative Spectroscopy and Radiative Transfer}
  264, 107553.
\newblock \doi{https://doi.org/10.1016/j.jqsrt.2021.107553}
\bibAnnoteFile{Zhang2021}

\bibitem[{{Ziurys}(1987)}]{Ziurys1987}
{Ziurys}, L.~M. (1987).
\newblock {Detection of Interstellar PN: The First Phosphorus-bearing Species
  Observed in Molecular Clouds}.
\newblock \emph{\apjl} 321, L81.
\newblock \doi{10.1086/185010}
\bibAnnoteFile{Ziurys1987}

\end{thebibliography}






\clearpage
\section*{Appendix A: Molecular spectroscopy}
\label{app:spectroscopy}

We list in Table \ref{tab:spectroscopy} the details of the spectroscopy used to fit the molecular emission of the different species analysed in this work. We have used entries from the Cologne Database for Molecular Spectroscopy (CDMS,\citealt{endres2016}) and the Jet Propulsion Laboratory (JPL; \citealt{pickett1998}), which are based on the laboratory works and theoretical calculations indicated in Table \ref{tab:spectroscopy}.

\begin{table}
\centering
\tabcolsep 3pt
\caption{Spectroscopy of the molecules analysed in this work. The molecular catalog, number and date of the entry, and the references for the line lists and dipole moments are listed.}
\begin{tabular}{l c c c c c }
\hline
Molecule & Catalog  &   Entry & Date & Line list ref. & Dipole moment ref. \\ 
\hline
PO$^+$ & JPL & 47005 & December 1996 & \citet{petrmichl1991} & \citet{peterson1990};   \\
 & & &   & & this work \\
 \hline
PO &  CDMS & 47507 & October 2019 &  \citet{Kawaguchi1983},   & \citet{KANATA1988} \\ 
 &   &  &  & \citet{BAILLEUX2002}  &  \\ 
\hline
SO$^+$ & JPL &  48010 & December 1996 & \citet{amano1991} &  \citet{turner1992}, from  \\ 
 & & &   & & Peterson \& Woods, priv. comm; \\
 & & &   & & this work \\
\hline
$^{34}$SO & CDMS & 50501  & August 1998 &  \citet{tiemann1974} &  \citet{powell1964},  \\
 & & &   & & \citet{lovas1992}$^{a}$ \\
 \hline
S$^{18}$O & CDMS & 50502  & August 1998 &  \citet{tiemann1974} & \citet{powell1964},   \\
 & & &   & & \citet{lovas1992}$^{a}$ \\
\hline
NO$^+$ & CDMS & 30512 & January 2017 & \citet{bowman1982} & \citet{POLAK2004}  \\ 
\hline
NO & CDMS & 30517  & January 2015 & \citet{MULLER2015} & \citet{neumann1970}  \\ 
\hline 
\end{tabular}
\label{tab:spectroscopy}
\vspace{0mm}
{\\ (a) The dipole moment of SO was used.
}
\end{table}

\clearpage

\section*{Appendix B: Calculation of the PO$^+$ dipole moment}
\label{app:po+dipole}

Due to the lack of experimental measurements of permanent dipole for the ion PO$^+$, quantum chemical calculations are needed to obtain it. There are previous theoretical calculations of the permanent dipole moment for the cation PO$^+$(${^1}\Sigma$) (\citealt{peterson1990}), who derived a value of $\mu$=3.44. This is the value used in the JPL molecular database entry (47005, December 1996) of PO$^+$ (see Table \ref{tab:spectroscopy}). In this work we have carried out new calculations for updating this value. Thus, we optimized the PO$^+$ (${^1}\Sigma$)  with the CCSD(T)-F12 method \citep{adler2007,Knizia2009}. With the inclusion of the explicit F12 electronic correlation, we can obtain results near the basis set limit with smaller basis sets. To do this, we selected de cc-pVTZ-F12 and cc-pVTZ-F12-CABS as orbital basis set and complete auxiliary basis set respectively. For the correlation and coulomb fitting we increase the size of the basis set to the augmented aug-cc-pVQZ. The coupled cluster calculations showed that the $T{_1}$ diagnostic obtained for this molecule is 0.025. The T1 diagnostic is an indicator of the multi-configurational character of a chemical species \citep{T1}. When this value exceeds 0.020 for a closed shell system or 0.045 for an open shell system, multi-configurational methods are needed to correctly describe the wavenfunction \citep{T12}. In this case, 0.025 lies above the limit for a closed shell system. Thus, our new, and previous coupled cluster calculations \citep{Moussaoui2003}, and those computed  with quadratic CI methods \citep{peterson1990} were not taking into account herein, since multi-configurational calculations are needed to correctly describe this molecule. Recently, high level multi-configurational calculations have been carried out to describe electronic spectroscopic properties of the PO$^+$ cation like transition dipole moments between different electronic states \citep{Zhang2021}. Previously, multiconfigurational CASSCF calculations have been carried out to compute the permanent dipole moment of PO$^+$ \citep{Moussaoui2003}. However dynamic correlation in the inactive orbitals were not taken into account.

In this sense, we optimized the geometry of PO$^+$(${^1}\Sigma$) with the fully internally contracted CASPT2 method \citep{CASPT2a,CASPT2b}. For the selection of the active orbitals we chose a full valence active space, namely, the four 4$s$ and 4$p$ electrons of phosphorous cation and the six 3$s$ and 3$p$ electrons of oxygen, leaving a (10e,8o) active space. This system has a non-negligible basis set dependence, then, we started the calculation of CASPT2(10e,8o) in combination with the correlation-consistent basis set aug-cc-pVTZ \citep{dunningTZ}, up to the 6Z \citep{dunningQZ,dunning5Z,dunning6Z}. 

Table \ref{tab:calcPO+} shows the  computed equilibrium distance ($r_\mathrm{P-O}$), harmonic frequency ($\nu$), rotational constant ($B{_e}$) and the dipole moment ($\mu$) for the cation PO$^+$(${^1}\Sigma$). The value of the permanent dipole ($\mu$) used in this work is 3.131 D, computed at the full valence CASPT2(10e,8o)/aug-cc-pV6Z. From previous calculations, the most similar result was obtained with the CASSCF(10e,8o) \citep{Moussaoui2003}.

\begin{table}
\centering
\tabcolsep 5pt
\caption{Equilibrium distance ($r_\mathrm{P-O}$), harmonic frequency ($\nu$), rotational constant ($B{_e}$), and the dipole moment ($\mu$) for the cation PO$^+$(${^1}\Sigma$).}
\begin{tabular}{ c c c c c  }
\hline
 Basis set  & $r_\mathrm{P-O}$ (\(\text{\r{A}}\)) & $\nu$ (cm$^{-1}$)  & $B{_e}$ (MHz) & $\mu$ (Debye) \\
\hline
CCSD(T)-F12/cc-pVTZ-F12$^a$ &  1.4283  &  1415.18  &  23481.93937  &  3.043  \\
CASPT2(10e,8o)/aug-cc-pVTZ   &  1.4418  &  1379.44  &  23043.79083  &  3.107  \\ 
CASPT2(10e,8o)/aug-cc-pVQZ   &  1.4339  &  1389.61  &  23300.21209  &  3.120  \\ 
CASPT2(10e,8o)/aug-cc-pV5Z   &  1.4274  &  1402.29  &  23510.57499  &  3.129  \\ 
CASPT2(10e,8o)/aug-cc-pV6Z   &  1.4256  &  1412.27  &  23570.90768  &  3.131  \\ 
Exp. \citep{dyke1982}    &  1.4249  &  1411.50  &----&----\\ 
\hline
\end{tabular}
\label{tab:calcPO+}
\vspace{0mm}
{\\ (a) For the correlation and coulomb fitting the augmented aug-cc-pVQZ was used.}
\end{table}

\clearpage

\section*{Appendix C: Calculation of the SO$^+$ dipole moment}
\label{app:so+dipole}

The SO$^+$ entry (48010, December 1996) of the JPL molecular database assumed a value of 1.0 for the dipole moment. Therefore, the derived column density has to be corrected using the true value of the dipole moment. As in the case of PO$^+$ \citep{Zhang2021}, transition dipole moments between different electronic states has been computed, obtaining theoretical results in agreement with experiments \citep{Houria2006,wei2012}. However, permanent dipole moment measurement for the SO$^+$(${^2}\Pi$) is needed. \citet{turner1992} used a value of the dipole moment $\mu$=2.30 D obtained by Peterson \& Woods (1992).  However, no data about the followed methodology have been found. To obtain a reliable value of the dipole moment, we optimized the SO$^+$ cation in its doublet ground state (${^2}\Pi$) with the above mentioned coupled cluster method but using the resolution-of-identity (RI) approach for the correlation integrals \citep{feyereisen1993} and with the cc-pVQZ-F12 and cc-pVQZ-F12-CABS basis sets. Although the cation SO$^+$(${^2}\Pi$) is an open shell system, the $T{_1}$ diagnostic found for this molecule is 0.022, far from the limit for an open shell system (0.045). Table \ref{tab:calcSO+} shows the comparison of our coupled cluster calculations with the available experimental results. The derived dipole moment for the SO$^+$(${^2}\Pi$) cation obtained and used in this work is 2.016 D, 0.284 D lower than the previous calculation of Peterson \& Woods (1992).

\begin{table}
\centering
\tabcolsep 5pt
\caption{Equilibrium distance ($r_\mathrm{S-O}$), harmonic frequency ($\nu$), rotational constant ($B{_e}$), and the dipole moment ($\mu$) for the cation SO$^+$(${^2}\Pi$).}
\begin{tabular}{ c c c c c  }
\hline
 Method & $r_\mathrm{S-O}$ (\(\text{\r{A}}\)) & $\nu$ (cm$^{-1}$) & $B{_e}$ (MHz) & $\mu$ (Debye) \\
\hline
CCSD(T)-F12/RI/cc-pVQZ-F12${^a}$   & 1.4256 &  1323.19  &  23297.53460  &  2.016\\ 
Exp. (ref)                     & 1.4245 &  1306.78  &  23341.60095  &  ----\\ 
\hline
\end{tabular}
\label{tab:calcSO+}
\vspace{0mm}
{\\ (a) For the correlation and coulomb fitting the augmented aug-cc-pV5Z was used.}
\end{table}

\end{document}